\documentclass{aa}

\usepackage{graphicx}
\usepackage{txfonts}
\usepackage{lipsum}
\usepackage{color}
\usepackage{aalongtable}
\usepackage{lscape}
\usepackage{amstext}
\usepackage[dvipsnames]{xcolor}
\usepackage{tikz}
\usetikzlibrary{positioning}
\usepackage{soul}
\usepackage[
    colorlinks=true,
    linkcolor=black,
    urlcolor=blue,
    citecolor=blue
]{hyperref}

\begin{document} 

\title{CARMENES input catalogue of M dwarfs}
\subtitle{V. Luminosities, colours, and spectral energy distributions\thanks{Table A.3 (summary table) is available at the CDS via anonymous ftp to \url{cdsarc.u-strasbg.fr (130.79.128.5)} or via \url{http://cdsarc.u-strasbg.fr/viz-bin/qcat?J/A+A/MMM/NNN}. An extended version of this table can be downloaded from the GitHub repository \url{https://github.com/ccifuentesr/CARMENES-V}.}}

\author{
    C.~Cifuentes\inst{\ref{cab-vil}}
    \and J.\,A.~Caballero\inst{\ref{cab-vil}}
        \and M.~Cort\'es-Contreras\inst{\ref{cab-vil},\ref{svo}}
        \and D.~Montes\inst{\ref{ucm}}
        \and F.\,J.~Abell\'an\inst{\ref{ucm}}
        \and R.~Dorda\inst{\ref{iac}} 
        \and G.~Holgado\inst{\ref{cab-vil}}
        \and M.\,R.~Zapatero~Osorio\inst{\ref{cab-tor}}
        \and J.\,C.~Morales\inst{\ref{ice},\ref{ieec}} 
        \and P.\,J.~Amado\inst{\ref{iaa}}       
        \and V.\,M.~Passegger\inst{\ref{hs},\ref{ok}}
        \and A.~Quirrenbach\inst{\ref{lsw}}
        \and A.~Reiners\inst{\ref{iag}}
        \and I.~Ribas\inst{\ref{ice},\ref{ieec}} 
        \and J.~Sanz-Forcada\inst{\ref{cab-vil}}
        \and A.~Schweitzer\inst{\ref{hs}}
        \and W.~Seifert\inst{\ref{lsw}}
        \and E.~Solano\inst{\ref{cab-vil},\ref{svo}}
}
        
\institute{ 
        Centro de Astrobiolog\' ia (CSIC–INTA), ESAC, Camino Bajo del Castillo s/n, E-28691 Villanueva de la Ca\~nada, Madrid, Spain \label{cab-vil} \email{ccifuentes@cab.inta-csic.es}
    \and
    Spanish Virtual Observatory, Spain \label{svo}     
    \and
    Departamento de F\'isica de la Tierra y Astrof\'isica \& IPARCOS-UCM (Instituto de F\'isica de Part\'iculas y del Cosmos de la UCM), Facultad de Ciencias F\'isicas, Universidad Complutense de Madrid, E-28040 Madrid, Spain \label{ucm}
    \and
    Instituto de Astrofísica de Canarias (IAC), Calle V\'ia L\'actea, s/n, E-38205 San Crist\'obal de La Laguna, Tenerife, Spain \label{iac}
    \and
        Centro de Astrobiolog\'ia (CSIC-INTA), Carretera de Ajalvir km 4, E-28850 Torrej\'on de Ardoz, Madrid, Spain \label{cab-tor}
    \and      
        Institut de Ci\`encies de l’Espai (CSIC-IEEC), Can Magrans s/n, Campus UAB, E-08193 Bellaterra, Barcelona, Spain \label{ice}
    \and
        Institut d’Estudis Espacials de Catalunya (IEEC), E-08034 Barcelona, Spain \label{ieec}
    \and
        Instituto de Astrof\'isica de Andaluc\'ia (IAA-CSIC), Glorieta de la Astronom\'ia s/n, E-18008 Granada, Spain \label{iaa}
        \and
        Hamburger Sternwarte, Gojenbergsweg 112, D-21029 Hamburg, Germany \label{hs}
        \and
        Homer L. Dodge Department of Physics and Astronomy, University of Oklahoma, 440 West Brooks Street, Norman, OK-73019 Oklahoma, United States of America \label{ok}
    \and
        Landessternwarte, Zentrum f\"ur Astronomie der Universt\"at Heidelberg, K\"onigstuhl 12, D-69117 Heidelberg, Germany \label{lsw}
    \and
        Institut f\"ur Astrophysik, Georg-August-Universit\"at, Friedrich-Hund-Platz 1, D-37077 G\"ottingen, Germany \label{iag}
}

\date{Received 29 April 2020 / Accepted 10 July 2020}


\abstract
{
The relevance of M dwarfs in the search {for} potentially habitable Earth-sized planets has grown significantly in the last years.
}
{
In our on-going effort to comprehensively and accurately characterise confirmed and potential {planet-hosting} M dwarfs, in particular for the CARMENES survey, we have carried out a comprehensive multi-band photometric analysis involving spectral energy distributions, luminosities, absolute magnitudes, colours, and spectral types, from which we have derived basic astrophysical parameters.
}
{
We have carefully compiled photometry in 20 passbands from the ultraviolet to the mid-infrared, and combined it with the latest parallactic distances and close-multiplicity information, mostly from {\em Gaia} DR2, of a sample of 2479 K5\,V to L8 stars and ultracool dwarfs, including 2210 nearby, bright M dwarfs.
For this, we made extensive use of Virtual Observatory tools.
}
{We have homogeneously computed accurate bolometric luminosities and effective temperatures of 1843 single stars, derived their radii and masses, {studied the impact of metallicity}, and compared our results with the literature.
The over 40\,000 individually inspected magnitudes, together with the basic data and derived parameters of the stars, individual and averaged by spectral type, have been made public to the astronomical community.
{In addition}, we {have reported} 40 new close multiple systems and candidates ($\rho <$ 3.3\,arcsec) and 36 overluminous stars that are assigned to young Galactic populations.
}
{
In the new era of exoplanet searches around M dwarfs via transit (e.g. {\em TESS}, {\em PLATO}) and radial velocity (e.g. CARMENES, NIRPS+HARPS), this work is of fundamental importance for stellar and therefore planetary parameter determination.
}

\maketitle


\section{Introduction}
\label{section:introduction}

Low-mass stars are remarkably abundant and long-lived objects in the Galaxy.
Among them, M dwarfs are by far the most common type of star in the solar neighbourhood, vastly outnumbering their more massive counterparts \citep{Hen94,Hen06,Rei04,Boc10,Win15}.
In their mainly convective interiors, the fusion process is slow and, therefore, the lifespan is long, as they remain on the main sequence for tens of billions of years \citep{Ada97,Bar98}.
Such abundance and prevalence make low-mass stars very attractive targets for multiple areas of astrophysical research.

Collectively, M dwarfs are excellent probes for the examination of the Galactic structure \citep{Bah80,Sca86,Rei97,Cha03b,Pir05,Cab08,Fer17}, {and are also} very convenient tracers of Galactic kinematics and evolution \citep{Rei95,Giz02,Wes06,Boc07}.
Individually, M dwarfs have proven to be interesting targets for the discovery of low-mass exoplanets, and a sizable body of current literature pays special attention to them \citep[e.g.][]{Bos06,Tar07,Zec09,Bon13,Man13b,Cla14,Dre15,Fis16,Kop17,Rei18b}.
In particular, low-mass, small-sized stars are particularly suited to the search for close-in terrestrial planets because their detection becomes easier with decreasing stellar size and planetary orbital period
\citep{Ang16,Gil17,Zec19}. 

Our understanding of how planets form and evolve rests fundamentally on the characterisation of their host stars. 
As an example, the luminosity of the star determines the equilibrium temperature of its planet and delimits the habitable zone, which is the circumstellar region where water can be liquid \citep[][but see \citealt{Tar07} for the particular M-dwarf case]{Kas93,Kop17}.  
Determining precise stellar parameters of M dwarfs and how their uncertainties propagate to those of their planets is, therefore, of paramount importance.
Many efforts have been undertaken in this respect, including empirical determination of masses, radii, and their relation to luminosity 
\citep{Vee74,Hen93,Cha97,Del00,Bon05,Man15,Man19,Ter15,Ben16,Sch19}, 
effective temperature, surface gravity, and metallicity \citep{Cas08,Roj13,Mon18,Pas18,Pas19,Raj18b}, or activity and rotation periods \citep{Sta86,Rei95,Haw96,Mor08,Haw14,New15,Jef18,Die19,Sch19b}.

This work is part of a series of papers devoted to describing the CARMENES input catalogue of M dwarfs.
CARMENES, the Calar Alto high-Resolution search for M dwarfs with Exoearths with Near-infrared and optical Echelle Spectrographs\footnote{\url{http://carmenes.caha.es}} \citep{Qui14}, is the name of an instrument specifically designed for discovering M-dwarf planets with the radial-velocity method, the consortium that built it, and of the science project that is being carried out during guaranteed time observations \citep[GTO;][]{Qui18,Rei18b}. 
Here we continue the work started by \citet{Alo15a} on spectral typing from low-resolution spectroscopy of M dwarfs (I), and followed up by \citet{Cor17} on multiplicity from high-resolution lucky imaging (II), \citet{Jef18} on activity from high-resolution spectroscopy (III), and \citet{Die19} on rotation periods from photometric time series~(IV).

In this fifth item of the series, we focus on the analysis of  multi-wavelength photometry, from the far ultraviolet to the mid infrared, of a large sample of nearby, bright M dwarfs, including those monitored by CARMENES, as well as some late K dwarfs and early and mid L dwarfs.
We derive accurate bolometric luminosities, identify new close binaries, members in young stellar kinematic groups, and other outliers in colour-colour, colour-magnitude, and colour-spectral type diagrams.
We also explore different relationships between colours, absolute magnitudes, spectral types, luminosities, masses, and radii. 
For that, we make extensive use of the second data release of {\em Gaia} astrometry and photometry \citep[{\em Gaia} DR2;][]{Gaia18bro}, numerous public all-sky surveys from the ground and space, and Virtual Observatory tools such as the Aladin interactive sky atlas \citep{Bon00}, the Tool for OPerations on Catalogues And Tables \citep[TOPCAT;][]{Tay05}, and the Virtual Observatory Spectral energy distribution Analyser \citep[VOSA;][]{Bay08}.

Our work is also connected to that of \citet{Sch19}, who derived masses and radii from effective temperatures (determined from  spectral synthesis) and luminosities (measured exactly as in the present paper) for 293 M dwarfs monitored by CARMENES.
As a result, here we complement the description of the calculation of stellar masses and radii of all planet hosts detected by CARMENES \citep[e.g.][ to cite a few]{Rei18,Rib18,Tri18,Zec19,Luq19,Mor19}. 


\section{Data}
\label{section:data}
    
In this Section we describe the building process of our sample, as well as the compilation of their photometric and astrometric data from public catalogues.

\subsection{Sample}

\label{ssection:sample}

\begin{figure}[]
    \centering
    \includegraphics[width=1 \linewidth]{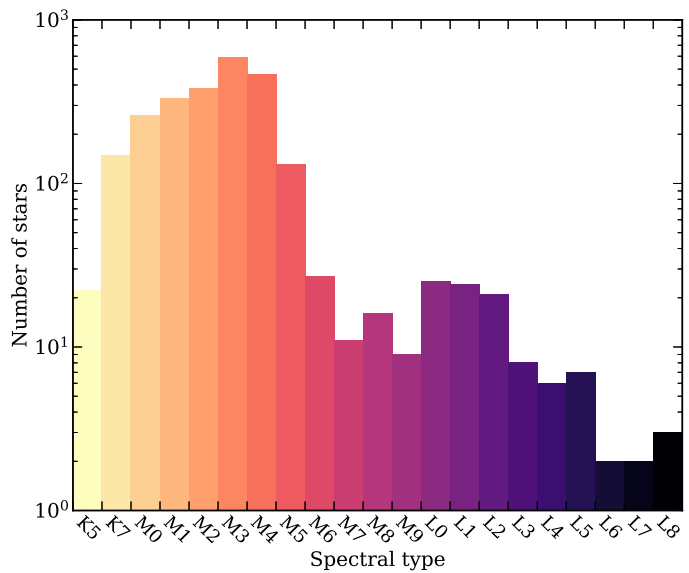}    
    \caption{Distribution of spectral types in our sample.}
    \label{fig:spectraltypes}
\end{figure}

Our sample is based mainly on Carmencita, the CARMENES input catalogue \citep{Alo15a,Cab16}.
Currently, Carmencita contains 2191 M dwarfs and 3 K dwarfs, namely J04167--120 (\object{LP~714--47}), \object{J11110+304E} (\object{HD~97101~A}), and \object{J18198--019} (\object{HD~168442}), which satisfied simple selection criteria based on $J$-band magnitude and spectral type regardless of multiplicity, age, or metallicity \citep[cf.][]{Alo15a}. 
Except for the three K dwarfs, Carmencita includes M dwarfs visible from the Calar Alto Observatory in Southern Spain ($\delta \gtrsim$ --23\,deg) with spectral types from M0.0\,V to M9.0\,V and near-infrared brightnesses between $J$ = 4.2\,mag and 11.5\,mag. 
The spectral types, compiled by \citet{Cab16}, came from a number of sources.
However, the spectral types of {2028} M dwarfs (92.5\,\%) were taken from only three references: \citet{Haw02}, \citet{Lep13}, and \citet{Alo15a}, which are equivalent among them according to the latter authors.
Of the remaining {163} M dwarfs, most spectral types also came from reliable, equivalent sources \citep[e.g.][]{Gra03,Sch05,Ria06}, which assures a relative homogeneity in our sample.

{As described in the references above, Carmencita is unbiased except for the fact that it may include overluminous and lack underluminous stars in the $J$ band at a fixed spectral type.
This fact probably translates into a larger fraction of (overluminous) close multiples and young active stars, and a lower fraction of (underluminous) very low metallicity M-type dwarfs \citep[subdwarfs and extreme subdwarfs;][]{Giz97,Lep07b}.
From the distribution of the $\zeta$ index, a metallicity proxy measured in low-resolution spectra of a large number of Carmencita stars \citep[cf.][]{Alo15a}, we extrapolated that most of our M dwarfs have solar-like metallicities, but that there could be a significant number of them with [Fe/H] $<$ --1.0.
}

\begin{figure}[]
    \centering
    \includegraphics[width=.99\linewidth]{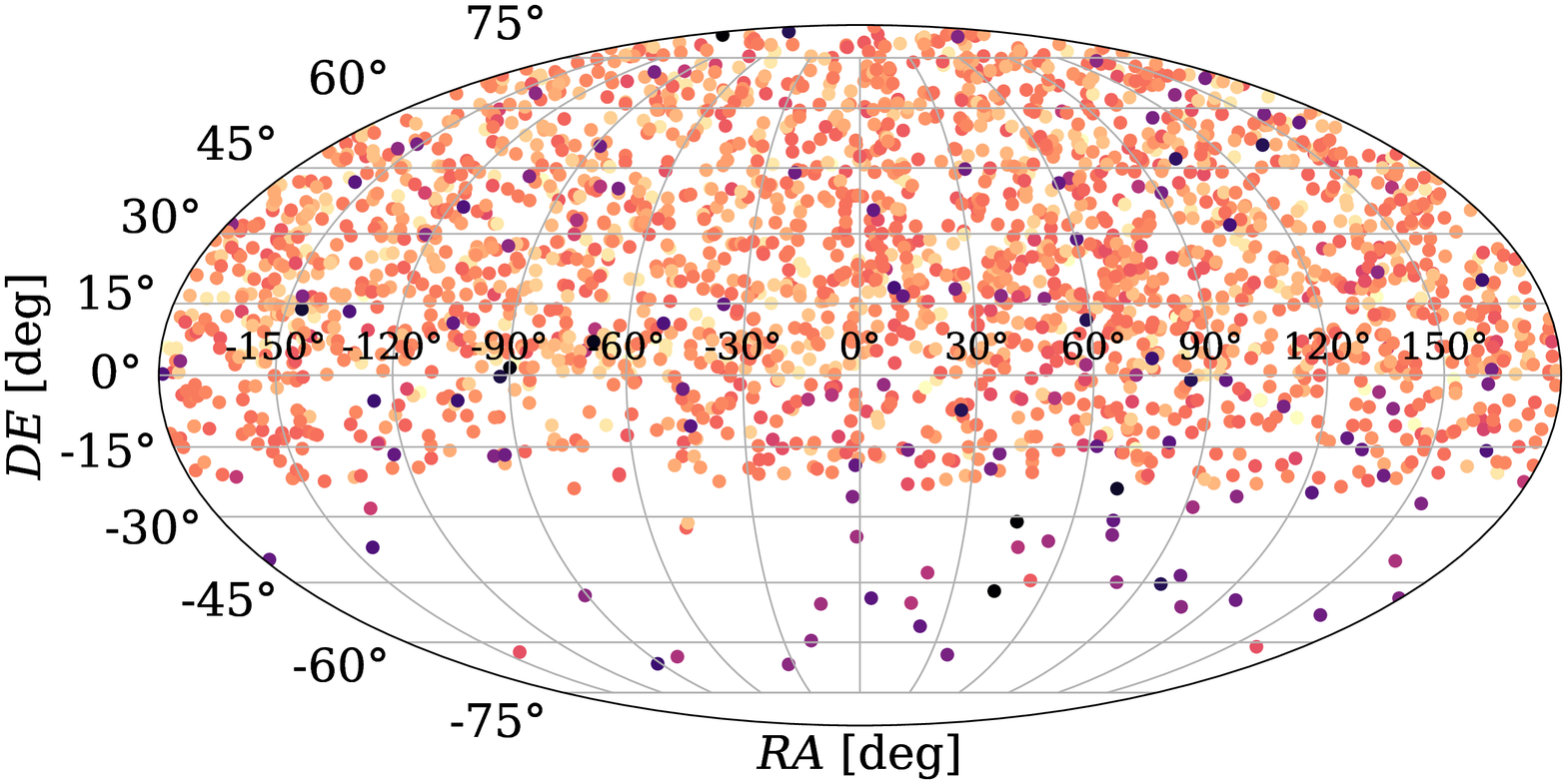}
    \includegraphics[width=.99\linewidth]{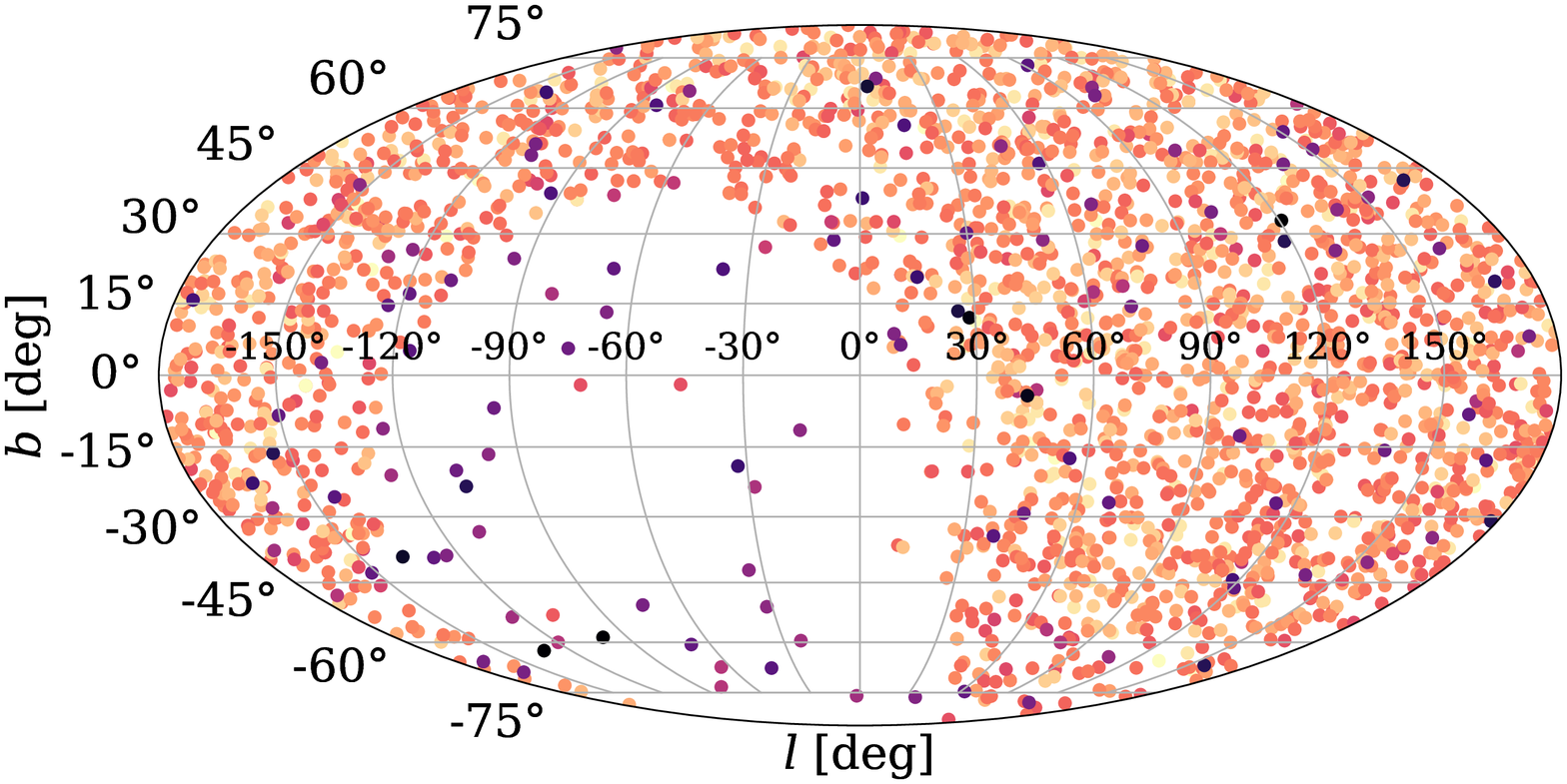}
    \includegraphics[width=.99\linewidth]{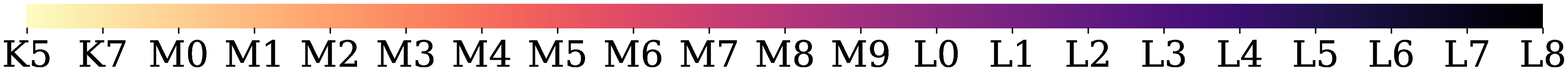}
    \caption{Location in the sky of the 2479 targets in our sample, colour-coded by spectral type, in equatorial ({top}) and Galactic coordinates ({bottom}).
    We note the absence of Carmencita M dwarfs with declinations lower than $\delta$ = --23\,deg.
    }
    \label{fig:RA_DE}
\end{figure}

In order to extend the photometric sample to a wider spectral range and to avoid any boundary value problem, we complemented Carmencita with additional stars earlier than M0.0\,V, and with stars and brown dwarfs later than M6.5\,V.
The eventual distribution of spectral types is displayed in Fig.~\ref{fig:spectraltypes}.
On the warm side, we included 168 bright stars with spectral types between K5\,V and K7\,V from
\cite{Kir91},
\cite{Lep13},
and \cite{Alo15a},
and the RECONS list of the 100 nearest stars\footnote{\href{http://www.recons.org/TOP100.posted.htm}{\tt http://www.recons.org/TOP100.posted.htm}} \citep{Hen06}.
We did not include the very bright K stars \object{$\eta$~Cas~B}, \object{36~Oph~C}, \object{BD+01~3942~A}, \object{$\xi$~Cap~B},  \object{61~Cyg~A}, and \object{61~Cyg~B}, whose photometry is strongly affected by saturation or blending due to close multiplicity.

On the cool side, we first included seven M5.0--9.0\,V stars from the REsearch Consortium On Nearby Stars (RECONS) with declinations of $\delta <$ --23\,deg.
Next, we added 110 ultracool dwarfs from \citet{Sma17} with a Two Micron All-Sky Survey (2MASS) near-infrared counterpart \citep{Skr06} and relative error in {\em Gaia} DR2 parallaxes ($\delta \varpi / \varpi$) less than 1\,\%.
That addition made 12 M8.0--9.5\,V and 98 L0.0--8.0 ultracool dwarfs.  
We did not include four T-type brown dwarfs (\object{SIMP~J013656.57+093347.3}, \object{ULAS~J141623.94+134836.30}, \object{2MASS~15031961+2525196}, and \object{WISE~J203042.79+074934.7}) and one L dwarf, \object{HD~16270~B}, because of their poor 2MASS photometric quality (see Sect.~\ref{ssection:photometry}). 

As a result, the joint K-M-L spectro-photometric sample contained 2479 targets distributed among 171 late-K dwarfs, 2210 M dwarfs, and 98 L dwarfs. 
For all targets in the sample we employed and tabulated equatorial coordinates from {\em Gaia} DR2 except for the 58 stars (five K, 53 M) that were not catalogued by the ESA space mission.
For all 58 stars, we used the positions at the epoch of 2MASS projected to the epoch J2015.5 with proper motions from \cite{van07} and \cite{Zac12}, as compiled by \cite{Cab16}.  

The spatial distribution of the 2479 targets is illustrated in Fig.~\ref{fig:RA_DE}.
For the sake of simplicity, we will use hereafter the term ``stars'' for the {2479} objects in our sample, including the stellar and substellar objects later than M7\,V, also known as ultracool dwarfs \citep{Kir97}.


    \subsection{Photometry}
    \label{ssection:photometry}

\begin{figure}[]
    \centering
    \includegraphics[width=0.99\linewidth]{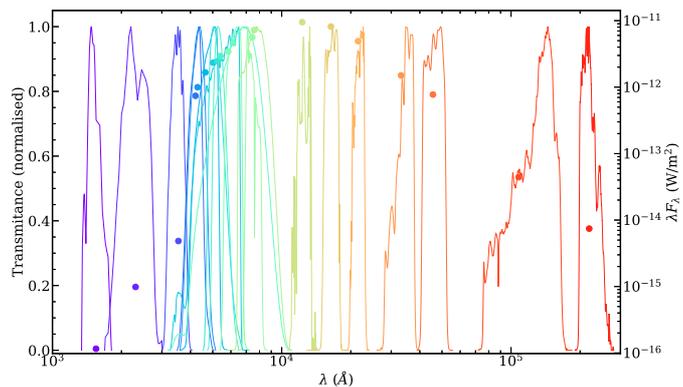}
    \caption{Normalised transmission curves of the 20 passbands employed for the compilation of photometry, {taken from the SVO Filter Profile Service.}
    For comparison, coloured filled circles depict the spectral energy distribution of \object{DS~Leo} (Karmn~J11026+219, {M1.0\,V}).}
    \label{fig:filters}
\end{figure}

\begin{table*}[!h]
\caption{Passbands employed for the compilation of photometry.}
\label{tab:filters}
\centering
\footnotesize{
\begin{tabular}{lcccll}
            \hline
            \hline 
            \noalign{\smallskip}
            Band & $\lambda_{\rm eff}$ & $W_{\rm eff}$ & $F_\lambda^0$ &  Survey$^{a}$ & { Description } \\
                & (\AA) & (\AA) & (W\,m$^{-2}$\,\AA$^{-1}$) & & \\
            \noalign{\smallskip}
            \hline
            \noalign{\smallskip}
        \textit{FUV}    & 1549.0   & 265.6 & $6.491 \times 10^{-12}$  & {\it GALEX} & { GALEX {\em FUV} } \\
        \textit{NUV}    & 2304.7  & 768.3 & $4.450 \times 10^{-12}$  & {\it GALEX}  & { GALEX {\em NUV} } \\
        $u'$    & 3594.9 & 558.4 & $ 3.639 \times 10^{-12}$ &  SDSS9 & { SDSS $u'$ full transmission } \\
        $B_T$   & 4206.4 & 708.4 & $ 6.598 \times 10^{-12}$   &  Tycho-2  & { Tycho $B$ } \\
        $B$     & 4297.2 & 843.1 & $6.491 \times 10^{-12}$ & UCAC4  & { UCAC4 $B$ filter, defined as identical to {\tt GCPD/Johnson.B\_Landolt} } \\   
                        & 4297.2 & 843.1 & $6.491 \times 10^{-12}$ &  APASS9  & { APASS $B$ filter, defined as identical to {\tt GCPD/Johnson.B\_Landolt} } \\
        $g'$    & 4640.4 & 1158.4 & $ 5.521 \times 10^{-12}$ &  UCAC4   & { UCAC4 $g'$ filter, defined as identical to {\tt MISC/APASS.sdss\_g} and {\tt SLOAN/SDSS.g} } \\
                        & 4640.4 & 1158.4 & $5.521 \times 10^{-12}$ &  SDSS9 & { SDSS $g'$ full transmission } \\
                        & 4640.4 & 1158.4 & $ 5.521 \times 10^{-12}$ &  APASS9  & { APASS $g'$ filter, defined as identical to {\tt SLOAN/SDSS.g} } \\ 
                        & 4810.8 & 1053.1 & $5.043 \times 10^{-12}$ &  PS1 DR1  & { PS1 $g'$ filter } \\ 
        $G_{B_P}$       & 5020.9  & 2279.5 & $ 4.035 \times 10^{-12}$ & {\it Gaia} DR2  & { {Gaia} $G_{BP}$ filter, DR2 revised curve } \\ 
        $V_T$   & 5243.9 & 1005.7 & $ 3.984 \times 10^{-12}$   & Tycho-2  & { Tycho $V$} \\      
        $V$     & 5394.3 & 870.6 & $3.734 \times 10^{-12}$ & UCAC4 & {  UCAC4 $V$ filter, defined as identical to {\tt GCPD/Johnson.V\_Landolt} } \\ 
                        & 5394.3 & 870.6 & $3.734 \times 10^{-12}$ & APASS9 & { APASS $V$ filter, defined as identical to {\tt GCPD/Johnson.V\_Landolt} } \\
        $r'$    & 6122.3 & 1111.2 & $2.529 \times 10^{-12}$ & UCAC4 & { UCAC4 $r'$ filter, defined as identical to {\tt MISC/APASS.sdss\_r} and {\tt SLOAN/SDSS.r} } \\
                        & 6122.3 & 1111.2 & $2.529 \times 10^{-12}$ & SDSS9  & {  SDSS $r'$ full transmission } \\
                        & 6122.3 & 1111.2 & $2.529 \times 10^{-12}$ & APASS9  & {  APASS $r'$ filter, defined as identical to {\tt SLOAN/SDSS.r} } \\ 
                        & 6122.3 & 1318.1 & $2.529 \times 10^{-12}$ & CMC15 & {  SDSS $r'$ full transmission } \\
                        & 6156.4 & 1252.4 & $2.480 \times 10^{-12}$ &  PS1 DR1  & {  PS1 $r'$ filter } \\
        $G$     & 5836.3 & 4358.4 & $2.495 \times 10^{-12}$ & {\it Gaia} DR2  & { {Gaia} $G$ filter, DR2 revised curve } \\  
        $i'$    & 7439.5 & 1044.6 & $1.409 \times 10^{-12}$ & UCAC4  & { UCAC4 $i'$ filter, defined as identical to {\tt MISC/APASS.sdss\_i} and {\tt SLOAN/SDSS.i} } \\
                        & 7439.5 & 1044.6 & $1.409 \times 10^{-12}$ & SDSS9 & { SDSS $i'$ full transmission } \\
                        & 7439.5 & 1044.6 & $1.409 \times 10^{-12}$ & APASS9  & { APASS $i'$ filter, defined as identical to {\tt SLOAN/SDSS.i} } \\ 
                        & 7503.7 & 1206.6 & $1.372 \times 10^{-12}$ & PS1 DR1 & { PS1 $i'$ filter } \\
        $G_{R_P}$ & 7588.8 & 2943.7 & $ 1.294 \times 10^{-12}$ & {Gaia} DR2  & { {\it Gaia} $G_{RP}$ filter, DR2 revised curve } \\   
        $J$     & 12285.4 & 1624.2 & $3.143 \times 10^{-13}$  & 2MASS  & { 2MASS $J$ } \\
        $H$  & 16386.1 & 2509.4 & $1.144 \times 10^{-13}$  & 2MASS  & { 2MASS $H$ } \\
        $K_s$  & 21521.6 & 2618.9 & $4.306 \times 10^{-14}$  & 2MASS  & { 2MASS $K_s$ } \\
    $W1$& 33156.6 & 6626.4 & $8.238 \times 10^{-15}$  & AllWISE & { WISE $W1$ filter } \\
                        & 33156.6 & 6626.4 & $8.238 \times 10^{-15}$ & WISE  & { WISE $W1$ filter } \\
    $W2$& 45644.9 & 10422.7 & $2.431 \times 10^{-15}$  & AllWISE  & { WISE $W2$ filter } \\
                        &  45644.9 & 10422.7 & $2.431 \times 10^{-15}$  & WISE & { WISE $W2$ filter } \\
    $W3$& 107868.4 & 55055.7 & $6.570 \times 10^{-17}$  & AllWISE  & { WISE $W3$ filter } \\
                        & 107868.4 & 55055.7 & $6.570 \times 10^{-17}$  & WISE  & { WISE $W3$ filter } \\
    $W4$& 219149.6 & 41016.8 & $4.995 \times 10^{-18}$  & AllWISE  & { WISE $W4$ filter } \\
                        & 219149.6 & 41016.8 & $4.995 \times 10^{-18}$ & WISE  & { WISE $W4$ filter } \\
\noalign{\smallskip}
\hline
\end{tabular}
}
\tablefoot{
    \tablefoottext{a}{
        {\em GALEX} DR5: Galaxy Evolution Explorer, \cite{Bia11};
        SDSS DR9: Sloan Digital Sky Survey, \cite{Ahn12};
        UCAC4: The fourth U.S. Naval Observatory CCD Astrograph Catalog, \cite{Zac12};
        Pan-STARRS1: Panoramic Survey Telescope and Rapid Response System, \cite{Kai10}, \cite{Ton12}, and \cite{Cha16};
        CMC15: Carlsberg Meridian Catalogue, \cite{CMC15};
        APASS9: The AAVSO Photometric All-Sky Survey, \cite{Hen16};
        Tycho-2: \cite{Hog00};
        2MASS: Two Micron All-Sky Survey, \cite{Skr06};
        {\it Gaia} DR2: \cite{Gaia16bro,Eva18} with the revised response curves of \cite{Mai18}; 
        AllWISE: \cite{Cut14};
        WISE: Wide-field Infrared Survey Explorer, \cite{Cut12}.}
}
\end{table*}

\begin{figure}[]
    \centering
    \includegraphics[width=.99\linewidth]{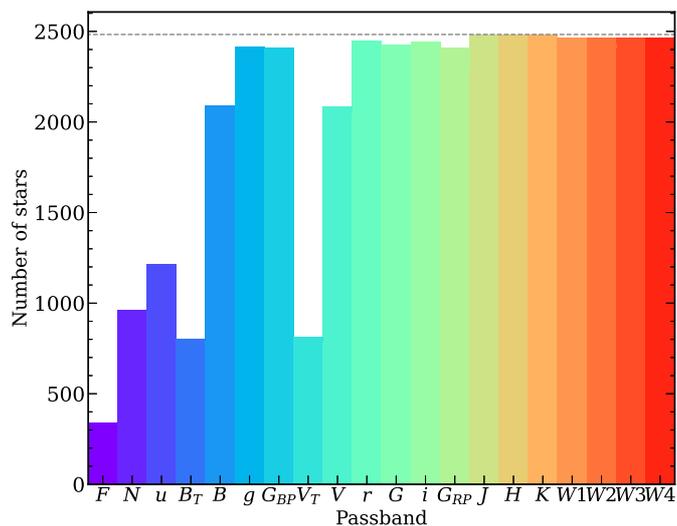}       
    \caption{Completeness in every passband. 
    Light shaded regions account for measurements with poor quality flags.
    The dashed horizontal line indicates the total number of stars in the sample.}
    \label{fig:hist_filters}
\end{figure}

\begin{figure}[]
    \centering
    \includegraphics[width=.99\linewidth]{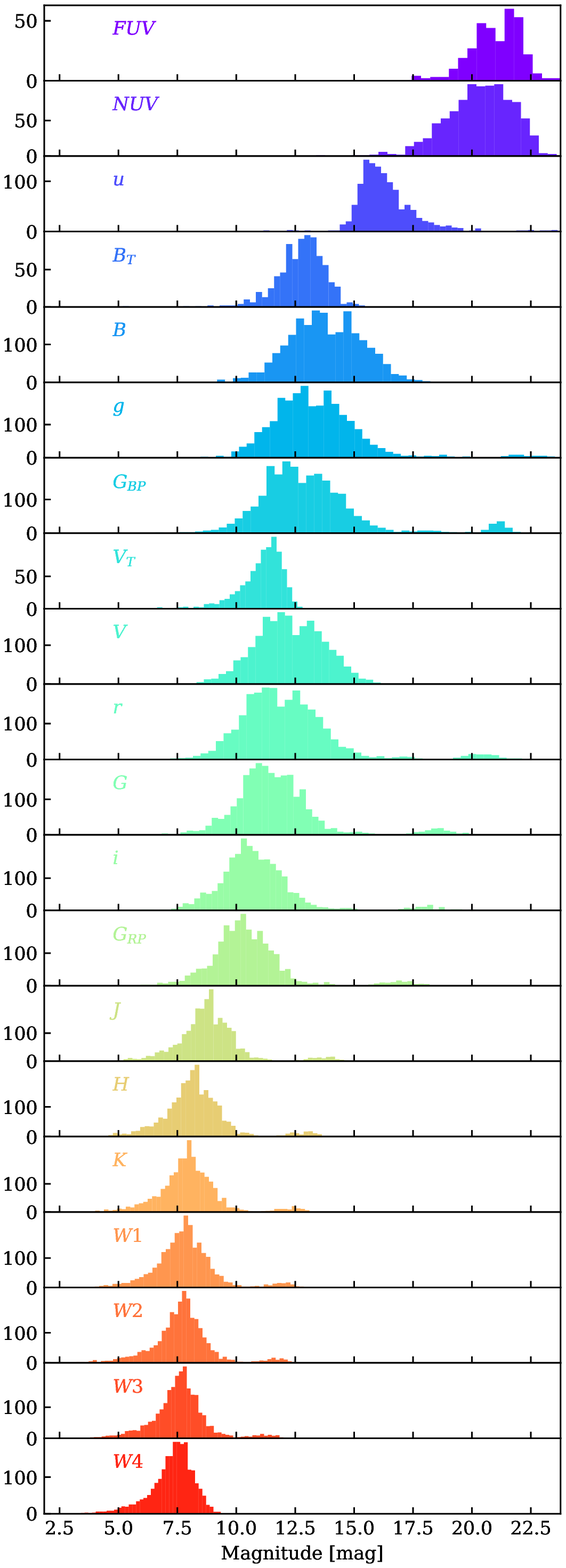}    
    \caption{Distribution of compiled magnitudes in every passband.
    The width of the bins follows the Freedman-Diaconis rule.}
    \label{fig:hist_magnitudes}
\end{figure}

For every star in the sample, we compiled multiwavelength broadband photometry covering a wide spectral range from the ultraviolet to the mid-infrared, as illustrated in Fig.~\ref{fig:filters}.
First of all, with Aladin we manually retrieved the 2MASS equatorial coordinates, $JHK_s$ magnitudes and uncertainties, and photometric quality flags of all 2479 stars (we had done this previously for the Carmencita stars; \citealt{Cab16}).
Next, we added photometric data from different public catalogues.
We started by adding {\em Gaia} DR2 $G$, $G_{BP}$, and $G_{RP}$ magnitudes, obtained with the query form available in the {\it Gaia} Archive\footnote{\tt \url{http://gea.esac.esa.int/archive/}}.
We followed by adding magnitudes and uncertainties of
the Galaxy Evolution Explorer ({\em GALEX}) $FUV$ and $NUV$, 
the Ninth Sloan Digital Sky Survey Data Release (SDSS9) $u'g'r'i'$, 
Tycho-2 $B_T$ and $V_T$, 
the AAVSO Photometric All-Sky Survey Data Release 9 (APASS9) $B$ and $V$, 
the Fourth US Naval Observatory CCD Astrograph Catalog (UCAC4) $BVg'r'i'$, 
the Carlsberg Meridian Catalogue 15 (CMC15) $r'$, 
and of the Wide-field Infrared Survey Explorer (AllWISE and WISE) $W1W2W3W4$ (and their quality flags when available).
For that, we used the TOPCAT automatic positional cross-match tool {\tt CDS X-match} with a search radius of 5\,arcsec and the ``All'' find option.
For a few high proper-motion stars, we enlarged the search radius to 10\,arcsec.
Next we used Aladin to:
($i$) visually inspect the automatic cross-matches of all sources (and correct them, especially in mismatched cases of high proper motion and close binary sources), and 
($ii$) compile, by hand, the most reliable photometry of Pan-STARRS1 DR1 only for the stars for which $g'$, $r'$, or $i'$ magnitudes were not available in other catalogues (PS1 DR1 delivered up to 60 multi-epoch observations for every star over three years in the five PS1 passbands).
The passband name, effective wavelength $\lambda_{\rm eff}$, effective width $W_{\rm eff}$, zero point flux $F_\lambda^0$, 
survey acronym, and corresponding references of the {20} compiled passbands are listed in Table~\ref{tab:filters}.
The passband parameters were calculated by VOSA with the latest filter transmission curves available at the Filter Profile Service\footnote{\tt \url{http://svo2.cab.inta-csic.es/theory/fps/}} of the Spanish Virtual Observatory. 
When there were several surveys providing photometric data in the same passband (e.g. $r'$ in UCAC4, SDSS9, APASS9, and PS1 DR1), we prioritised the surveys with the highest spatial resolution, sensitivity, and accuracy.
PanSTARRS1 DR1 has slightly different passband parameters from those of the other $g'r'i'$ surveys.
Virtually all our K and M dwarfs saturated or were in the non-linear regime in SDSS9 $z'$ and PS1 DR1 $z'y'$, so we did not compile data in these passbands.

{\em Gaia} $G$, $G_{B_P}$, and $G_{R_P}$ magnitude uncertainties were derived from the uncertainties in the fluxes, while UCAC4 {\em BVg'r'i'} magnitude uncertainties were collected from an additional TOPCAT {table access protocol query}. 
However, we chose APASS9 $BV$ over UCAC4 $BV$ when the UCAC4 uncertainties were 0.00\,mag, 0.99\,mag, or missing.
In the case of poor photometric quality in AllWISE {\em W1} to {\em W4} ({\tt Qflag} $\neq$ A,B), we chose the data available in WISE when it improved the quality of AllWISE data.
{We also identified possible flux excesses in the {\em Gaia} DR2 $G_{BP}$ and $G_{RP}$ photometric data with the keyword \texttt{phot\_bp\_rp\_excess\_factor}, following the guidelines of \cite{Eva18} to separate well-behaved single sources from spurious ones.}

$J$ band magnitudes are available for all the stars in the sample, and the completeness in passbands $g'$, $G_{BP}$,  $G$, $r'$, $i'$, $G_{RP}$, $H$, $K_s$, $W1$, $W2$, $W3$, and $W4$ is greater than 97\,\%.
For Johnson $B$ and $V$ the completeness is around 86\,\%, whereas for Tycho-2 $B_T$ and $V_T$ it is only 25\,\%.
At the blue end, $u'$ is complete for 50\,\% of the sample, and the ultraviolet passbands $FUV$ and $NUV$ are available for 39\,\% and 14\,\%, respectively.
This is graphically summarised in Fig.~\ref{fig:hist_filters}.

In total, we collected 40\,094 individual magnitudes.
Of them, 39\,896 have magnitude uncertainties and 33\,594 have good quality photometry, defined as: 2MASS {\tt Qflag} $=$ A (with signal-to-noise ratio $\geq$10), WISE {\tt Qflag} $=$ A,B, $G_{BP} <$ 19.5\,mag (see Sect.~\ref{sssection:diagrams_colour_colour}), and no flux excess in {\em Gaia} $G_{BP}$ and $G_{RP}$.
Figure~\ref{fig:hist_magnitudes} shows the distribution of magnitudes for each band, ordered by increasing $\lambda_{\rm mean}$.
The distributions of the bluest bands are broader than the reddest ones, while those of the most complete bands (e.g. $g'$, $r'$, $G$, $J$, $W1$) exhibit small secondary peaks at fainter magnitudes, which correspond to late M and early L dwarfs.


    \subsection{Distances}
    \label{ssection:distances}

\begin{table}[]
\caption{Reference of the 2425 parallactic distances in the sample.}
\label{tab:distances}
\centering
\begin{tabular}{lcl}
            \hline
            \hline 
            \noalign{\smallskip}
            Acronym$^a$ & Number of stars & Reference \\
            \noalign{\smallskip}
            \hline
            \noalign{\smallskip}
            Gaia2 & 2306 & \citealt{Gaia18bro} \\
            HIP2 & 41 & \citealt{van07} \\
            Dit14 & 34 & \citealt{Dit14} \\
            vAl95 & 16 & \citealt{van95} \\
            FZ16 & 14 & \citealt{Fin16} \\
            Galli18 & 2 & \citealt{Gal18} \\
            Hen06 & 2 & \citealt{Hen06} \\
            Jao05       & 2 &  \citealt{Jao05} \\
            Wein16 & 2 & \citealt{Wei16} \\  
            Dahn17 & 1 & \citealt{Dah17} \\                            
            GC09   & 1 & \citealt{Gat09} \\
            Jen52 & 1 & \citealt{Jen52} \\
            Lep09 & 1 & \citealt{Lep09} \\          
            Ried10 & 1 & \citealt{Rie10} \\                       
            TGAS & 1 & \citealt{Gaia16bro} \\         
\noalign{\smallskip}
\hline
\end{tabular}
\tablefoot{
\tablefoottext{a}{Acronyms used in on-line table.}
}
\end{table}

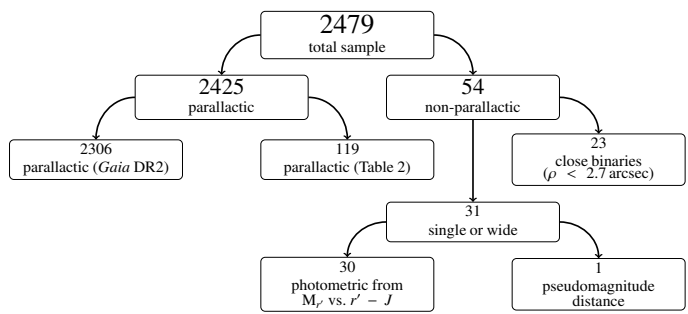
\begin{figure}
    \centering
    \resizebox{1\columnwidth}{!}{%
        \tikzstyle{block} = [rectangle, draw, 
        text width=15em, text centered, rounded corners, minimum height=2em]
        \tikzstyle{line} = [draw, -latex'new]
        \begin{tikzpicture}[node distance = 2cm, auto]\label{ams1}
        \node [block] at (0, 0) (c1) {{\Huge {2479}} \\ \vspace{0.2cm} {\Large total sample}};
        \node [block] at (-4, -2) (c2) {{\huge {2425}}\\ \vspace{0.2cm} {\Large parallactic}};
        \node [block] at (4, -2) (c3) {{\huge {54}}\\ \vspace{0.2cm} {\Large non-parallactic}};
        \node [block] at (-8, -4) (c4) {{\Large {2306}}\\ \vspace{0.2cm} {\Large parallactic ({\em Gaia} DR2)}};
        \node [block] at (0, -4) (c5) {{\Large {119}}\\  \vspace{0.2cm} {\Large parallactic 
        (Table~\ref{tab:distances})}};
        \node [block] at (4, -6) (c6) {{\Large {31}}\\ \vspace{0.2cm} {\Large single or wide}};       
        \node [block] at (0, -8) (c7) {{\Large {30}}\\ \vspace{0.2cm} {\Large photometric from M$_{r'}$ vs. $r'-J$}}; 
        \node [block] at (8, -8) (c8) {{\Large {1}}\\ \vspace{0.2cm} {\Large pseudomagnitude distance}};     
        \node [block] at (8, -4) (c9) {{\Large {23}}\\ \vspace{0.2cm} {\Large close binaries ($\rho<2.7$\,arcsec)}};
        \draw[->, line width=1.5pt] (c1.west) to[in=90, out=180] (c2.north);
        \draw[->, line width=1.5pt] (c1.east) to[in=90, out=0] (c3.north);
        \draw[->, line width=1.5pt] (c2.west) to[in=90, out=180] (c4.north);
        \draw[->, line width=1.5pt] (c2.east) to[in=90, out=0] (c5.north);
        \draw[->, line width=1.5pt] (c3.south) to[in=90, out=270] (c6.north);
        \draw[->, line width=1.5pt] (c3.east) to[in=90, out=0] (c9.north);
        \draw[->, line width=1.5pt] (c6.west) to[in=90, out=180] (c7.north);
        \draw[->, line width=1.5pt] (c6.east) to[in=90, out=0] (c8.north);
        \end{tikzpicture}
        }
        \caption{Schematic diagram of sources of heliocentric distances.}
        \label{fig:schema}
 \end{figure}

\begin{figure}[]
    \centering
    \includegraphics[width=1 \linewidth]{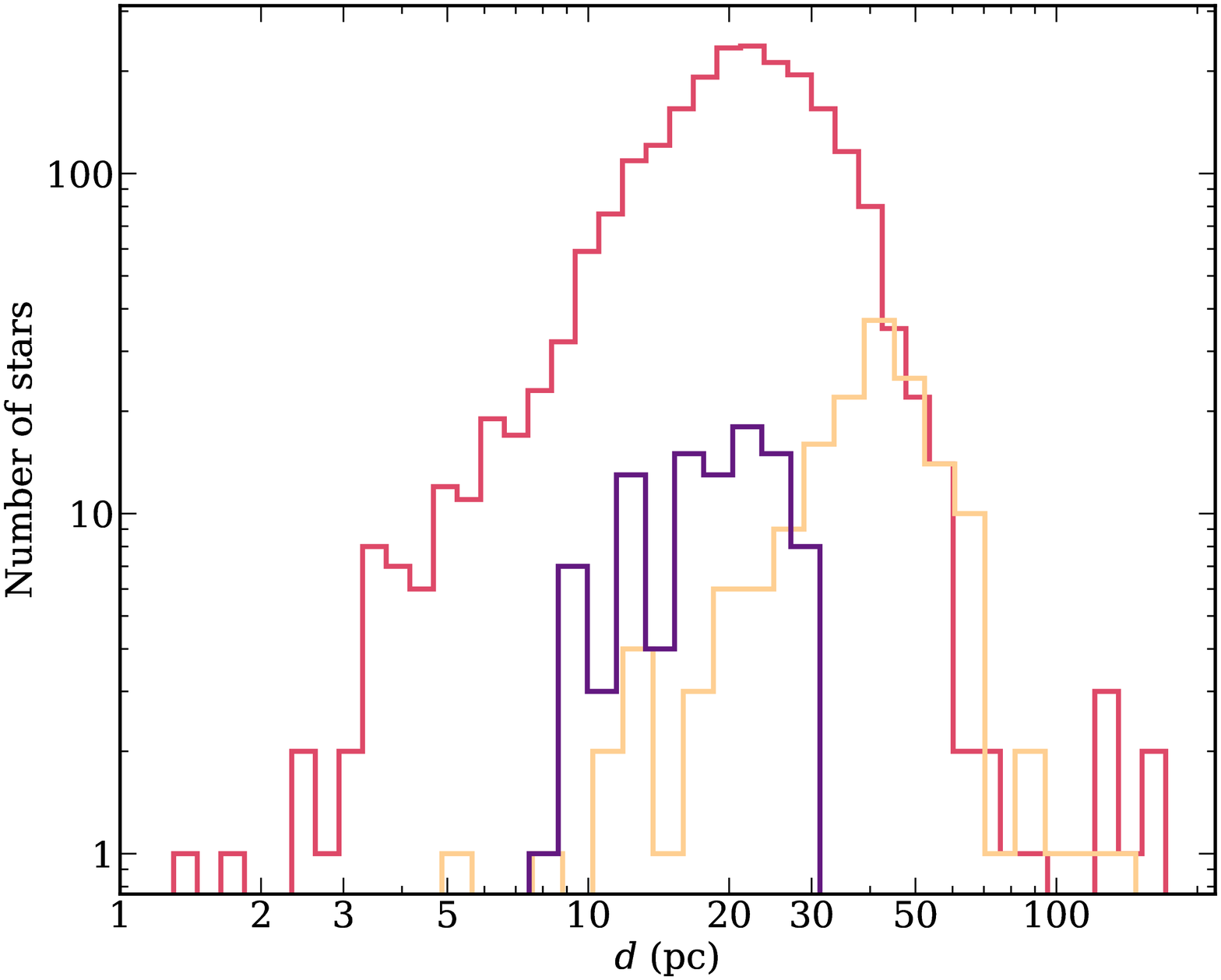}
    \includegraphics[width=1 \linewidth]{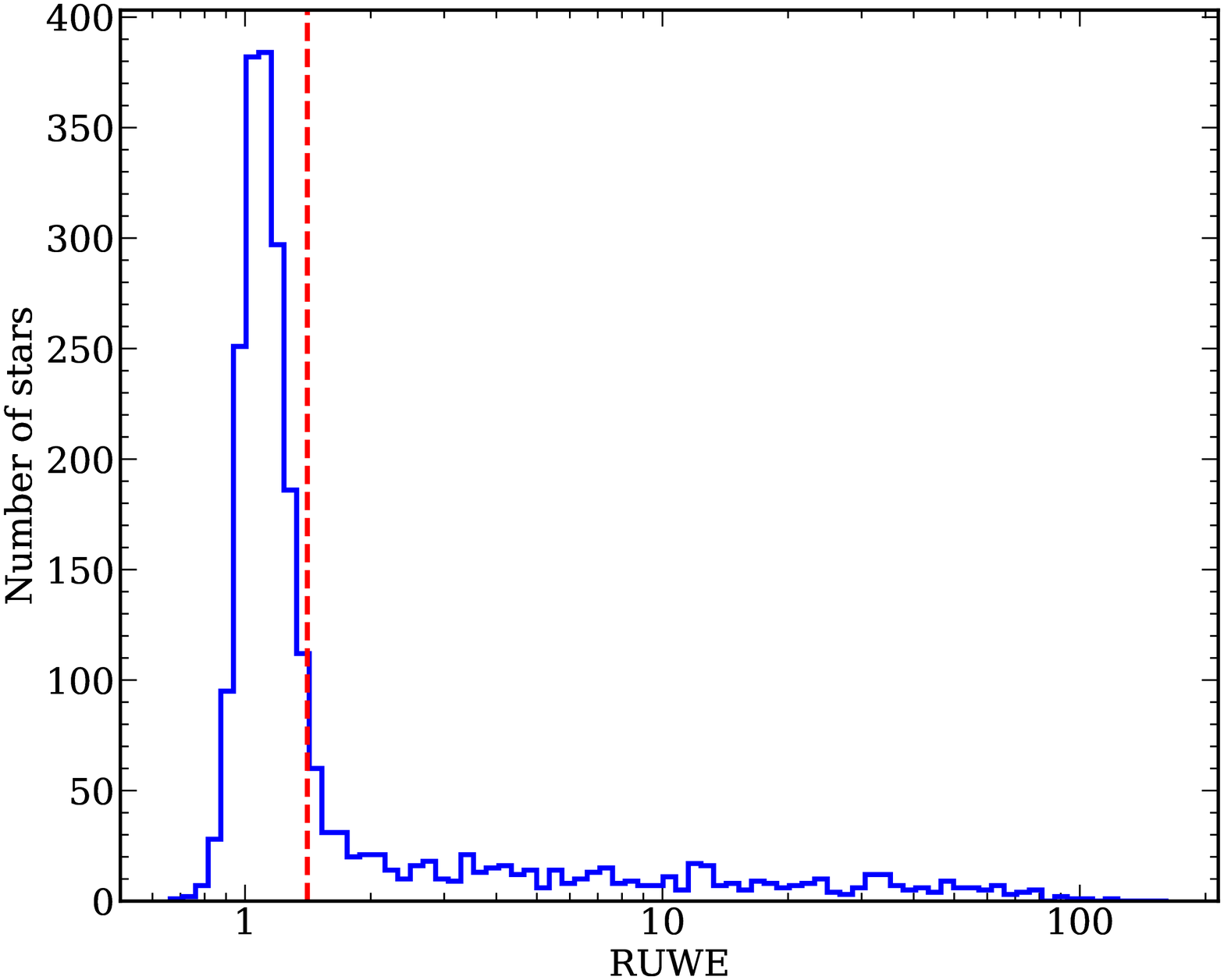}   
    \caption{
    Histogram of distances for all stars in the sample, for K (yellow), M (red), and L (violet) dwarfs ({top}), and {\tt RUWE} values for the stars identified in the {\em Gaia} DR2 catalogue ({bottom}).
    The vertical red dashed line in the bottom panel sets the threshold for well-behaved astrometric solutions at {\tt RUWE} = 1.41.}
     \label{fig:distances_RUWE}
\end{figure}

We compiled equatorial coordinates, proper motions, parallaxes, and astrometric quality indicators from {\em Gaia} DR2. 
Of the 2479 stars in our sample, 2425 (97.8\,\%) had parallactic distances.
Of them, 2306 parallaxes came from {\em Gaia} DR2 (93.0\,\%) and 119 
from a number of references, as detailed in Table~\ref{tab:distances}.
For 16 stars with unavailable parallactic distances, we used the trigonometric distances of their confirmed proper motion companions from {\em Gaia} DR2 (ten cases) and \citet[][six cases]{van07}.
As a result, there were 54 stars without parallactic distance, of which 23 are close binaries: four spectroscopic binaries from \citet{Rei12} and \citet{Jef18}, and 19 resolved binaries (16 with $\rho \lesssim$ 0.8\,arcsec, and three at $\rho$ = 1.1--2.7\,arcsec; see Sect.~\ref{ssection:multiplicity}). 
The remaining 31 stars are single or have wide companions at angular separations of $\rho >$ 16\,arcsec.
For 30 of them, we derived photometric distances from $r'-J$ colours following the prescription in Sect.~\ref{sssection:diagrams_absmag_colour}.
For the remaining star, a Pleiades member with an $r'-J$ colour outside the validity range, we adopted the ``pseudomagnitude'' distance to the open cluster of \cite{Che16}. 
As a result, we compiled or derived distances for 2456 stars (i.e. all but the 23 close binaries without parallax).
Figure~\ref{fig:schema} shows a schematic summary of the origin of all compiled distances.

Our sample spans a distance range from 1.30\,pc (\object{Proxima Centauri}) to 171\,pc (\object{Haro 6--36}).
However, ignoring late K dwarfs, overluminous young M dwarfs (in Taurus, Upper Scorpius, and the $\beta$~Pictoris moving group; Sect.~\ref{ssection:young}), and one star with a large parallax uncertainty ($\delta \varpi / \varpi \sim$ 8\,\%), 
the most distant ``regular'' M dwarf is \object{LP~415--17}, at 73.0\,pc \citep{Die18,Hir18}.
Actually, 92\,\% of the stars are at less than 40\,pc, with only {half a dozen} objects further than 100\,pc.
{The top panel in Fig.~\ref{fig:distances_RUWE} shows the distance distribution of our K, M, and L sub-samples.}

{\em Gaia} DR2 provides statistical parameters to assess the quality of the astrometric data for each source.
The a posteriori mean error of unit weight ({\tt UWE}) is a goodness-of-fit indicator that is implicit in the {\em Gaia} DR2 solution.
Because of its strong dependence on colour and magnitude, a re-normalised {\tt UWE}, or {\tt RUWE}, is a more convenient indicator of well-behaved astrometric solutions \citep{Are18, Lin18}.
The latter authors set a threshold on {\tt RUWE} at 1.4, based on the empirical distribution of a large sample of stars, under which they retained 70\,\% of their sources. 
We derived the {\tt RUWE} values for all stars with {\em Gaia} DR2 measurements in our sample (2421; there are 125 {\em Gaia} DR2 stars without parallax), and display the corresponding {\tt RUWE} histogram in the bottom panel of Fig.~\ref{fig:distances_RUWE}.
In our case, by retaining 70\,\% of our sources we re-defined a cut in {\tt RUWE} = 1.41, which is equivalent to the 1.4 value.

Our sample is not volume limited.
First, its basis, the Carmencita catalogue, is not complete.
Carmencita contains all known M dwarfs in the solar neighbourhood that are further north than $\delta$ = --23\,deg with published ``spectroscopic'' (i.e. non-photometric) spectral types that are brighter than the completeness magnitudes shown in \citet{Alo15a}, meaning they are magnitude limited by spectral subtype: M0.0--0.5\,V with $J <$ 7.3\,mag, M1.0--1.5\,V with $J <$ 7.8\,mag, M2.0--2.5\,V with $J <$ 8.3\,mag, and so on.
{We refer the reader to the consequences of these selection criteria on the metallicity properties of the sample described in Sect.~\ref{ssection:sample}.}
Next, the K dwarf and ultracool dwarf additions are not complete either, because, for example, we discarded known K and L dwarf binaries. 
However, from the distribution of distances, our sample in the Calar Alto sky is complete for M0.0\,V, M4.0\,V, and M6.0\,V stars at approximate distances of 25\,pc, 15\,pc, and 5\,pc, respectively.


    \subsection{Multiplicity}
    \label{ssection:multiplicity}
        
\begin{figure}[]
   \centering
   \includegraphics[width=1 \columnwidth]{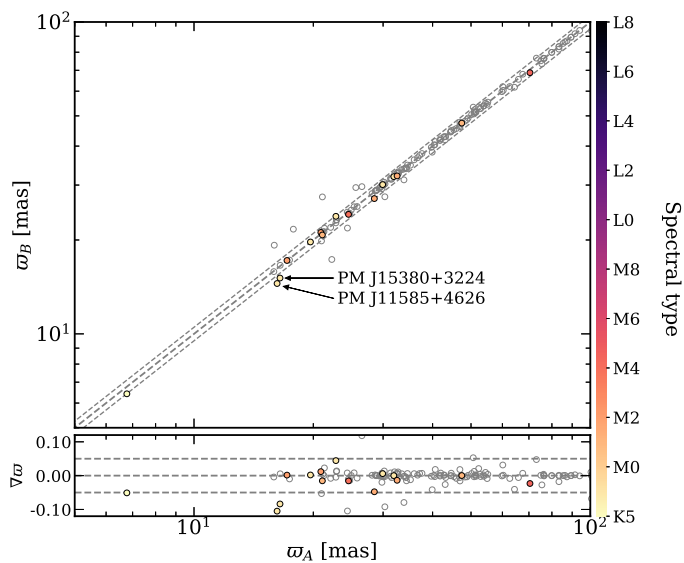}  
   \caption{Parallax diagram of the primary (A) and secondary (B) components of the 15 new binary systems in Table~\ref{tab:multiplicity} with parallactic information in both components, colour-coded by spectral type.
   {The bottom panel shows the normalised difference between both parallaxes, i.e. $\nabla\varpi = (\varpi_B-\varpi_A)/\varpi_A$.}
   Grey empty circles are the 134 previously known pairs in our sample with parallactic information for both components and angular separation of $\rho <$ 5\,arcsec.
   The black dash-dotted and dashed lines mark the 1:1 and 1:1\,$\pm$\,0.05 (i.e. 5\,\% difference), respectively.
   Two slight outliers from our list of binary candidates are labelled with their common names.
   }
\label{fig:binaries_distance}
\end{figure}

\begin{figure}[]
    \centering
    \includegraphics[width=1\columnwidth]{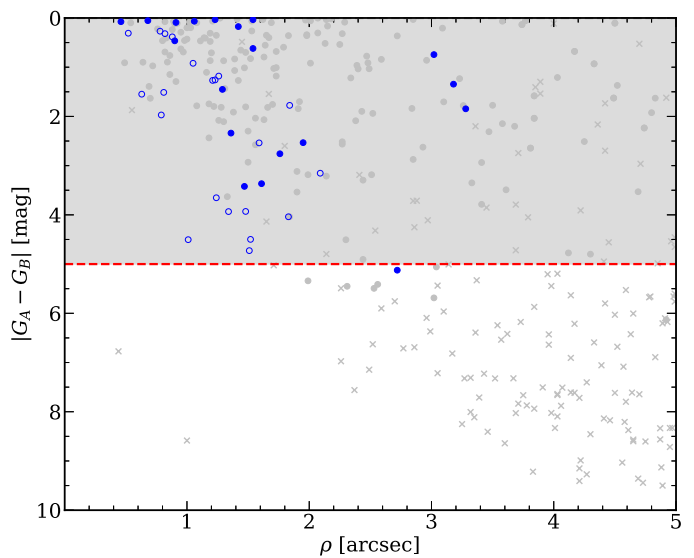}   
    \caption{Difference in the {\em Gaia} $G$ magnitude values for the 359 stars with their closest companion within 5\,arcsec as a function of angular separation at epoch J2015.5.
    Known binaries and background stars are depicted with grey filled circles and grey crosses, respectively. 
    New binaries are represented with blue circles, filled if they are confirmed by common parallactic distance, and open if only one component has a measured parallax.
    The red dashed line marks the boundary at $\Delta G$ = 5\,mag for contaminated sources.
    }
    \label{fig:binaries}
\end{figure}

We searched for additional {\em Gaia} DR2 sources within 5\,arcsec of our target stars at epoch J2015.5 using the ADQL\footnote{\tt \url{http://www.ivoa.net/documents/ADQL/}} query form in the {\it Gaia} Archive.
According to \citet{Gaia18bro} and, especially, \citet{Are18}, {\it Gaia} can resolve equal-brightness sources separated by down to 0.4\,arcsec, which were not resolved in most previous all-sky surveys, such as 2MASS or AllWISE 
(see \citealt{Cab19} for a practical example of close binaries resolved for the first time by {\it Gaia}). 
For the 2421 stars in our sample that were catalogued by {\em Gaia}, the search provided 388 additional sources around 353  stars at $\rho<$ 5\,arcsec.
Of them, 324 stars had only one additional source, 24 stars had two sources, 4 stars had three sources, and 1 star had four sources.
Besides, for the 58 stars in our sample not tabulated in the {\em Gaia} catalogue, we used the projected positions as explained in Sect.~\ref{ssection:sample}, which resulted in 11 additional sources around 6 stars. 
The cases of three or more additional sources corresponded to stars in crowded regions at low Galactic latitudes.

Of the 359 stars with close {\it Gaia} companion candidates,
166 were already tabulated as members in known physical pairs in the Washington Double Star catalogue \citep{Mas01}, 4 in \cite{Ans15}, and 1 in \cite{Hei87}.
Next, we analysed in detail the remaining 188 systems.
Of these, we classified 148 faint sources as background stars and point-like galaxies based on astrometric and photometric criteria:
96 sources have parallaxes $\varpi <$ 2\,mas and so are located at more than 0.5\,kpc;
four sources have parallaxes 2\,mas $< \varpi <$ 7\,mas and turned to be unrelated sources at 47--225\,pc (Bayesian distances computed by \citealt{Bai18});
one source with a parallax of 21.3\,mas is located twice as far as the main source;
and 47 sources do not have measured parallaxes, proper motions, or{} 2MASS near-infrared counterparts.
In spite of being more than 5\,mag fainter than the primary in $G$ band, all 47 sources are visible in digitisations of blue photographic plates of the 1950s (Digitised Sky Survey~I), implying that they are background sources much bluer than the stellar primaries\footnote{However, there are certain systems that deserve a high-resolution imaging follow-up, such as J02033--212 (\object{G~272--145}), J04429+189 (\object{HD~285968}), J05466+441 (\object{Wolf~237}), and J11311--149 (\object{LP~732--035}).}.

We investigated the remaining 40 sources not included in the two previous groups.
Of them, 15 are in physically bound systems with {\em Gaia} parallaxes for both components that agree within 1$\sigma$ errors except for two cases, marked in Fig.~\ref{fig:binaries_distance}.
The two systems are bona fide high proper motion pairs, for which we see that the tangential component of the orbital motion and the {\em Gaia} astrometric solution has not yet taken the close binarity into account.
All remaining 25 candidate companions are not visible in the Digitised Sky Survey~I and satisfy $\Delta G \lesssim$ 5\,mag ($\Delta G \sim$ 0.3\,mag in three cases with $G_{BP}$, $G$, and $G_{RP}$ photometry; see below).
In Table~\ref{tab:multiplicity} we list the {\em Gaia} DR2 equatorial coordinates, proper motions, parallaxes, $G$ magnitudes, angular separations $\rho$, and position angles $\theta$ of the 40 new binary systems and candidates.
Among them, there are three triple systems consisting of a spectroscopic binary and a fainter companion (see Table~\ref{tab:multiplicity} notes).
All systems are separated by 3.3\,arcsec at most, which explains why other surveys, such as 2MASS, were not able to resolve them.

In the presence of a close companion, either physically associated or not, photometric measurements of a star can be compromised, especially when their brightness is comparable.
This photometric contamination impacts negatively on the parameters derived from it, such as luminosity, distance, or colours.
In this work, we considered the photometry of a star as contaminated if the $G$ flux of any companion at $\rho <$ 5\,arcsec, regardless of physical binding, is more than 1\,\% of its flux, that is if $\Delta G < -2.5 \log(F_{G,{\rm B}}/F_{G,{\rm A}}) = 5$\,mag, where $F_{G,{\rm A}}$ and $F_{G,{\rm B}}$ represent the fluxes of the primary and secondary components in the $G$ band, respectively.
In Fig.~\ref{fig:binaries} we plot $\Delta G$ versus $\rho$ of the 359 pairs in our sample with $\rho <$ 5\,arcsec.
Of them, 238 meet the criteria above, and their photometry is therefore flagged as potentially contaminated.
To those 238 stars we added another 372 stars from \cite{Cab16} that are known to be very close physical systems unresolved by {\em Gaia} (but resolved with micrometers, speckle, lucky imaging, or adaptive optics systems) and spectroscopic binaries.
The 610 ``close binaries'' are plotted as a reference in most figures afterwards with grey dots, but will not be considered in the following analysis.


\section{Analysis and results}
\label{section:analysis}

In this Section we present the main products of the exploitation of the astrometric and photometric data in the sample, including luminosities, masses, radii, colours, and bolometric corrections.

    \subsection{Luminosities}
    \label{ssection:luminosities}

\begin{table}
\caption{Set of constraints for the spectral energy distribution modelling in VOSA$^a$.}
\label{tab:VOSA}
\centering
\begin{tabular}{lcc}
            \hline
            \hline 
            \noalign{\smallskip}
            Spectral    & $T_{\rm eff}$ & $\log{g}$ \\
                        types           & [K] & [dex] \\
            \noalign{\smallskip}
            \hline
            \noalign{\smallskip}
            K5\,V to M2.0\,V            & 3300--4600    & 4.5--5.0 \\
            M2.5\,V to M5.0\,V  &  2800--3700   & 4.5--5.5  \\  
            M5.5\,V to L8.0             & 1200--3200    & 5.0--5.5 \\       
\noalign{\smallskip}
\hline
\end{tabular}
\tablefoot{
\tablefoottext{a}{Iron abundance set to zero ([Fe/H] = 0.0).}
}
\end{table}

\begin{figure}[]
    \centering
    \includegraphics[width=.99\linewidth]{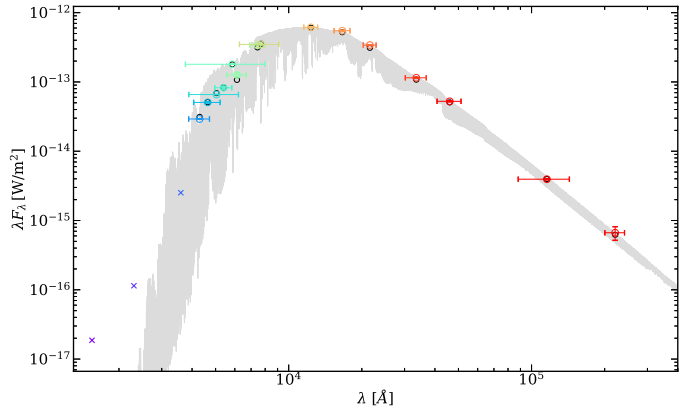}
    \caption{Spectral energy distribution of \object{LP~167--071} (J10384+485, M3.0\,V). 
    The empirical fluxes (coloured empty circles, following the same colour scheme as in Fig.~\ref{fig:hist_magnitudes}) are overimposed on the best-fitting BT-Settl CIFIST spectrum (grey; $T_{\rm eff}$ = 3300\,K and $\log{g}$ = 5.5).
    The modelled fluxes are depicted as grey empty circles.
    Photometric data in the ultraviolet are shown as crosses, and are not considered in the modelling.   
    Horizontal bars represent the effective widths of the bandpasses, while vertical bars (visible only for relatively large values) represent the flux uncertainty derived from the magnitude and parallax errors.
    }
    \label{fig:SED}
\end{figure}

\begin{figure}[]
    \centering
    \includegraphics[width=.95\linewidth]{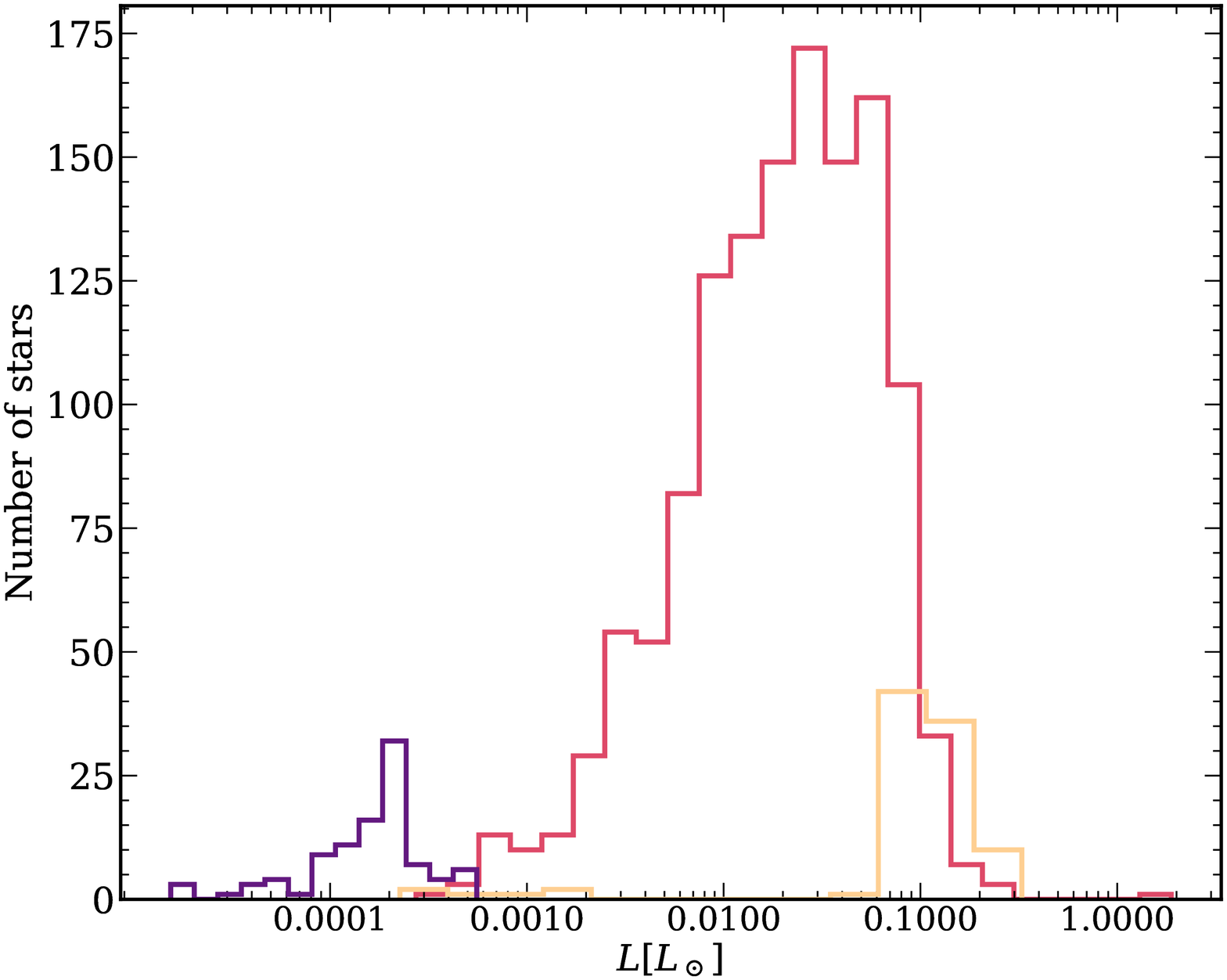} 
    \includegraphics[width=.95\linewidth]{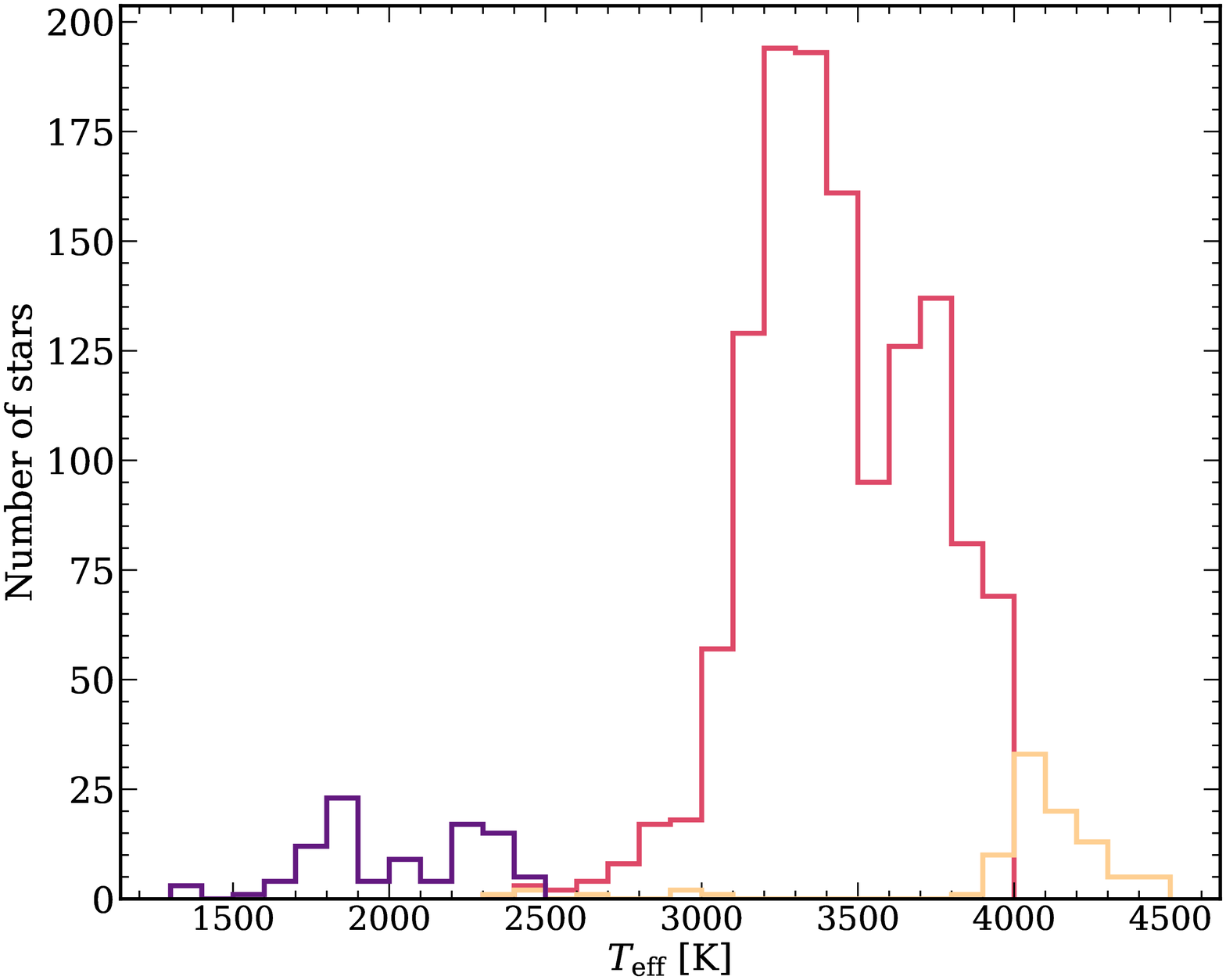} 
    \includegraphics[width=.95\linewidth]{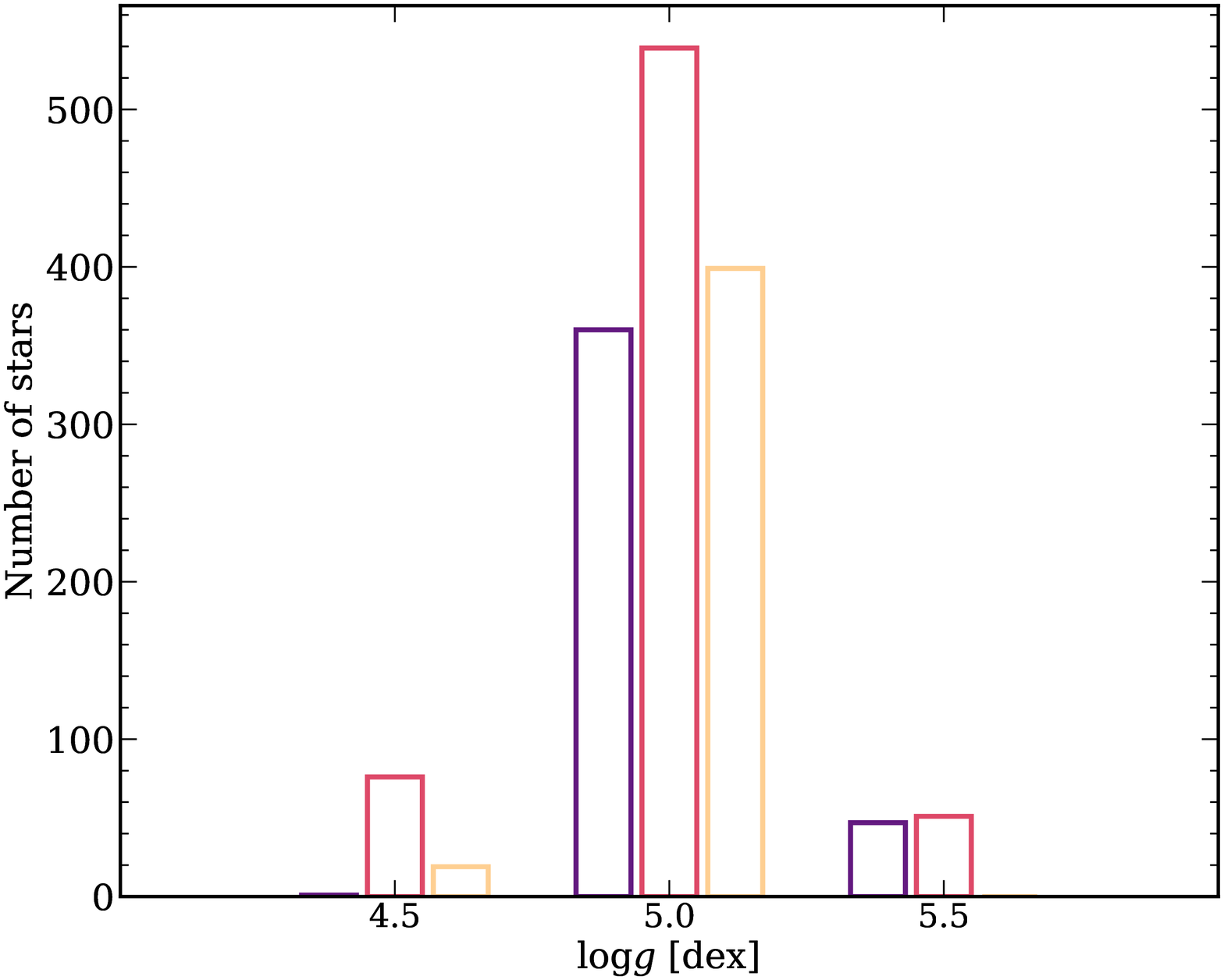}
    \caption{Distribution of bolometric luminosities ({top}), effective temperatures ({middle}), and surface gravities ({bottom}) for K (yellow), M (red), and L (violet) dwarfs.}
    \label{fig:luminosities}
\end{figure}

After discarding the 610 close binaries ($\rho <$ 5\,arcsec), we kept 1843 stars with parallax and whose photometry was not affected by close multiciplicity \citep[however, many of the latter are members of wide multiple systems;][]{Cor17}.
We used VOSA to compute their basic stellar parameters: 
bolometric luminosity, $L_{\rm bol}$,
effective temperature, $T_{\rm eff}$, 
and surface gravity, $\log{g}$.
Among the theoretical model grids available in VOSA for reproducing the observed spectral energy distribution (SED) of each target star,
we used the latest BT-Settl CIFIST grid \citep{Hus13, Bar15}.
We conservatively constrained the possible values of $T_{\rm eff}$ and $\log{g}$ as a function of spectral type as discussed by \citet{Pec13} and \citet{Pas18}, respectively, and summarised in Table~\ref{tab:VOSA}.
We fixed the metallicity to solar (BT-Settl CIFIST models are provided for [Fe/H] = 0.0 only) and visual extinction to zero ($A_V$ = 0\,mag, in view of the closeness of the overall sample; see Sect.~\ref{ssection:distances}).
For each star, the VOSA input was the compiled photometry in the passbands in Table~\ref{tab:filters}, parallactic distance, and their  uncertainties.

In the fitting process, we included the observed fluxes of up to 17 passbands, from optical Tycho-2 $B_T$ to mid-infrarred AllWISE $W4$.
Since we were only interested in the photospheric emission, we excluded from the fit the other three passbands (i.e. {\em GALEX} $FUV$ and $NUV$ and SDSS9 $u'$) because the chromospheric emission dominates in the bluest spectral range, especially in late-M dwarfs  \citep{Rei12,Ste13}.
At wavelengths bluewards of $B_T$ ($\lambda <$ 4280\,{\AA}) and redwards of $W4$ ($\lambda >$ 220883\,{\AA}) we followed the VOSA best-fit model {(see example in Fig.~\ref{fig:SED})}.
The uncertainty in this assumption was very small, as the estimated fraction of photospheric energy in BT-Settl CIFIST spectra bluewards of $B_T$ (in the Wien domain) ranges from 0.46\,\% to 0.0002\,\% for M0\,V and M8\,V, respectively, and redwards of $W4$ (in the Rayleigh-Jeans domain) ranges from 0.0036\,\% to 0.0087\,\% for M0\,V and M8\,V, respectively.

For the best fit, VOSA uses a $\chi^2$ metric, where each photometric point is weighted with its uncertainty.
If this uncertainty is blank or artificially set to zero, VOSA assumes a large value instead, which depends on the largest relative error on the SED, and assigns to the point a low weight\footnote{\href{http://svo2.cab.inta-csic.es/theory/vosa/helpw4.php?otype=star&action=help}{\texttt{http://svo2.cab.inta-csic.es/theory/vosa/}}}. 
The theoretical uncertainties of $T_{\rm eff}$ and $\log{g}$ are determined by the BT-Settl CIFIST model grid, which provides synthetic models in steps of 100\,K {(50\,K for spectra cooler than 2400\,K)} and 0.5\,dex, respectively.
VOSA estimates the error in the output parameters as half the grid step around the best-fit value.

Complementing the VOSA automatic identification {of photometric outliers in the SED}, we inspected all the 1843 individual SEDs and marked 7.1\,\% of all data points as `{\tt Bad}', as they had bad quality flags (Sect. \ref{ssection:photometry}) or clearly deviated from the SED trend in the optical and, therefore, were not included in the model fitting.
After a careful inspection, we also ignored the possible infrared excesses automatically detected by VOSA, even for the two single, very young stars in the Taurus-Auriga association (see Sect.~\ref{ssection:young}). 

In Fig.~\ref{fig:luminosities} we show the distributions of luminosities, effective temperatures, and surface gravities stacked by spectral type. 
We derived luminosity values ranging from 1.54~$10^{-5}$\,$L_\odot$ for the nearby L8 dwarf \object{DENIS-P~J0255-4700}, to 0.3276\,$L_\odot$ for the K7\,V dwarf \object{HD~196795}, except for a very young early M member of the $\beta$~Pictoris moving group, namely \object{StKM~1--1155}, which has an exceptional luminosity of  1.8817\,$L_\odot$.
Although very similar, our luminosities supersede those tabulated by \cite{Sch19} for the M dwarfs in the CARMENES GTO survey, as we updated some parallactic distances and APASS9 and PS1~DR1 optical magnitudes.


    \subsection{Young star candidates}
    \label{ssection:young}

\begin{figure}[]
    \centering
    \includegraphics[width=.99\linewidth]{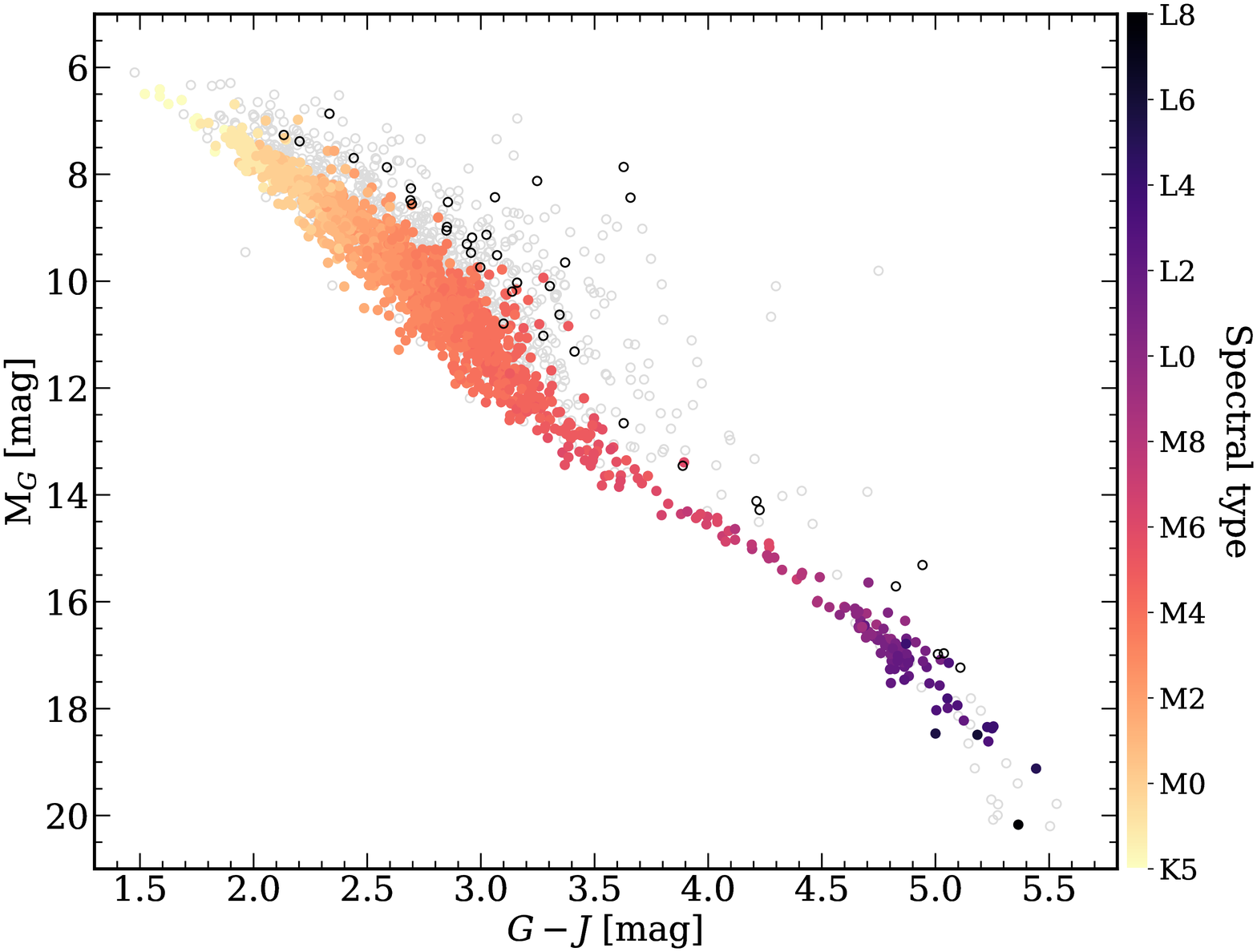}
    \includegraphics[width=.99\linewidth]{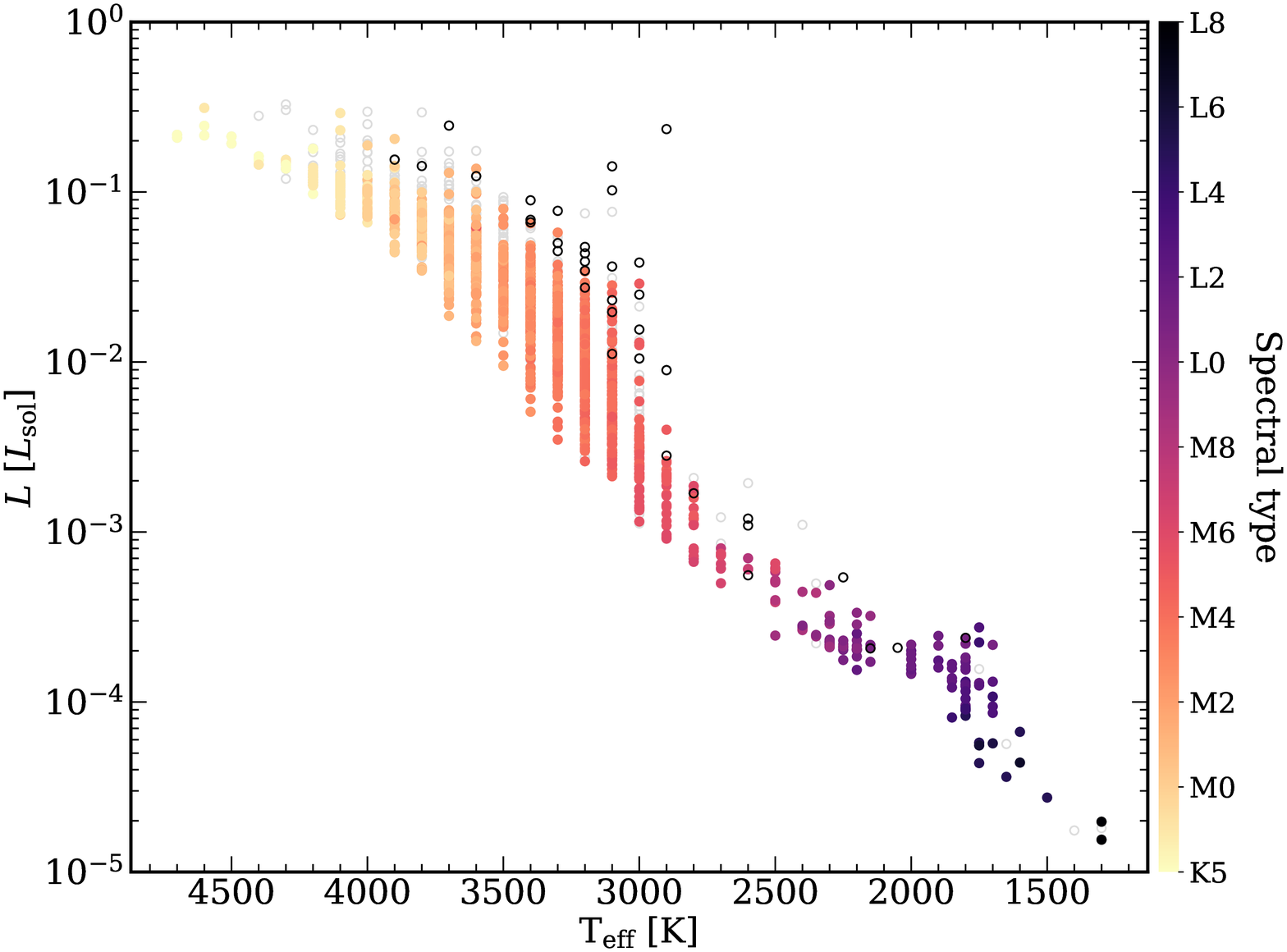}
    \caption{Absolute magnitude $M_G$ against $G-J$ colour ({top}), and {bolometric luminosity against effective temperature from VOSA} ({bottom}).
    In the top panel, empty grey circles represent stars with poor photometric quality data in $G$, $J$, or both passbands (Sect.~\ref{ssection:photometry}), poor astrometric quality data ({\tt RUWE} $>$1.41) or non-parallactic distances (Sect.~\ref{ssection:distances}), and close binary stars (Sect.~\ref{ssection:multiplicity}).
    In the bottom panel, empty grey circles represent stars with poor astrometric quality data or non-parallactic distances, and close binary stars.
    In both panels, black empty circles are the 36 known young overluminous stars identified in our sample.
    The remaining ``regular'' stars are colour-coded by spectral type.
    }
    \label{fig:HR}
\end{figure}

\begin{table*}
\caption{Overluminous young stars identified in our sample.}
\label{tab:young}
\centering
\begin{tabular}{llll}
            \hline
            \hline 
            \noalign{\smallskip}
            Karmn  & Name & Stellar kinematic group & Reference \\
            \noalign{\smallskip}
            \hline
            \noalign{\smallskip}
J0045+1634$^a$ & 2MUCD 20037 & Argus  & \citealt{Gag14} \\
J01352-072 & Barta 161 12 & $\beta$~Pictoris  & \citealt{Alo15b} \\
J02443+109W & MCC 401 &$\beta$~Pictoris  & \citealt{Jan17} \\
J03510+142 & 2MASS J03510078+1413398 & $\beta$~Pictoris  & \citealt{Gag15} \\
J03548+163 & LP 413--108 & Hyades  & \citealt{Cra86} \\
J03565+319 & HAT 214--02089 & Hyades?  & \citealt{Roe11} \\
J04206+272 & XEST 16--045 & Taurus  & \citealt{Sce07} \\
J04238+092 & LP 535--073 & Hyades  & \citealt{Wei79} \\
J04238+149 & IN Tau & Hyades  & \citealt{van34} \\
J04252+172 & V805 Tau & Hyades  & \citealt{van66} \\
J04369-162 & 2MASS J04365738--1613065 & Tuc-Hor  & \citealt{Mal14} \\
J04414+132 & TYC 694--1183--1 & Hyades  & \citealt{Joh62} \\
J0443+0002$^a$ & 2MUCD 10320 & $\beta$~Pictoris  & \citealt{Alo15b} \\
J04433+296 & Haro 6-36 & Taurus  & \citealt{Har53} \\
J04595+017 & V1005 Ori & $\beta$~Pictoris  & \citealt{Alo15b} \\
J05019+011 & 1RXS J050156.7+010845 & $\beta$~Pictoris  & \citealt{Alo15b} \\
J05084-210 & 2MASS J05082729--2101444 & $\beta$~Pictoris  & \citealt{Alo15b} \\
J0608-2753$^a$ & 2MASS 06085283-2753583 & $\beta$~Pictoris  & \citealt{Alo15b} \\
J07310+460 & 1RXS J073101.9+460030 & Columba  & \citealt{Mal13} \\
J07446+035 & YZ CMi & $\beta$~Pictoris & \citealt{Alo15b} \\
J09449-123 & G 161-071 & Argus  & \citealt{Bar17} \\
J11519+075 & RX J1151.9+0731 & $\beta$~Pictoris  & \citealt{Alo15b}\\
J12508-213 & DENIS J125052.6--212113 & Pleiades?  & \citealt{Cla10}  \\
J14200+390 & IZ Boo & Young?  & \citealt{Moc02} \\
J14259+142 & StKM 1--1155 & $\beta$~Pictoris  & \citealt{Alo15b} \\
J15079+762 & HD 135363 B & IC~2391  & \citealt{Mon01,Lep07a} \\
J15166+391 & LP 222--065 & Young disc  & \citealt{Jef18} \\
J1552+2948$^a$ & 2MASS J15525906+2948485 & $\sim$100\,Ma  &  \citealt{Cru09} \\
J15597+440 & RX J1559.7+4403 & AB Dor  & \citealt{Bin16} \\
J16102-193 & K2--33 & USco  & \citealt{Pre01} \\
J17572+707 & LP 044--162 & Argus?  & \citealt{Gag15} \\
J21100-193 & BPS CS 22898-0065 & $\beta$~Pictoris  & \citealt{Alo15b} \\
J22088+117 & 2MASS J22085034+1144131 & $\beta$~Pictoris  & \citealt{Shk17} \\
J23228+787 & NLTT 56725 & Columba  & \citealt{Mak07,Mon18} \\
J23301-026 & 2MASS J23301129--0237227 & $\beta$~Pictoris  & \citealt{Alo15b} \\
J23317-027 & AF Psc & $\beta$~Pictoris  & \citealt{Alo15b} \\
\noalign{\smallskip}
\hline
\end{tabular}
\tablefoot{
\tablefoottext{a}{Ultra-cool dwarfs from \cite{Sma17} not in the CARMENES catalogue of M dwarfs.}
}
\end{table*}

In the two panels of Fig.~\ref{fig:HR} we display two related plots: a Hertzsprung-Russell diagram with luminosities and effective temperatures from our VOSA analysis, and a colour-absolute magnitude diagram with {\em Gaia} and 2MASS data. 
After discarding stars with poor astro-photometric data or very close companions, we identified overluminous stars that departed from the main sequence defined by ``regular'' single stars in the $M_G$ versus $G-J$ diagram, as in the case of StKM~1--1155.
We searched the literature for information on their membership in known young kinematic groups \citep[i.e. younger than or of the age of the Hyades, $\tau \lesssim$ 0.6\,Ga -- ][]{Per98,Mon01,Zuc04}.
The 36 identified overluminous stars include members of very young associations and moving groups (Taurus-Auriga, Upper Scorpius, $\beta$~Pictoris), moderately young groups (Argus, Tucana-Horologium, Columba, IC~2391 supercluster), middle-aged open clusters and groups (Pleiades, AB Doradus, Hyades), and a miscellanea classification including one star of about 100\,Ma \citep{Cru09}, an active one that kinematically belongs to the young Galactic disc \citep{Jef18}, and an {ultra-fast-rotating, H$\alpha$-variable, X-ray-emitting, young star candidate} \citep[IZ~Boo --][]{Ste86,Fle98,Moc02,Jef18}.
The 36 stars and their respective references are listed in Table~\ref{tab:young}. 
As expected, these sources are also overluminous in the Hertzsprung-Russell diagram.
Besides, there are a dozen stars neither tabulated by us nor classified as young star candidates in the literature that are also overluminous, which will deserve attention in forthcoming works.


    \subsection{Diagrams}
    \label{ssection:diagrams}

We present and discuss several diagrams involving colours, absolute magnitudes, and bolometric corrections.
    
    \subsubsection{Colour-spectral type}
    \label{sssection:diagrams_colour_SpT}

We computed 20 average colour indices for adjacent filters and their standard deviation for late-K to late-L dwarfs, using only the good quality photometric data.
We list them in Table \ref{table.colours}.
The size of the sample for each colour index and spectral type is shown in parentheses.
Colour indices computed from samples with less than four elements are included for completeness, albeit with a word of caution.
As expected, the amount of data available in the ultraviolet and optical blue passbands decreases for later spectral types (see again Fig.~\ref{fig:hist_magnitudes}).
In particular, for spectral types M4\,V and earlier we have all possible colour combinations, and for spectral types {later than M4\,V and up to L5} we have all possible colour combinations only between $G$ and $W3$.
This colour compilation complements, and most of the time supersedes, previous determinations  \citep[][]{Bes98,Dah02,Haw02,Kna04,Wes05,Cov07,Zha09,Boc10,Lep13,Pec13,Raj13,Dav14,Fil15,Man15,Bes17}.

\begin{figure}[]
    \centering
    \includegraphics[width=.99\linewidth]{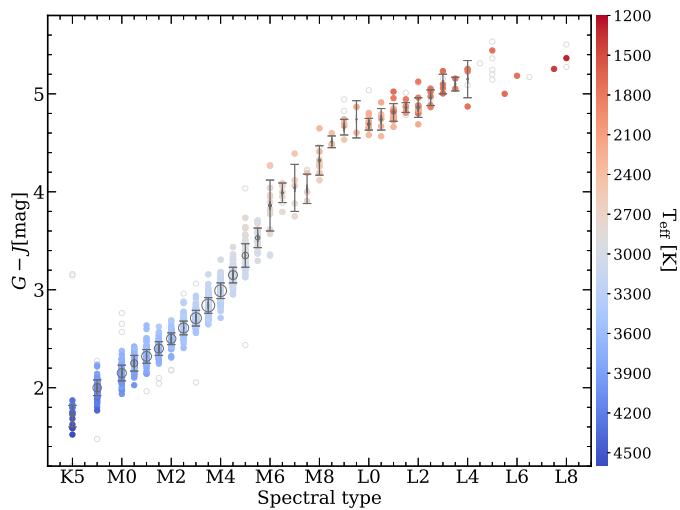}
    \caption{$G-J$ colour against spectral type. 
    Black empty circles mark the average colour for each spectral type with a size proportional to the number of stars and vertical bars account for their standard deviation in spectral types with more than one valid colour value.
        Empty grey circles depict bad photometric data, as explained in Sect.~\ref{ssection:photometry}, and their values are not considered in the calculations of the average colours.}
    \label{fig:GJ_SpT}
\end{figure}

From all the possible combinations, the {\em Gaia} DR2-2MASS colour $G-J$ provides one of the most solid estimators of spectral type from late-K to mid-L dwarfs.
This is illustrated in Fig.~\ref{fig:GJ_SpT}. 
Firstly, $G-J$ covers a wide range in colour of about 3.6\,mag between K5\,V and L8, with a slight flattening restricted to the late L objects.
Secondly, it exhibits one of the smallest dispersions in late-M and L dwarfs among all analysed colours, with a median deviation of 0.08\,mag.
Thirdly, the $G$ and $J$ passbands offer a high availability in this spectral type range, with 97.7\,\% and 100\,\% completeness in $G$ and $J$, respectively.
Also, faint objects benefit from the reliability of 2MASS and {\em Gaia} DR2 photometry.
This colour index is superior to previous colour indices used to discriminate late spectral types, such as $i'-J$ \citep{Rei01,Haw02,Wes05,Cov08}, and 
finds a compromise between completeness, photometric data quality ({\em Gaia} and 2MASS), scatter of the data, and colour interval spanned by the sequence.
On the contrary, the use of {\em Gaia}-only and 2MASS-only colours for spectral typing presents some serious caveats, from the degeneracy of $G_{B_P}-G_{R_P}$ for spectral types M8\,V and later, to the narrow interval of 1\,mag of $G-G_{R_P}$ from late-K to late-M dwarfs and its pronounced flattening from late-M to mid-L dwarfs, to the blueing of $J-H$ in the M-dwarf domain.

In Fig.~\ref{fig:colourSpT}, we plot six additional colour-spectral type diagrams that show the behaviour of other passbands from the near-ultraviolet to mid-infrared, and their adequacy for spectral type estimation.
In all cases, data with poor photometric quality are included as empty grey circles, but not considered for any calculation.
Firstly, the optical-mid-infrared $G-W3$ colour serves as a useful complement for the $G-J$ colour, especially in the late-M and early-L regime.
The $G-W3$ colour also exhibits a monotononic, low-scatter, steady increase from K5\,V to L8, although the median of the dispersion is 0.17\,mag, twice the value obtained with the $G-J$ index.
Additionally, it benefits from the widest interval in colour of all the diagrams, with approximately 7\,mag separating K5\,V and L8.

The purely optical colour $r'-i'$,  extensively used in the literature, can help to determine spectral types of late-K to late-M dwarfs, but it fails to discriminate the types for cooler objects.
It peaks at about 2.8\,mag (around M7--8\,V), and becomes bluer beyond this point, as shown by for example \cite{Haw02} and \cite{Lie06}.

The purely infrared colour $J-W2$ exhibits a remarkably low dispersion from M0\,V to M8\,V (less than 0.06\,mag), but it covers a colour interval of only 0.5\,mag. 
The colour $G_{R_P}-W1$ offers an adequate alternative, with a dispersion slightly larger in the same range (0.09\,mag), but spanning five times the colour interval.
Furthermore, colours including the $W4$ passband suffer from poor quality data for spectral types M8\,V and later.

The $NUV - G_{RP}$ colour is sensitive to both spectral type and ultraviolet flux excesses, which may be caused by chromospheric activity and/or interaction between close binaries.
The first case includes ``regular'' stars later than M3--4\,V at the boundary of stellar full convection.
The second case comprises, according to \citet{Ans15}, young stars (including all our overluminous young stars except one Hyades member) and unidentified binaries, which include unresolved background ultraviolet sources, unresolved old binaries with white dwarf companions, and short-period ($P < $ 10\,d) tidally interacting binaries that induce ongoing activity on each other.
These phenomena give rise to a distinguishable population appended to the main sequence.

Finally, the optical $B-V$ colour became a commonly used index in the literature, including \cite{Bes98}, \cite{Ram05}, \cite{Cas08}, \cite{Smi18}, \cite{Sun18}, \cite{DeB18}, or \cite{Coc19}, just to name a few.
However, the $B-V$ colour has some disadvantages in the M-dwarf domain:
\begin{itemize}
\item Both $B$ and $V$ lack the completeness in the optical range that other passbands, such as $G_{BP}$, $r'$, $i'$, or $G_{RP}$, deliver. 
\item $B-V$ fails to produce a photometric sample statistically that is consistent beyond M5\,V, while the {\em Gaia} DR2, 2MASS, or AllWISE passbands succeed.
\item $B-V$ does not correlate with spectral type beyond M5\,V.
\item The width of the colour interval from late K to mid M is 1\,mag, only a few times the scatter of the main sequence (0.12\,mag), with a striking flattening between M0\,V and M3\,V.
\item The mean uncertainties of $B$ and $V$ in our sample are 0.056\,mag and 0.048\,mag, respectively.
For comparison, the same parameters for $G$ and $J$ are 0.0012\,mag and 0.029\,mag, respectively.
\end{itemize}

Therefore, we discourage the use of $B-V$ as an estimator of spectral type for stars cooler than K5\,V.
This is especially applicable when the {\em Gaia} DR2 (and 2MASS or AllWISE) magnitudes are available.
The same reasoning above also applies to the $B_T$ and $V_T$ Tycho-2 passbands, which are even less complete.

        \subsubsection{Colour-colour}
    \label{sssection:diagrams_colour_colour}

\begin{figure}[]
    \centering
    \includegraphics[width=.99\linewidth]{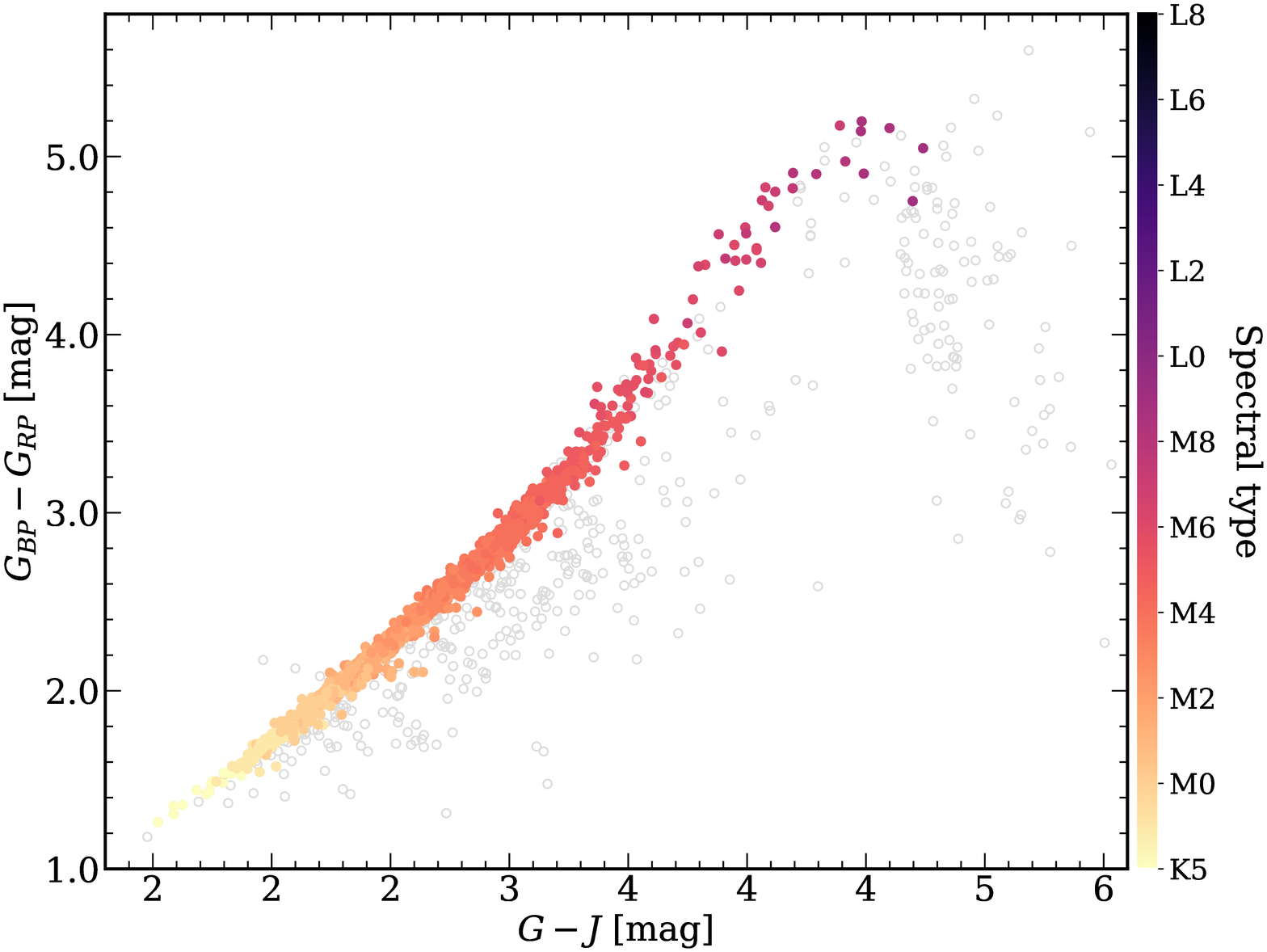}
    \includegraphics[width=.99\linewidth]{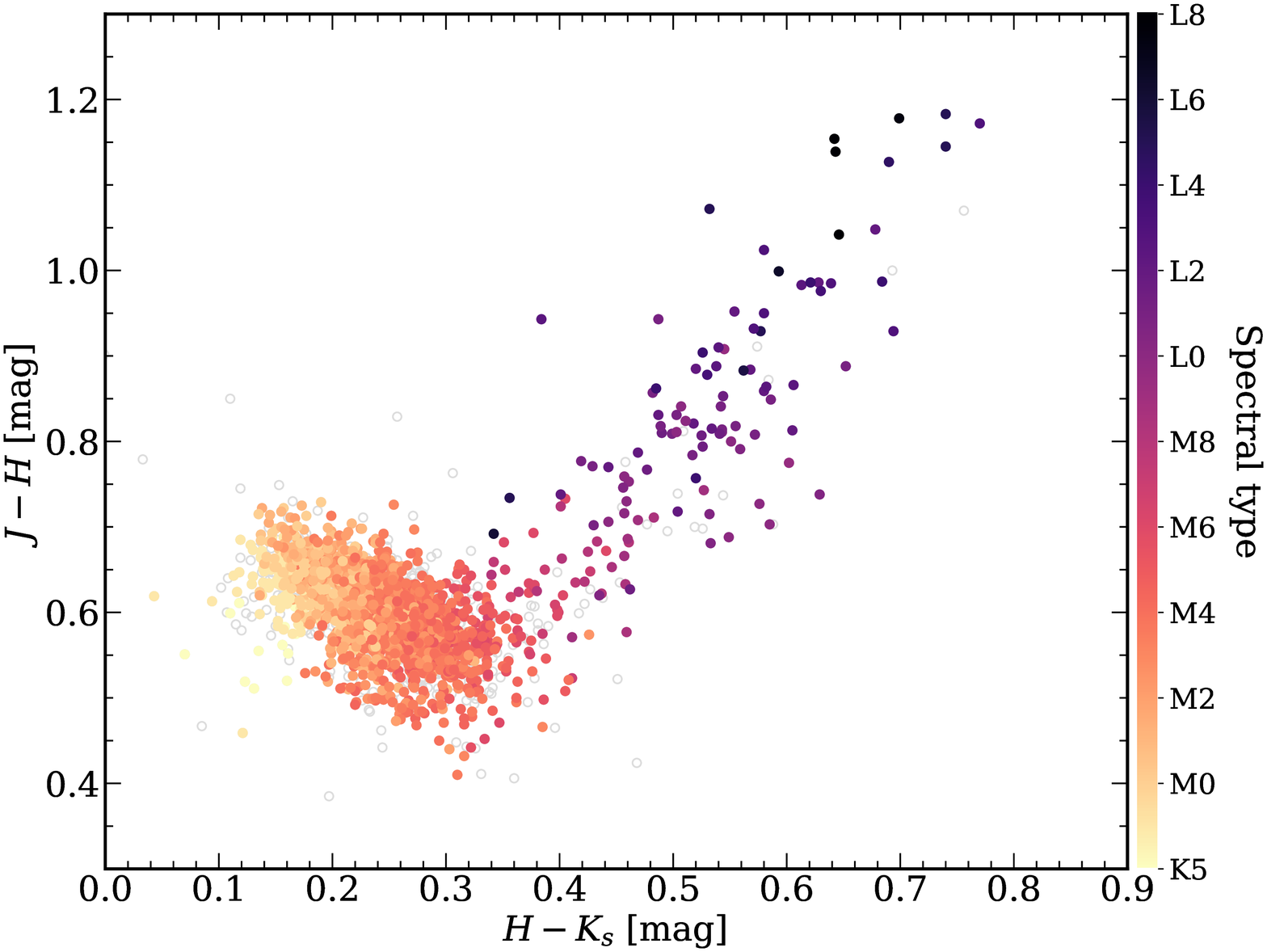}
    \caption{Colour-colour diagrams representing $G_{BP}-G_{RP}$ vs. $G-J$ ({top panel}) and $J-H$ vs. $H-K_s$ ({bottom panel}).
In both panels, empty grey circles represent stars with poor photometric quality data in any of the involved passbands, close binaries, or young stars.
The remaining ``regular'' stars are colour-coded by spectral type.
      }
    \label{fig:colour-colour}
\end{figure}

As in the colour-spectral type diagrams, main sequence stars occupy a well-defined locus in  colour-colour diagrams.
In spite of the degeneracy beyond M8\,V, the narrowest main sequence is observed in the 2MASS-{\em Gaia} $G_{BP}-G_{RP}$ versus $G-J$ colour-colour diagram shown in the top panel of Fig.~\ref{fig:colour-colour}. Outliers in the diagram are mostly unresolved binaries and young stars for colours bluer than $G-J \sim$ 4.5\,mag (spectral types earlier than M8\,V), albeit other possibilities  also exist.
For example, \object{G~78--3} (J02455+449), at 57.8\,pc \citep{Gaia18bro}, is an M5\,V star \citep{Haw96}, which exhibits a $G-J$ colour typical of an early-M dwarf.
For colours redder than $G-J \sim$ 4.5\,mag, we confirm the findings of \cite{Sma19}, who reported an unreliability in {\em Gaia} blue-band photometry of very late objects with $G_{BP} >$ 19.5\,mag due to background underestimation by the {\em Gaia} automatic pipeline \citep{Gaia18bab,Eva18,Sma19}.
As a comparison, in Fig.~\ref{fig:colour-colour} we also show a widely used, 2MASS-only, colour-colour diagram \citep{Kir99,Kna04,Lep05,Hew06,Cov07}.
There, late-K to late-M dwarfs occupy a compact region that ranges from $H-K_s \sim$ 0.15\,mag, $J-H \sim$ 0.65\,mag to $H-K_s \sim$ 0.30\,mag, $J-H \sim$ 0.55\,mag, while later stars and brown dwarfs become redder
\citep[][and references above]{Kir99}.
We did not {notice} any near-infrared flux excess, as found in young T~Tauri M-type stars and brown dwarfs with warm circumstellar discs \citep{Car01,Cab04,Her08}.

In Fig.~\ref{fig:colourcolour} we display a selection of six additional colour-colour diagrams.
In all cases we plot far-ultraviolet to mid infrared-colours against $G-J$.
Apart from the stars with poor photometric quality, we also discarded the 2MASS magnitudes of the extraordinarily red \object{2MUCD~20171} \citep[J03552+113;][]{Fah13} and blue \object{SDSS~J141624.08+134826.7} \citep[J1416+1348A;][]{Bur10}
ultracool dwarfs, which were clear outliers in many colour-colour diagrams involving 2MASS magnitudes.
As in the colour-spectral type diagrams, the two colour-colour diagrams involving the bluest colours illustrate the two populations of ultraviolet active and inactive sources ($NUV - G_{RP}$) and the poor spectral sequence based on $B-V$ colour.   
The two diagrams involving UCAC4/SDSS9/APASS9/CMC15/PS1\,DR1 $g'r'i'$ passbands, which will also be used at the Vera C. Rubin Observatory for the Legacy Survey of Space and Time (LSST), show a slightly larger spread than {\em Gaia} data and the double slope of the $r'-i'$ colour also found by \citet{Haw02} and \citet{Lie06}.
Interestingly, $g'-i'$ has a smaller dispersion in the late K and M dwarf domain than $r'-i'$, but a much larger dispersion at $G-J \gtrsim$ 4.5\,mag.
This extra scatter at the reddest colours is more likely due to the intrinsic spectral variations at the M/L boundary (\`a la \citealt{Haw02}; e.g. metallicity) than due to data analysis systematics or Poissonian error at the survey magnitude limits (\`a la \citealt{Sma19}; e.g. background).
Finally, the colour-colour diagrams with near-infrared 2MASS and AllWISE data {(specially $W3$ and $W4$)} are very sensitive to $T_{\rm eff}$ variations at the L spectral types, but quite insensitive in the late-K and M dwarf domain.
However, their sensitivity to metallicity must be investigated in detail with, for example, resolved photometry of M-dwarf wide common proper motion companions to FGK-type stars with well-determined stellar astrophysical parameters \citep{Mon18,Esp19}.

    \subsubsection{Absolute magnitude-colour}
    \label{sssection:diagrams_absmag_colour}

\begin{table*}
\caption{Fit parameters for the empirical relations$^a$.}
\label{tab:eqn_parameters}
\centering
\footnotesize{
\begin{tabular}{ll cccccc cc}
        \hline
        \hline 
        \noalign{\smallskip}
        $Y$ & $X$  & ${a}$ & ${b}$ & ${c}$ & ${d}$ & ${e}$ & $R^2$ & $\Delta X$ \\
        \relax
        (mag) & (mag) & (mag) &  (mag$^{-1}$) & (mag$^{-2}$) &  (mag$^{-3}$) & (mag$^{-4}$)   & & (mag) \\
        \noalign{\smallskip}
        \hline
        \noalign{\smallskip}
        $M_{r'}$ & $r'-J$ & +8.38 $\pm$ 2.68  & --2.74 $\pm$ 2.36 & +1.47 $\pm$ 0.68 & --0.132 $\pm$ 0.063 & 0 & 0.9398  & [2.0, 5.1]  \\
        $M_G$ & $G-J$ & +16.24 $\pm$ 4.57 & --13.04 $\pm$ 4.80 & +5.64 $\pm$ 1.66  & --0.622 $\pm$ 0.188  & 0 & 0.9308  & [2.0, 4.0] \\ 
        \noalign{\smallskip}
        \hline
        \noalign{\smallskip} 
        & (mag) & (mag) & (mag$^{-1}$) & (mag$^{-2}$) & (mag$^{-3}$) &  (mag$^{-4}$)  & & (mag) \\
        \noalign{\smallskip} 
        \hline
        \noalign{\smallskip} 
        $\log{L/L_\odot}$ & $M_J$ & +2.051 $\pm$ 0.075 & --0.662 $\pm$ 0.030 & +0.0267 $\pm$ 0.0039 & --0.00102 $\pm$ 0.00016  & 0 & 0.9923 & [4.4, 11.2)  \\
        &  & --3.906 $\pm$ 0.998 &  +0.334 $\pm$ 0.156 & --0.0263 $\pm$ 0.0061 &  0 & 0 & 0.9477 & [11.2, 14.8] \\
        $\log{L/L_\odot}$ & $M_G$  & +0.145 $\pm$ 0.201 & +0.074 $\pm$ 0.060 & --0.0382 $\pm$ 0.0060  & +0.00119 $\pm$ 0.00019 & 0  &  0.9901 & [6.4, 14.0)  \\
        & & --2.329 $\pm$ 0.687 & +0.092 $\pm$ 0.084 & --0.0103 $\pm$ 0.0025 & 0  & 0 &  0.9782 & [14.0, 20.2] \\ 
        \noalign{\smallskip}
        \hline
        \noalign{\smallskip} 
        (mag) & (mag) & (mag) & (mag$^{-1}$) & (mag$^{-2}$) & (mag$^{-3}$) &  (mag$^{-4}$)  &  & (mag) \\
        \noalign{\smallskip} 
        \hline
        \noalign{\smallskip} 
        $BC_G$ & $G-J$  & +0.404 $\pm$ 0.187 & +0.161 $\pm$ 0.239  & --0.465 $\pm$ 0.112 & +0.1159 $\pm$ 0.0225 & --0.0115 $\pm$ 0.0017 &  0.9960 & (1.5, 5.4]  \\ 
        $BC_{r'}$ & $r'-J$  & +0.557 $\pm$ 0.085 & --0.036 $\pm$ 0.091 & --0.318 $\pm$ 0.035 &  +0.0552 $\pm$ 0.0056 & --0.0037 $\pm$ 0.0003  &  0.9983  & (1.5, 7.5]   \\ 
       $BC_J$ & $G-J$ & +0.576 $\pm$ 0.094 & +0.735 $\pm$ 0.104 & --0.132 $\pm$ 0.038 & +0.0115 $\pm$ 0.0045 & 0 & 0.9547 & (0.5, 4.0]  \\ 
       $BC_{W2}$ & $G-J$ & --2.592 $\pm$ 0.667 & +5.845 $\pm$ 1.005 & --2.611 $\pm$ 0.559 & +0.586 $\pm$ 0.136 & --0.0496 $\pm$ 0.0122 & 0.9727 & (1.5, 4.0]  \\ 
\noalign{\smallskip}
\hline
\end{tabular}}
\tablefoot{
\tablefoottext{a}{In all cases, the polynomial fits follow the form $Y = a + b X + c X^{2} + d X^{3} + e X^{4} $ and are applicable in the range $\Delta X$. In all cases, $R^2$ is the correlation coefficient from the Pearson product-moment matrix.
{These relations should be applied to solar-metallicity stars only (Sect.~\ref{section:discussion}).}
More significant figures are available at GitHub (Sect.~\ref{section:summary}).
}
}
\end{table*}

\begin{figure}[]
    \centering
    \includegraphics[width=.99\linewidth]{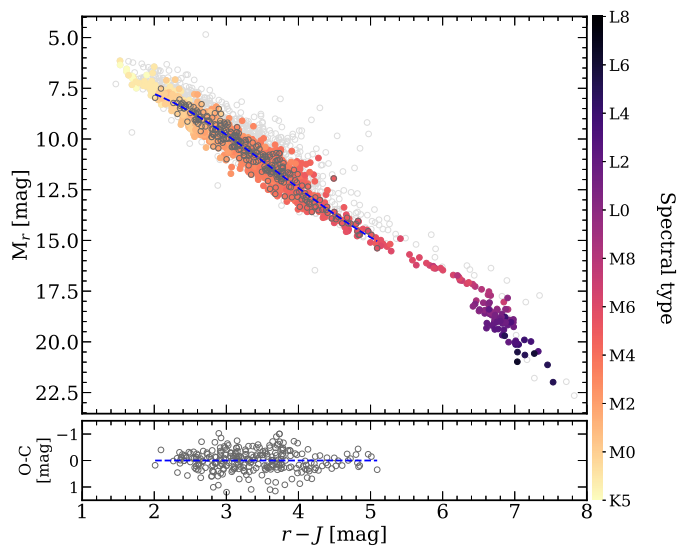}
    \caption{Same as Fig.~\ref{fig:HR} but for $M_{r'}$ vs. $r'-J$. 
    The GTO stars in the sample are shown in dark grey. 
    The blue dashed line represents the polynomial fit given in Table~\ref{tab:eqn_parameters}.
    {The fit residuals are shown in the small bottom panel.} 
    }
    \label{fig:Mabs-color_fit}
\end{figure}

In Fig.~\ref{fig:HR} we show the $M_G$ versus $G-J$ diagram. 
In Fig.~\ref{fig:Mabs-color_fit} we show a similar diagram \citep[see more examples in e.g.][]{Dup12}, but for $r'$ instead of $G$, and we overplot a quadratic polynomial fit to 278 CARMENES GTO target stars with spectral types ranging from K7\,V to M9\,V  \citep{Rei18b}.
All of them have well-behaved {\em Gaia} astrometric solutions (i.e. {\tt RUWE} $<$ 1.41; Fig.~\ref{fig:distances_RUWE}) and do not have close companions \citep{Cor17,Bar18}, extreme values of metallicity \citep{Alo15a,Pas18,Pas19,Pas20}, young ages \citep{Tal18}, or large-amplitude photometric variability \citep{Die19}. 
We also fitted another quadratic polynomial to the $M_G$ versus $G-J$ data of the GTO stars.
Thus, with the parameter fits in Table~\ref{tab:eqn_parameters} and only $r'$ or $G$ and $J$ magnitudes, one can estimate a stellar distance with a median accuracy of  36\,\% for stars in the colour ranges listed in the column $\Delta X$.
From our knowledge of the CARMENES GTO stars, the most important contributor to the fit uncertainty is not the parallax or magnitude error, stellar variability, or unresolved multiplicity, but the intrinsic scatter of the M-dwarf colour sequence due to different metallicity.

The $M_G$ versus $G-J$ relation is particularly helpful because, although there are about 420 million sources with known {\it Gaia} DR2 and 2MASS magnitudes \citep{Mar19}, there are several million near-infrared sources that lack a parallax determination. 
However, for the 31 single stars in our sample without published trigonometric parallaxes, we estimated photometric distances homogeneously from the $M_{r'}$ versus $r'-J$ relation assuming null extinction. 
Because of the relatively large uncertainty in the estimates, we did not use these photometric distances throughout our work, but only tabulated them in the on-line summary table described below.
{In general, we only recommend the use of these relations for estimating photometric distances for stars with solar-like metallicity, as well as good photometric quality
\citep[e.g.][{and see Sect.~\ref{section:discussion}}]{Boc07}.
}


    \subsection{Absolute magnitudes and bolometric corrections}
    \label{ssection:diagrams_absmag_luminosity}

\begin{figure}[]
    \centering
    \includegraphics[width=0.49\textwidth]{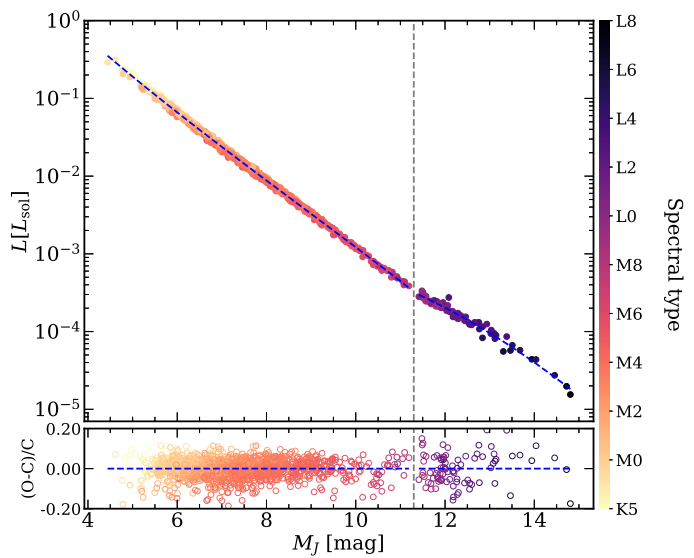}
    \caption{Same as Fig.~\ref{fig:HR}, but for $L_{\rm VOSA}$ vs. $M_J$ {and normalised fit residuals in the small bottom panel}.
    The vertical dashed line separates the three-degree (late-K and M dwarfs) and two-degree (L dwarfs) fit ranges.
        }
    \label{fig:Mabs_Lbol}
\end{figure}

The absolute magnitude of a star is directly related to its bolometric luminosity.
In our sample, we found that the $J$-band absolute magnitude, $M_J$, provides the correlation with VOSA luminosity that is most complete and that has the smallest scatter.
Figure~\ref{fig:Mabs_Lbol} shows $L_{\rm VOSA}$ (in solar units) versus $M_J$ fitted in the late-K- to late-M- and L-dwarf domains, with three-degree and two-degree polynomials, respectively.
Although in Table~\ref{table.colours} we list the fit parameters for both luminosities from 2MASS $J$ and {\em Gaia} $G$, we preferred $J$ over $G$ because the larger effective width of the broad {\em Gaia} passband introduces more dispersion in the data, quantified by $R^2$.
With these relationships in the M-dwarf domain, it is possible to estimate bolometric luminosities from absolute magnitudes $M_J$ and $M_G$ with a relative precision of 4.2\,\% and 4.5\,\%, respectively.

\begin{figure}[]
    \centering
    \includegraphics[width=.99\linewidth]{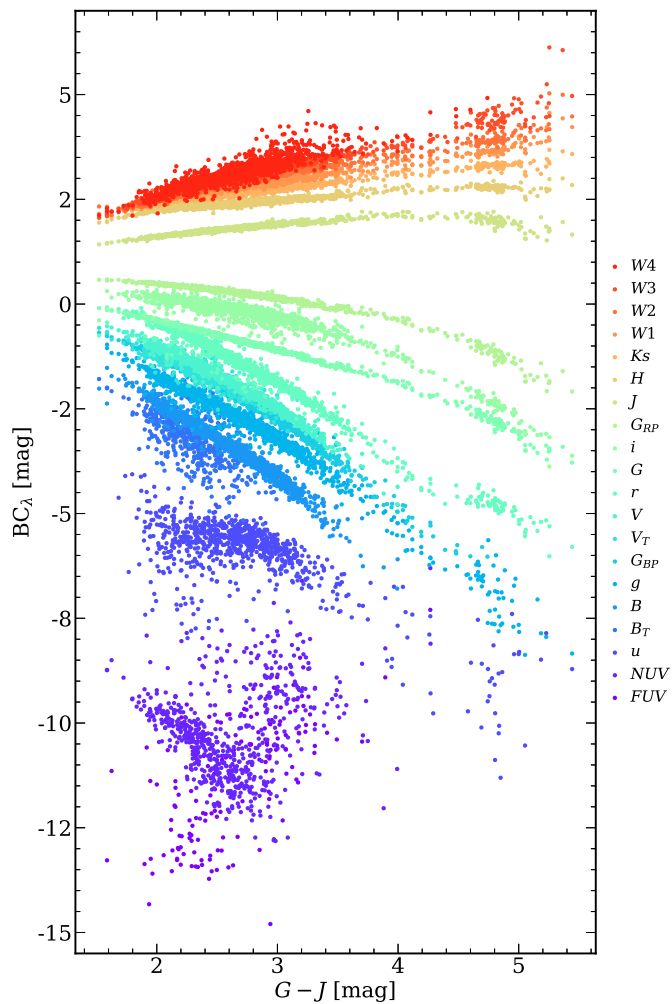}
    \caption{Bolometric corrections {for every star and passband} vs. $G-J$ colour.
    {The coloured sets contain all stars with available photometry in $G$, $J$, and the respective passband.}}
    \label{fig:BCall_GJ}
\end{figure}

\begin{figure}[]
    \centering
    \includegraphics[width=.99\linewidth]{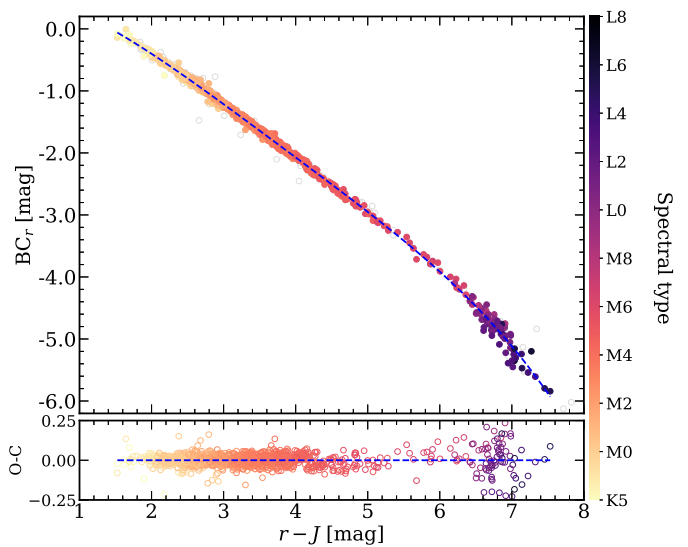}
    \caption{Same as Fig.~\ref{fig:HR} but for
    BC$_{r'}$ vs. $r'-J$.
    A polynomial fit is shown as a blue dashed line.
        }
    \label{fig:BC_colour}
\end{figure}

We calculated bolometric corrections, $BC_\lambda = M_{\rm bol} - M_\lambda$, for each investigated passband and plot them in Fig.~\ref{fig:BCall_GJ}.
For the calculation, we followed the sign criterion of \citet{Boh89} and the definition of the absolute bolometric magnitude $M_{\rm bol}$ by IAU Resolution B2 \citep{Mam15}, which is independent of the solar luminosity,
    \begin{equation}
    M_{\rm bol} = -2.5 \log_{10} {\frac{L_\star}{L_0}} = -2.5 \log_{10} L_\star + M_{\rm bol,0},
    \end{equation}
\noindent where $L_\star$ and $L_0$ are the luminosity of the star and the zero point of the absolute bolometric magnitude scale, respectively, and $M_{\rm bol,0} \equiv$ 71.197425\,mag.

From the sample of 2479 stars, for the following analysis we discarded:
($i$) stars with poor photometric or astrometric behaviour based on quality indicators (Sects.~\ref{ssection:photometry} and~\ref{ssection:distances}), 
($ii$) close binaries and stars with photometry contaminated by bright nearby companions ($\rho <$ 5\,arcsec; Sect.~\ref{ssection:multiplicity}), 
($iii$) overluminous objects known to belong to young associations and moving groups (Sect.~\ref{ssection:young}), and
($iv$) stars with extraordinarily anomalous colours or absolute magnitudes.

Of the different $BC_\lambda$ versus $G-J$ combinations in Fig.~\ref{fig:BCall_GJ}, the narrowest sequence is that of $BC_G$.
However, as illustrated by Fig.~\ref{fig:BC_colour}, the $BC_{r'}$ versus $r'-J$ sequence is even less scattered and spans wider ranges in 
X ((1.5,~7.5]\,mag in $r'-J$ versus (1.5,~5.4]\,mag in $G-J$) and
Y ([--5.8,~0.0]\,mag in $BC_{r'}$ versus [--3.8,~--0.1]\,mag in $BC_G$), probably due again to the broad $G$ effective width.
We fitted polynomials to the relations $BC_G$ versus $G-J$, $BC_{r'}$ versus $r'-J$, $BC_J$ versus $G-J$, and $BC_{W2}$ versus $G-J$, and provide the corresponding parameters and correlation coefficients in Table~\ref{tab:eqn_parameters}.
All in all, these relationships are complementary and can help to estimate relatively precise luminosities of M dwarfs with only a handful of widely available data ($G$ and $\varpi$ from {\em Gaia}, $J$ from 2MASS, $r'$ from a number of surveys including the forthcoming LSST).


    \subsection{Masses and radii}
    \label{ssection:masses}

Finally, we derived radii $\mathcal{R}$ and masses $\mathcal{M}$ of the well-behaved stars.
For $\mathcal{R}$, we used the Stefan-Boltzmann law $L = 4 \pi \mathcal{R}^2 \sigma T_{\rm eff}^4$ and $L$ and $T_{\rm eff}$ from VOSA.
For $\mathcal{M}$ we used the $\mathcal{M}$-$\mathcal{R}$ relation in Eq.~6 of \cite{Sch19}, which came from a compilation of detached, double-lined, double-eclipsing, main-sequence, M-dwarf binaries from the literature\footnote{$\mathcal{M} = \alpha + \beta \mathcal{R}$, with $\alpha$ = --0.0240\,$\pm$\,0.0076\,$M_\odot$, $\beta$ = 1.055\,$\pm$\,0.017\,$M_\odot / R_\odot$, and $\mathcal{M}$ and $\mathcal{R}$ in solar units.}.
This relation is applicable in a wide range of metallicities for M dwarfs older than a few hundred million years.
VOSA also computes two stellar radii, one from a {model dependent} dilution factor and $d$, the other {using the Stefan-Boltzmann law}, but we did not use them.


\section{Discussion}
\label{section:discussion}

\begin{table*}
\caption{Average astrophysical parameters for K5\,V to L2.0 objects.}
\label{tab:parameters-1}
\centering
\begin{tabular}{l cccccccc c}
\hline
\hline 
\noalign{\smallskip}
Spectral & $BC_G$ & $BC_J$ &  $L$  & $T_{\rm eff}$ & $\mathcal{R}$ & $\mathcal{M}$  & $N$ \\
type & (mag) & (mag) & ($10^{-4} L_\odot$) &  (K) & ($\mathcal{R_\odot}$) & ($\mathcal{M_\odot}$) &  \\
\noalign{\smallskip}
\hline
\noalign{\smallskip}
K5\,V & --0.206 $\pm$ 0.065 & 1.490 $\pm$ 0.047 & 1800 $\pm$ 420 & 4400 $\pm$ 180 & 0.693 $\pm$ 0.054  & 0.707 $\pm$ 0.057 & 13 \\
K7\,V & --0.393 $\pm$ 0.046 & 1.615 $\pm$ 0.027 & 960 $\pm$ 210 & 4050 $\pm$ 100 & 0.635 $\pm$ 0.046  & 0.646 $\pm$ 0.048 & 75 \\
M0.0\,V & --0.469 $\pm$ 0.082 & 1.654 $\pm$ 0.042 & 757 $\pm$ 230 & 3900 $\pm$ 140 & 0.613 $\pm$ 0.060  & 0.622 $\pm$ 0.063 & 104 \\
M0.5\,V & --0.570 $\pm$ 0.067 & 1.690 $\pm$ 0.038 & 585 $\pm$ 210 & 3800 $\pm$ 110 & 0.571 $\pm$ 0.076  & 0.578 $\pm$ 0.080 & 60 \\
M1.0\,V & --0.605 $\pm$ 0.054 & 1.719 $\pm$ 0.033 & 496 $\pm$ 150 & 3700 $\pm$ 85 & 0.550 $\pm$ 0.075  & 0.556 $\pm$ 0.079 & 112 \\
M1.5\,V & --0.664 $\pm$ 0.066 & 1.741 $\pm$ 0.037 & 409 $\pm$ 160 & 3600 $\pm$ 90 & 0.519 $\pm$ 0.082  & 0.524 $\pm$ 0.086 & 96 \\
M2.0\,V & --0.746 $\pm$ 0.077 & 1.769 $\pm$ 0.035 & 306 $\pm$ 130 & 3500 $\pm$ 105 & 0.473 $\pm$ 0.088  & 0.475 $\pm$ 0.093 & 99 \\
M2.5\,V & --0.825 $\pm$ 0.082 & 1.796 $\pm$ 0.034 & 228 $\pm$ 96 & 3400 $\pm$ 97 & 0.433 $\pm$ 0.086  & 0.432 $\pm$ 0.090 & 118 \\
M3.0\,V & --0.915 $\pm$ 0.085 & 1.827 $\pm$ 0.040 & 161 $\pm$ 74 & 3300 $\pm$ 87 & 0.389 $\pm$ 0.085  & 0.386 $\pm$ 0.090 & 144 \\
M3.5\,V & --0.985 $\pm$ 0.096 & 1.863 $\pm$ 0.043 & 111 $\pm$ 57 & 3300 $\pm$ 92 & 0.343 $\pm$ 0.082  & 0.338 $\pm$ 0.087 & 193 \\
M4.0\,V & --1.043 $\pm$ 0.093 & 1.890 $\pm$ 0.043 & 87 $\pm$ 47 & 3200 $\pm$ 88 & 0.309 $\pm$ 0.079  & 0.302 $\pm$ 0.083 & 170 \\
M4.5\,V & --1.160 $\pm$ 0.103 & 1.920 $\pm$ 0.041 & 50 $\pm$ 27 & 3100 $\pm$ 88 & 0.263 $\pm$ 0.069  & 0.253 $\pm$ 0.073 & 88 \\
M5.0\,V & --1.236 $\pm$ 0.122 & 1.951 $\pm$ 0.039 & 28 $\pm$ 13 & 3100 $\pm$ 58 & 0.207 $\pm$ 0.041  & 0.195 $\pm$ 0.043 & 52 \\
M5.5\,V & --1.420 $\pm$ 0.116 & 1.997 $\pm$ 0.057 & 20.1 $\pm$ 8.3 & 3000 $\pm$ 85 & 0.173 $\pm$ 0.032  & 0.159 $\pm$ 0.034 & 22 \\
M6.0\,V & --1.572 $\pm$ 0.178 & 2.062 $\pm$ 0.066 & 11.1 $\pm$ 3.9 & 2900 $\pm$ 108 & 0.138 $\pm$ 0.020  & 0.121 $\pm$ 0.021 & 14 \\
M6.5\,V & --1.837 $\pm$ 0.223 & 2.096 $\pm$ 0.065 & 7.2 $\pm$ 1.7 & 2750 $\pm$ 124 & 0.123 $\pm$ 0.011  & 0.106 $\pm$ 0.011 & 6 \\
M7.0\,V & --1.854 $\pm$ 0.129 & 2.105 $\pm$ 0.050 & 6.3 $\pm$ 1.1 & 2700 $\pm$ 94 & 0.119 $\pm$ 0.009  & 0.101 $\pm$ 0.010 & 4 \\
M7.5\,V & --2.169 $\pm$ 0.141 & 2.078 $\pm$ 0.044 & 5.8 $\pm$ 1.2 & 2500 $\pm$ 82 & 0.121 $\pm$ 0.008  & 0.104 $\pm$ 0.009 & 2 \\
M8.0\,V & --2.192 $\pm$ 0.163 & 2.082 $\pm$ 0.049 & 5.1 $\pm$ 1.6 & 2500 $\pm$ 91 & 0.121 $\pm$ 0.014  & 0.104 $\pm$ 0.014 & 7 \\
M8.5\,V & --2.342 $\pm$ 0.169 & 2.119 $\pm$ 0.052 & 3.4 $\pm$ 1.5 & 2400 $\pm$ 88 & 0.107 $\pm$ 0.015  & 0.088 $\pm$ 0.016 & 4 \\
M9.0\,V & --2.520 $\pm$ 0.158 & 2.137 $\pm$ 0.071 & 2.69 $\pm$ 0.35 & 2350 $\pm$ 86 & 0.096 $\pm$ 0.013  & 0.077 $\pm$ 0.014 & 5 \\
M9.5\,V & --2.627 $\pm$ 0.118 & 2.060 $\pm$ 0.073 & 2.35 $\pm$ 0.43 & 2300 $\pm$ 45 & 0.096 $\pm$ 0.007  & 0.077 $\pm$ 0.008 & 2 \\
L0.0 & --2.648 $\pm$ 0.102 & 2.031 $\pm$ 0.070 & 2.30 $\pm$ 0.43 & 2275 $\pm$ 59 & 0.097 $\pm$ 0.003  & 0.079 $\pm$ 0.004 & 12 \\
L0.5 & --2.746 $\pm$ 0.128 & 2.024 $\pm$ 0.081 & 2.17 $\pm$ 0.15 & 2250 $\pm$ 61 & 0.098 $\pm$ 0.004  & 0.079 $\pm$ 0.004 & 6 \\
L1.0 & --2.817 $\pm$ 0.123 & 1.974 $\pm$ 0.072 & 2.08 $\pm$ 0.26 & 2150 $\pm$ 165 & 0.101 $\pm$ 0.005  & 0.083 $\pm$ 0.006 & 15 \\
L1.5 & --2.868 $\pm$ 0.112 & 1.958 $\pm$ 0.070 & 1.81 $\pm$ 0.35 & 2000 $\pm$ 172 & 0.112 $\pm$ 0.015  & 0.094 $\pm$ 0.016 & 7 \\
L2.0 & --2.930 $\pm$ 0.101 & 1.951 $\pm$ 0.083 & 1.55 $\pm$ 0.24 & 1850 $\pm$ 92 & 0.116 $\pm$ 0.013  & 0.098 $\pm$ 0.014 & 14 \\
\noalign{\smallskip}
\hline
\end{tabular}
\end{table*}
\begin{table*}
\caption{Average absolute magnitudes for K5\,V to L2.0 objects.}
\label{tab:parameters-2}
\centering
\begin{tabular}{l cccccccc c}
\hline
\hline 
\noalign{\smallskip}
Spectral & $M_{B}$ & $M_{g'}$ &  $M_{G_{BP}}$  & $M_V$ & $M_{r'}$ & $M_{G}$  & $M_{i'}$ \\ 
type & (mag) & (mag) & (mag) & (mag) & (mag) & (mag) & (mag) \\ 
\noalign{\smallskip}
\hline
\noalign{\smallskip}
K5\,V & 8.33 $\pm$ 0.37 & 7.84 $\pm$ 0.35 & 7.32 $\pm$ 0.39 & 7.23 $\pm$ 0.28 & 6.92 $\pm$ 0.31 & 6.82 $\pm$ 0.35 & 6.52 $\pm$ 0.31 \\
K7\,V & 9.65 $\pm$ 0.44 & 9.07 $\pm$ 0.40 & 8.51 $\pm$ 0.35 & 8.32 $\pm$ 0.38 & 7.75 $\pm$ 0.41 & 7.66 $\pm$ 0.30 & 7.11 $\pm$ 0.37 \\
M0.0\,V & 10.11 $\pm$ 0.56 & 9.47 $\pm$ 0.50 & 8.93 $\pm$ 0.49 & 8.74 $\pm$ 0.49 & 8.14 $\pm$ 0.48 & 7.98 $\pm$ 0.41 & 7.42 $\pm$ 0.36 \\
M0.5\,V & 10.57 $\pm$ 0.52 & 9.89 $\pm$ 0.49 & 9.36 $\pm$ 0.49 & 9.11 $\pm$ 0.48 & 8.59 $\pm$ 0.48 & 8.33 $\pm$ 0.41 & 7.72 $\pm$ 0.40 \\
M1.0\,V & 10.90 $\pm$ 0.49 & 10.20 $\pm$ 0.45 & 9.70 $\pm$ 0.45 & 9.47 $\pm$ 0.47 & 8.87 $\pm$ 0.44 & 8.61 $\pm$ 0.39 & 7.92 $\pm$ 0.37 \\
M1.5\,V & 11.24 $\pm$ 0.57 & 10.48 $\pm$ 0.53 & 9.99 $\pm$ 0.53 & 9.76 $\pm$ 0.52 & 9.15 $\pm$ 0.51 & 8.86 $\pm$ 0.45 & 8.17 $\pm$ 0.45 \\
M2.0\,V & 11.72 $\pm$ 0.65 & 10.98 $\pm$ 0.61 & 10.47 $\pm$ 0.60 & 10.19 $\pm$ 0.58 & 9.63 $\pm$ 0.59 & 9.23 $\pm$ 0.53 & 8.55 $\pm$ 0.53 \\
M2.5\,V & 12.19 $\pm$ 0.65 & 11.40 $\pm$ 0.63 & 10.97 $\pm$ 0.62 & 10.67 $\pm$ 0.62 & 10.10 $\pm$ 0.61 & 9.65 $\pm$ 0.54 & 8.94 $\pm$ 0.54 \\
M3.0\,V & 12.77 $\pm$ 0.71 & 11.98 $\pm$ 0.66 & 11.52 $\pm$ 0.65 & 11.22 $\pm$ 0.66 & 10.67 $\pm$ 0.66 & 10.08 $\pm$ 0.57 & 9.38 $\pm$ 0.58 \\
M3.5\,V & 13.28 $\pm$ 0.80 & 12.47 $\pm$ 0.78 & 12.01 $\pm$ 0.77 & 11.71 $\pm$ 0.75 & 11.16 $\pm$ 0.75 & 10.51 $\pm$ 0.67 & 9.79 $\pm$ 0.67 \\
M4.0\,V & 13.75 $\pm$ 0.87 & 12.91 $\pm$ 0.84 & 12.47 $\pm$ 0.83 & 12.12 $\pm$ 0.81 & 11.61 $\pm$ 0.81 & 10.88 $\pm$ 0.72 & 10.13 $\pm$ 0.72 \\
M4.5\,V & 14.39 $\pm$ 0.94 & 13.62 $\pm$ 0.90 & 13.15 $\pm$ 0.89 & 12.77 $\pm$ 0.87 & 12.26 $\pm$ 0.86 & 11.44 $\pm$ 0.77 & 10.67 $\pm$ 0.77 \\
M5.0\,V & 15.30 $\pm$ 1.03 & 14.52 $\pm$ 1.05 & 14.10 $\pm$ 1.04 & 13.63 $\pm$ 1.00 & 13.15 $\pm$ 0.99 & 12.25 $\pm$ 0.81 & 11.42 $\pm$ 0.83 \\
M5.5\,V & 16.45 $\pm$ 1.03 & 15.49 $\pm$ 0.98 & 15.04 $\pm$ 0.98 & 14.59 $\pm$ 0.89 & 14.05 $\pm$ 0.97 & 12.88 $\pm$ 0.55 & 12.09 $\pm$ 0.61 \\
M6.0\,V & 17.21 $\pm$ 0.70 & 16.59 $\pm$ 0.99 & 16.07 $\pm$ 0.96 & 15.25 $\pm$ 0.39 & 15.14 $\pm$ 0.93 & 13.64 $\pm$ 0.64 & 12.86 $\pm$ 0.64 \\
M6.5\,V & 18.65 $\pm$ 0.78 & 17.76 $\pm$ 0.91 & 17.40 $\pm$ 0.89 & 16.92 $\pm$ 0.81 & 16.36 $\pm$ 0.83 & 14.42 $\pm$ 0.58 & 13.73 $\pm$ 0.58 \\
M7.0\,V & 18.92 $\pm$ 0.36 & 18.18 $\pm$ 0.40 & 17.66 $\pm$ 0.36 & 17.16 $\pm$ 0.23 & 16.76 $\pm$ 0.51 & 14.55 $\pm$ 0.22 & 14.05 $\pm$ 0.27 \\
M7.5\,V & \ldots & 18.62 $\pm$ 0.57 & 18.11 $\pm$ 0.54 & \ldots & 17.07 $\pm$ 0.51 & 15.01 $\pm$ 0.39 & 14.33 $\pm$ 0.36 \\
M8.0\,V &  \ldots  & 18.96 $\pm$ 0.65 & 18.53 $\pm$ 0.81 &  \ldots  & 17.41 $\pm$ 0.60 & 15.19 $\pm$ 0.51 & 14.50 $\pm$ 0.57 \\
M8.5\,V &  \ldots  & 19.28 $\pm$ 0.60 & 19.39 $\pm$ 0.69 &  \ldots  & 17.72 $\pm$ 0.52 & 15.76 $\pm$ 0.55 & 15.07 $\pm$ 0.65 \\
M9.0\,V &  \ldots  & 20.11 $\pm$ 0.32 & 19.59 $\pm$ 0.32 &  \ldots  & 18.11 $\pm$ 0.24 & 16.11 $\pm$ 0.25 & 15.67 $\pm$ 0.26 \\
M9.5\,V &  \ldots  & 20.23 $\pm$ 0.30 & 19.64 $\pm$ 0.25 &  \ldots  & 18.41 $\pm$ 0.17 & 16.39 $\pm$ 0.20 & 15.83 $\pm$ 0.29 \\
L0.0 &  \ldots  & 20.26 $\pm$ 0.33 &  \ldots  &  \ldots  & 18.41 $\pm$ 0.23 & 16.51 $\pm$ 0.21 & 16.06 $\pm$ 0.22 \\
L0.5 &  \ldots  & 20.56 $\pm$ 0.68 &  \ldots  &  \ldots  & 18.61 $\pm$ 0.26 & 16.69 $\pm$ 0.24 & 16.28 $\pm$ 0.26 \\
L1.0 &  \ldots  & 21.13 $\pm$ 0.68 &  \ldots  &  \ldots  & 18.85 $\pm$ 0.19 & 16.80 $\pm$ 0.19 & 16.38 $\pm$ 0.21 \\
L1.5 &  \ldots  & 21.20 $\pm$ 0.55 &  \ldots  &  \ldots  & 18.94 $\pm$ 0.19 & 16.97 $\pm$ 0.25 & 16.54 $\pm$ 0.24 \\
L2.0 &  \ldots  & 21.17 $\pm$ 0.56 &  \ldots  &  \ldots  & 19.07 $\pm$ 0.17 & 17.15 $\pm$ 0.22 & 16.68 $\pm$ 0.20 \\
\noalign{\smallskip}
\hline
\noalign{\smallskip}
Spectral & $M_{G_{RP}}$ & $M_J$ &  $M_H$  & $M_{K_s}$ & $M_{W1}$ & $M_{W2}$  & $M_{W3}$ \\ 
type & (mag) & (mag) & (mag) & (mag) & (mag) & (mag) & (mag) \\ 
\noalign{\smallskip}
\hline
\noalign{\smallskip}
K5\,V & 6.06 $\pm$ 0.31 & 5.13 $\pm$ 0.26 & 4.55 $\pm$ 0.24 & 4.44 $\pm$ 0.22 & 4.33 $\pm$ 0.22 & 4.40 $\pm$ 0.32 & 4.33 $\pm$ 0.35 \\
K7\,V & 6.78 $\pm$ 0.27 & 5.66 $\pm$ 0.31 & 5.02 $\pm$ 0.29 & 4.84 $\pm$ 0.28 & 4.76 $\pm$ 0.27 & 4.78 $\pm$ 0.30 & 4.72 $\pm$ 0.28 \\
M0.0\,V & 7.07 $\pm$ 0.36 & 5.88 $\pm$ 0.30 & 5.22 $\pm$ 0.30 & 5.04 $\pm$ 0.28 & 4.93 $\pm$ 0.27 & 4.91 $\pm$ 0.22 & 4.88 $\pm$ 0.23 \\
M0.5\,V & 7.36 $\pm$ 0.38 & 6.09 $\pm$ 0.32 & 5.44 $\pm$ 0.34 & 5.24 $\pm$ 0.32 & 5.11 $\pm$ 0.30 & 5.07 $\pm$ 0.27 & 5.03 $\pm$ 0.26 \\
M1.0\,V & 7.59 $\pm$ 0.36 & 6.26 $\pm$ 0.33 & 5.63 $\pm$ 0.35 & 5.41 $\pm$ 0.33 & 5.26 $\pm$ 0.32 & 5.23 $\pm$ 0.28 & 5.18 $\pm$ 0.28 \\
M1.5\,V & 7.83 $\pm$ 0.43 & 6.44 $\pm$ 0.40 & 5.81 $\pm$ 0.43 & 5.60 $\pm$ 0.42 & 5.46 $\pm$ 0.41 & 5.34 $\pm$ 0.33 & 5.32 $\pm$ 0.37 \\
M2.0\,V & 8.16 $\pm$ 0.50 & 6.72 $\pm$ 0.45 & 6.10 $\pm$ 0.48 & 5.87 $\pm$ 0.46 & 5.73 $\pm$ 0.46 & 5.62 $\pm$ 0.41 & 5.57 $\pm$ 0.40 \\
M2.5\,V & 8.54 $\pm$ 0.52 & 7.00 $\pm$ 0.48 & 6.41 $\pm$ 0.50 & 6.17 $\pm$ 0.49 & 6.03 $\pm$ 0.48 & 5.90 $\pm$ 0.45 & 5.82 $\pm$ 0.44 \\
M3.0\,V & 8.94 $\pm$ 0.55 & 7.35 $\pm$ 0.50 & 6.77 $\pm$ 0.52 & 6.50 $\pm$ 0.51 & 6.33 $\pm$ 0.51 & 6.19 $\pm$ 0.48 & 6.11 $\pm$ 0.47 \\
M3.5\,V & 9.33 $\pm$ 0.64 & 7.67 $\pm$ 0.60 & 7.10 $\pm$ 0.61 & 6.83 $\pm$ 0.60 & 6.66 $\pm$ 0.58 & 6.48 $\pm$ 0.56 & 6.39 $\pm$ 0.55 \\
M4.0\,V & 9.68 $\pm$ 0.69 & 7.97 $\pm$ 0.64 & 7.40 $\pm$ 0.65 & 7.13 $\pm$ 0.64 & 6.95 $\pm$ 0.63 & 6.78 $\pm$ 0.61 & 6.65 $\pm$ 0.60 \\
M4.5\,V & 10.19 $\pm$ 0.75 & 8.41 $\pm$ 0.67 & 7.81 $\pm$ 0.69 & 7.56 $\pm$ 0.67 & 7.36 $\pm$ 0.66 & 7.18 $\pm$ 0.64 & 7.04 $\pm$ 0.63 \\
M5.0\,V & 10.98 $\pm$ 0.74 & 9.08 $\pm$ 0.59 & 8.53 $\pm$ 0.58 & 8.19 $\pm$ 0.56 & 7.99 $\pm$ 0.56 & 7.81 $\pm$ 0.55 & 7.63 $\pm$ 0.53 \\
M5.5\,V & 11.53 $\pm$ 0.56 & 9.48 $\pm$ 0.48 & 8.88 $\pm$ 0.49 & 8.58 $\pm$ 0.47 & 8.37 $\pm$ 0.46 & 8.17 $\pm$ 0.45 & 8.02 $\pm$ 0.42 \\
M6.0\,V & 12.10 $\pm$ 0.54 & 10.04 $\pm$ 0.42 & 9.47 $\pm$ 0.41 & 9.16 $\pm$ 0.40 & 8.88 $\pm$ 0.39 & 8.65 $\pm$ 0.40 & 8.56 $\pm$ 0.35 \\
M6.5\,V & 12.92 $\pm$ 0.52 & 10.47 $\pm$ 0.19 & 9.85 $\pm$ 0.20 & 9.48 $\pm$ 0.19 & 9.25 $\pm$ 0.20 & 9.04 $\pm$ 0.24 & 8.80 $\pm$ 0.20 \\
M7.0\,V & 13.06 $\pm$ 0.20 & 10.58 $\pm$ 0.13 & 9.97 $\pm$ 0.12 & 9.66 $\pm$ 0.18 & 9.42 $\pm$ 0.19 & 9.22 $\pm$ 0.17 & 9.01 $\pm$ 0.14 \\
M7.5\,V & 13.46 $\pm$ 0.35 & 10.82 $\pm$ 0.23 & 10.16 $\pm$ 0.21 & 9.76 $\pm$ 0.19 & 9.55 $\pm$ 0.19 & 9.32 $\pm$ 0.18 & 9.06 $\pm$ 0.15 \\
M8.0\,V & 13.62 $\pm$ 0.49 & 10.92 $\pm$ 0.34 & 10.25 $\pm$ 0.35 & 9.86 $\pm$ 0.32 & 9.60 $\pm$ 0.27 & 9.39 $\pm$ 0.24 & 9.10 $\pm$ 0.20 \\
M8.5\,V & 14.18 $\pm$ 0.53 & 11.30 $\pm$ 0.37 & 10.62 $\pm$ 0.37 & 10.14 $\pm$ 0.35 & 9.85 $\pm$ 0.30 & 9.61 $\pm$ 0.26 & 9.25 $\pm$ 0.18 \\
M9.0\,V & 14.51 $\pm$ 0.23 & 11.51 $\pm$ 0.03 & 10.91 $\pm$ 0.14 & 10.41 $\pm$ 0.24 & 10.06 $\pm$ 0.22 & 9.77 $\pm$ 0.19 & 9.38 $\pm$ 0.12 \\
M9.5\,V & 14.72 $\pm$ 0.19 & 11.68 $\pm$ 0.17 & 10.99 $\pm$ 0.18 & 10.51 $\pm$ 0.19 & 10.19 $\pm$ 0.16 & 9.94 $\pm$ 0.14 & 9.39 $\pm$ 0.14 \\
L0.0 & 14.88 $\pm$ 0.21 & 11.83 $\pm$ 0.18 & 11.07 $\pm$ 0.20 & 10.55 $\pm$ 0.21 & 10.25 $\pm$ 0.10 & 9.98 $\pm$ 0.13 & 9.41 $\pm$ 0.21 \\
L0.5 & 15.07 $\pm$ 0.24 & 11.94 $\pm$ 0.12 & 11.14 $\pm$ 0.05 & 10.61 $\pm$ 0.06 & 10.28 $\pm$ 0.09 & 10.04 $\pm$ 0.14 & 9.51 $\pm$ 0.17 \\
L1.0 & 15.18 $\pm$ 0.20 & 11.97 $\pm$ 0.15 & 11.14 $\pm$ 0.08 & 10.62 $\pm$ 0.09 & 10.32 $\pm$ 0.17 & 10.07 $\pm$ 0.17 & 9.60 $\pm$ 0.20 \\
L1.5 & 15.32 $\pm$ 0.25 & 12.08 $\pm$ 0.19 & 11.30 $\pm$ 0.19 & 10.78 $\pm$ 0.19 & 10.44 $\pm$ 0.19 & 10.17 $\pm$ 0.19 & 9.68 $\pm$ 0.20 \\
L2.0 & 15.51 $\pm$ 0.22 & 12.27 $\pm$ 0.24 & 11.47 $\pm$ 0.24 & 10.96 $\pm$ 0.22 & 10.63 $\pm$ 0.12 & 10.34 $\pm$ 0.12 & 9.77 $\pm$ 0.28 \\
\noalign{\smallskip}
\hline
\end{tabular}
\end{table*}

\begin{figure*}[]
    \centering
    \includegraphics[width=.49\textwidth]{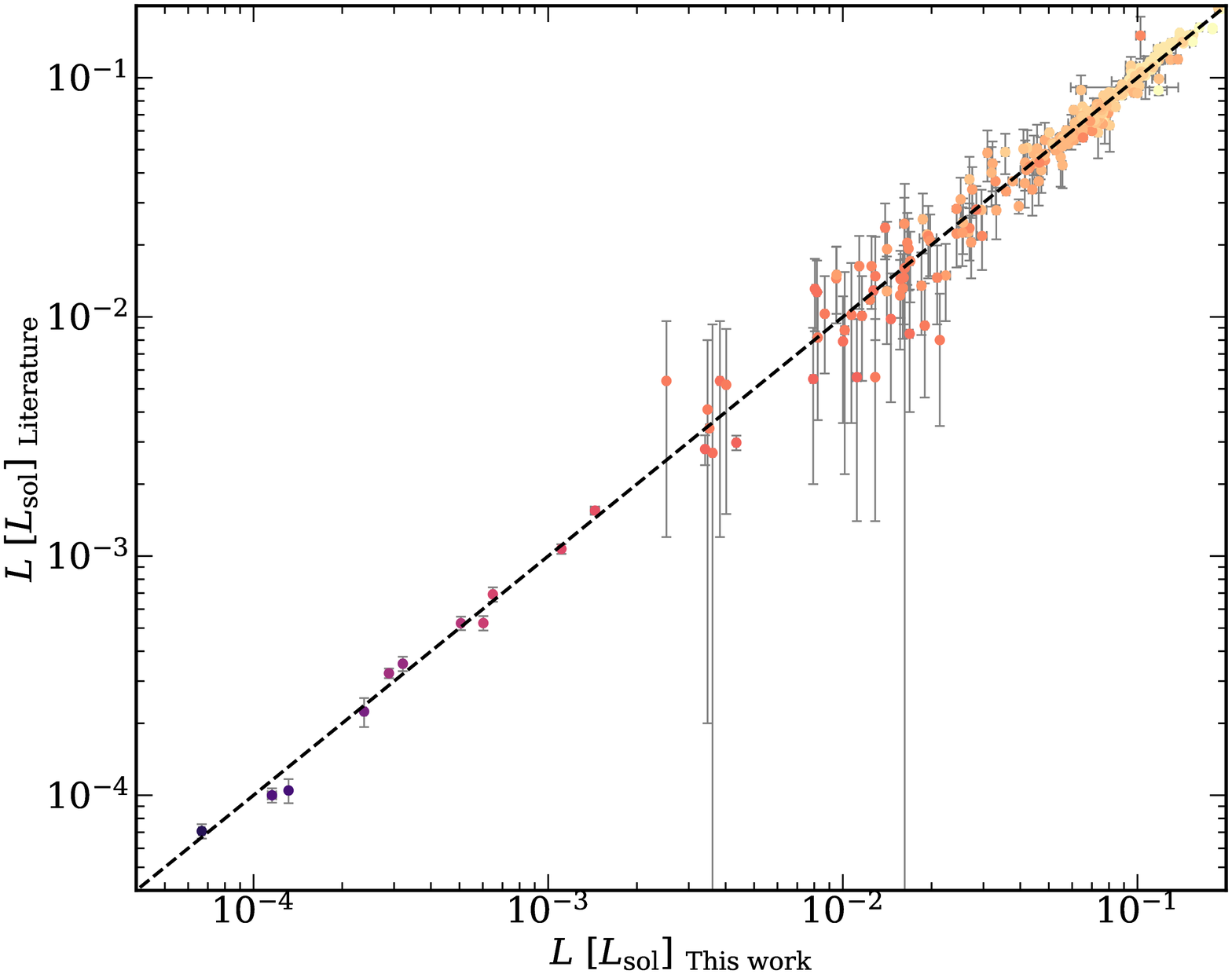}
    \includegraphics[width=.49\textwidth]{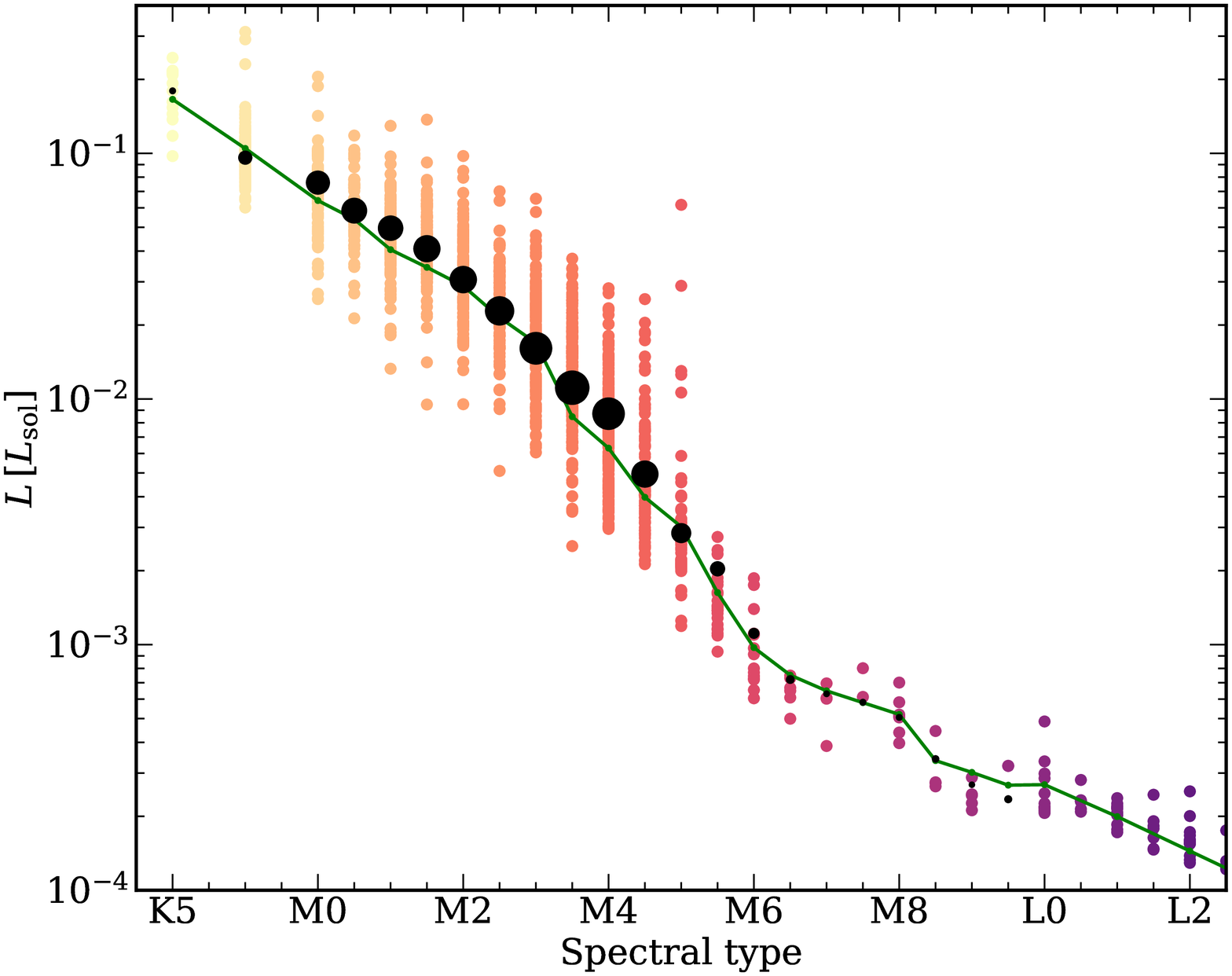} 
    \includegraphics[width=.90\linewidth]{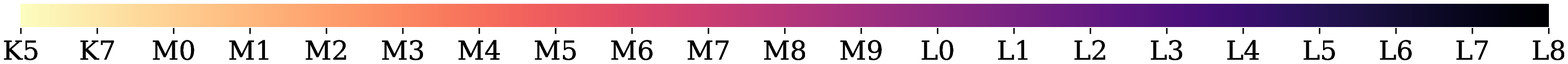}
    \caption{Comparison of $L$ from VOSA and from the literature ({left}) and individual (coloured points) and median $L$ (black circles) as a function of spectral type as in Table~\ref{tab:parameters-1} ({right}).
    In the right panel, the green line outlines the empirical $L$-spectral type sequence of \citet[][with updated values for M0.0\,V to M9.5\,V; E. E. Mamajek, priv. comm.]{Pec13} and the size of the black points is proportional to the number of stars per spectral type.
    }
    \label{fig:Discussion.L}
\end{figure*}

\begin{figure*}[]
    \centering
    \includegraphics[width=.49\textwidth]{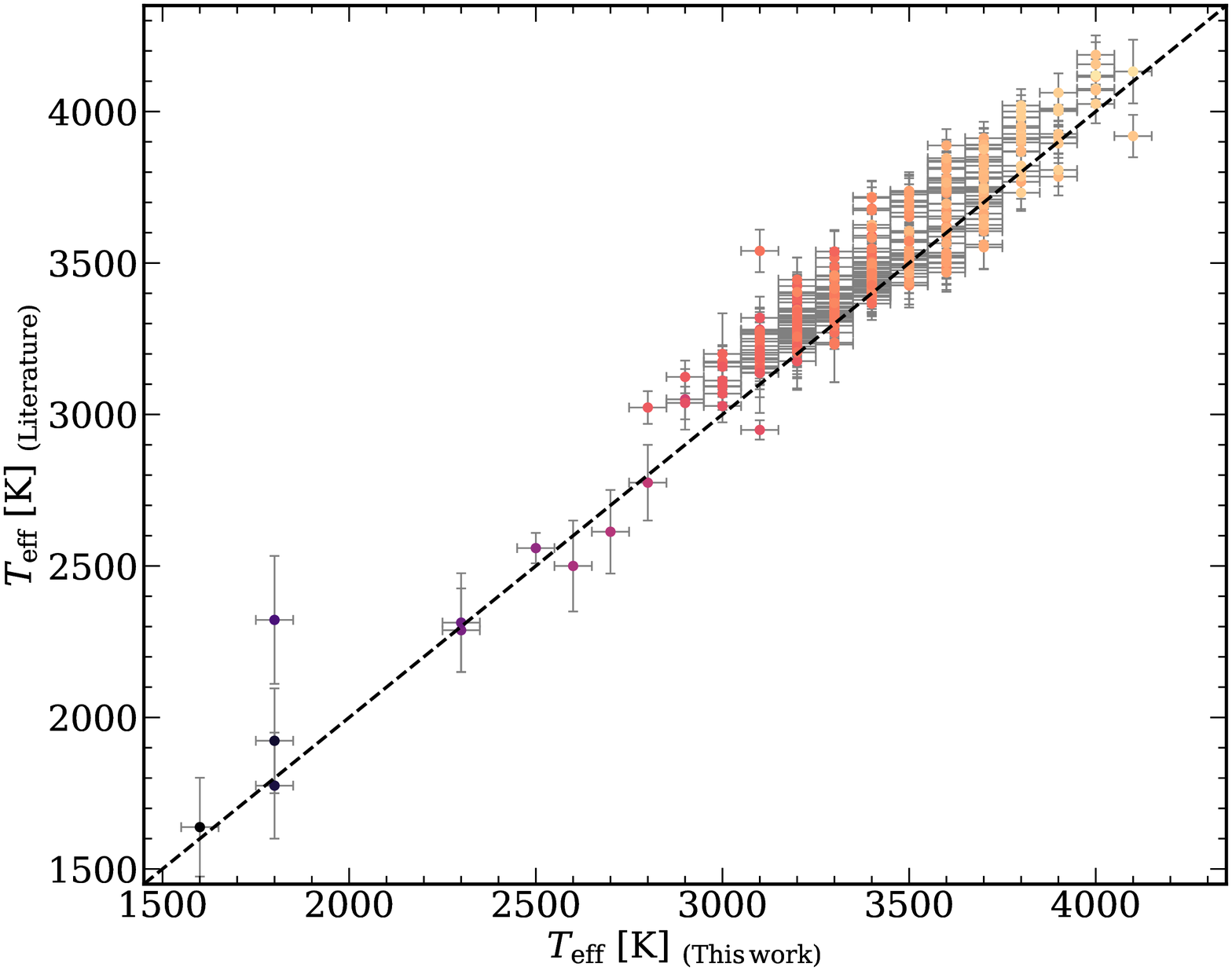}
    \includegraphics[width=.49\textwidth]{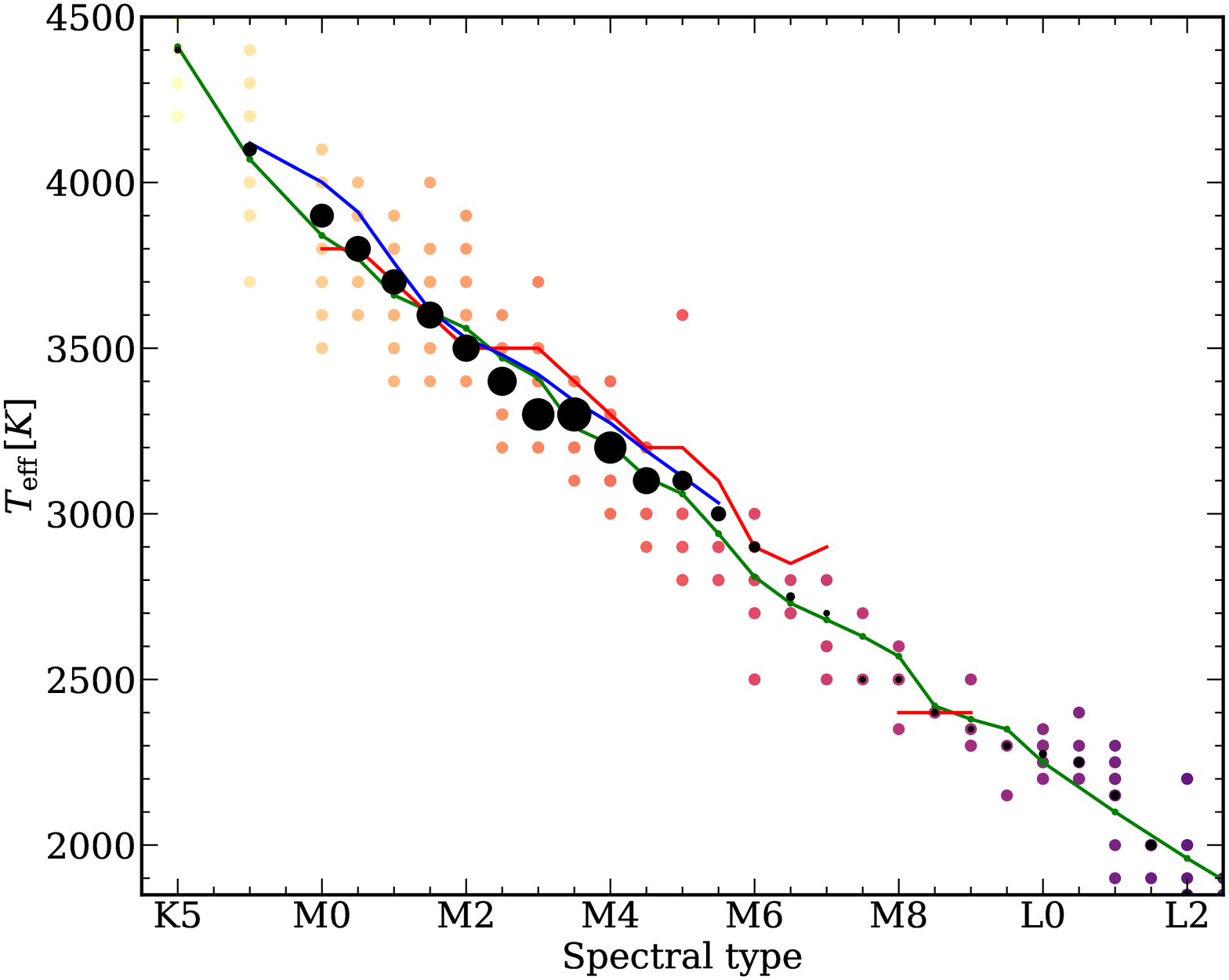}
    \includegraphics[width=.49\textwidth]{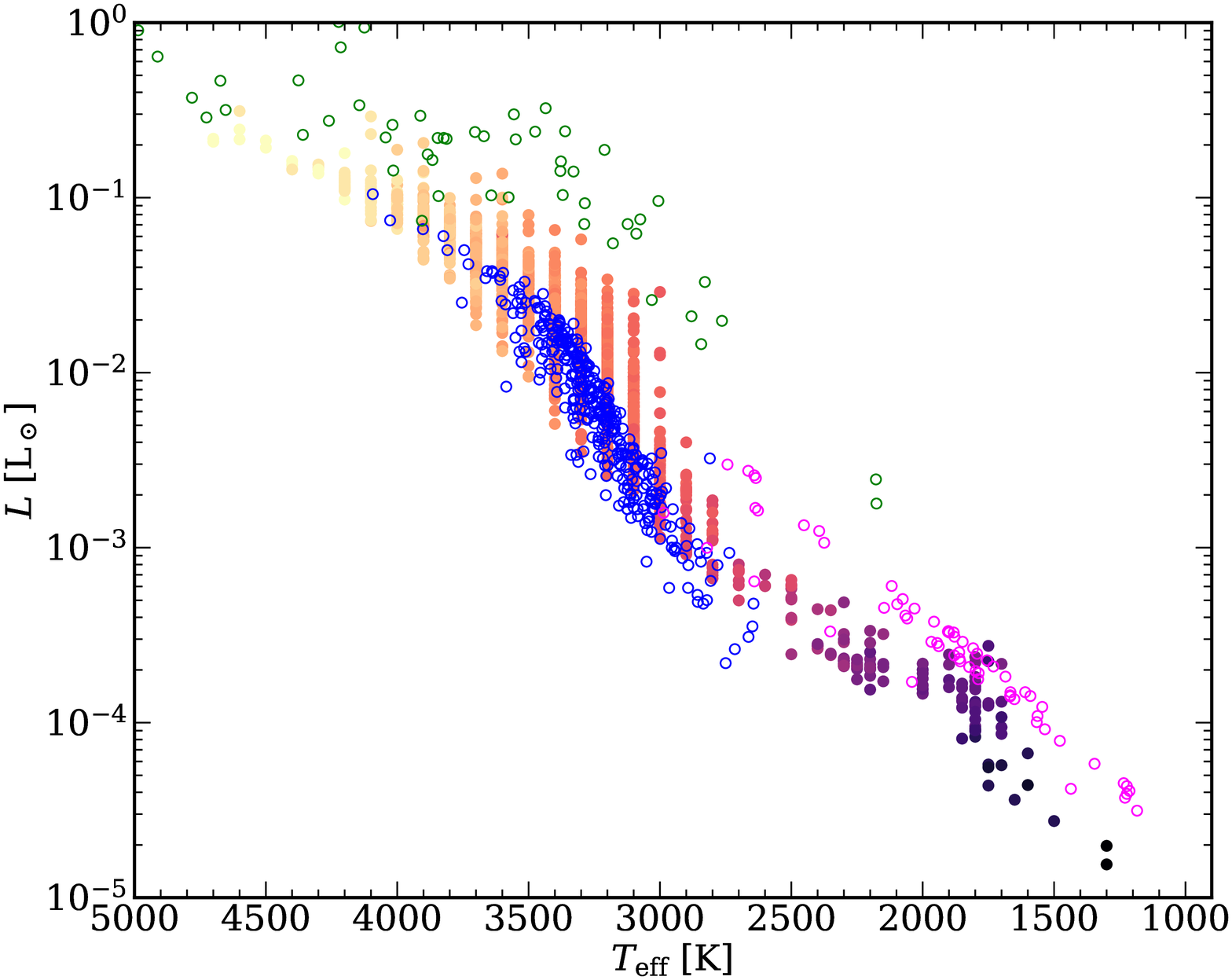}
    \includegraphics[width=.49\textwidth]{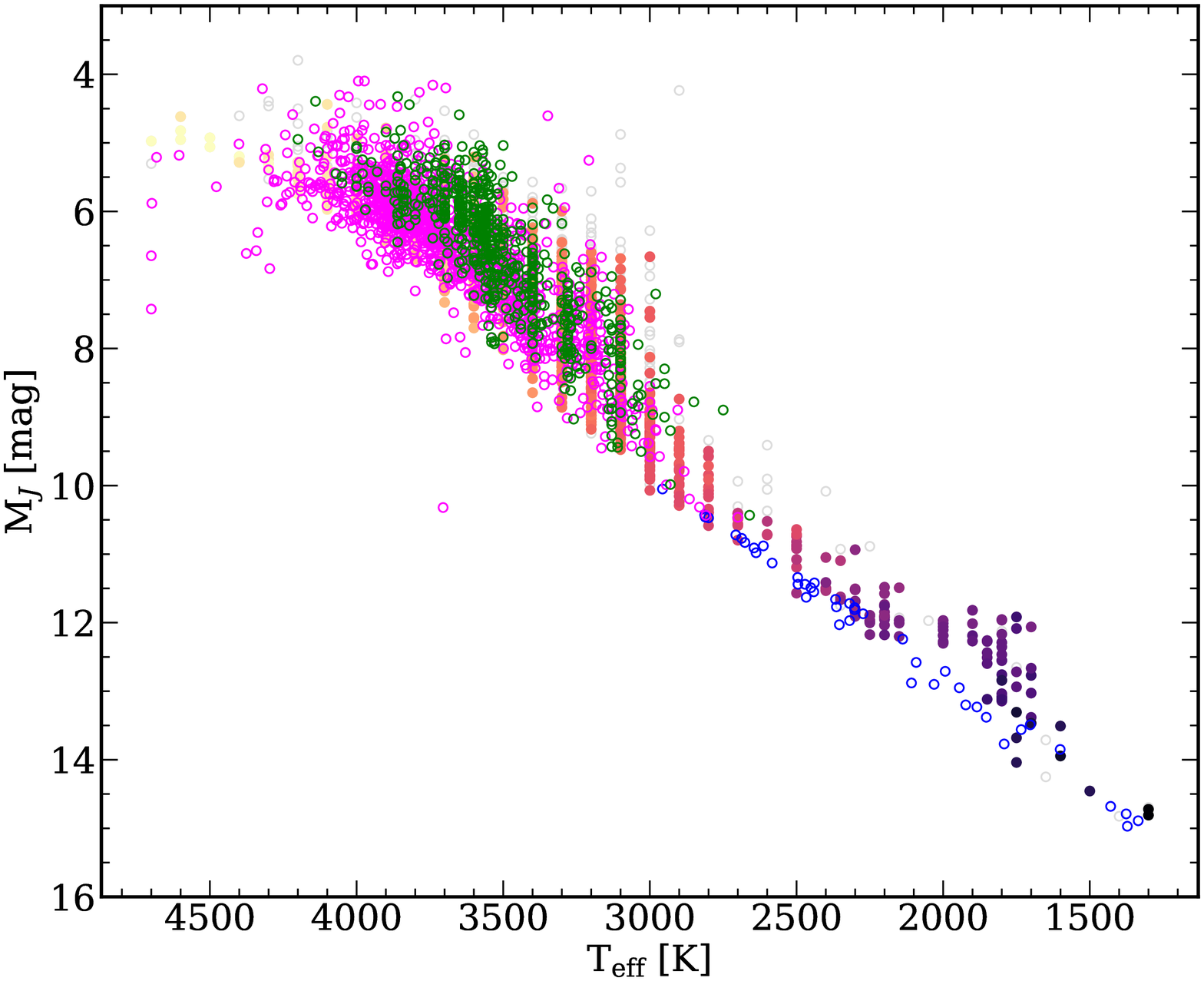}
    \includegraphics[width=.90\linewidth]{cif20_colorbar_SpT_bis.eps}
    \caption{Four representative diagrams involving $T_{\rm eff}$.
    In the four panels our investigated stars are represented with filled circles colour-coded by spectral type.
    {Top left:} 
    Comparison of $T_{\rm eff}$ from this work and from the literature.
    {Top right:} 
    Individual (coloured points) and median (black circles) values of $T_{\rm eff}$ as a function of the spectral sequence shown in Table~\ref{tab:parameters-1}.
    The size of the black circles is proportional to the number of stars per spectral type.
    The green, red, and blue lines mark the mean values tabulated by \cite{Pec13}, \cite{Raj18b}, and \cite{Pas19}, respectively.
    {Bottom left:} 
    $L$ vs. $T_{\rm eff}$.
    As a comparison we also plot pre-main sequence stars with BT-Settl model fitting from \citet[][green empty circles]{Pec13}, 
    M dwarfs in the MEarth sample with stellar parameters from \citet[][blue empty circles, inferred from the pseudo-equivalent width of Mg~{\sc i} near-infrared lines]{New15}, and high-confidence moving group members from \citet[][magenta empty circles]{Fah16} with parameters computed as in \citet{Fil15}.
    {Bottom right:} 
    $J$-band absolute magnitude vs. $T_{\rm eff}$.
    As a comparison we also plot the samples of \citet[][blue open circles]{Dah02}, \citet[][green open circles]{Lep13}, and \citet[][magenta empty circles]{Gai14a}.
    }
    \label{fig:Discussion.Teff}
\end{figure*}

\begin{figure*}[]
    \centering
    \includegraphics[width=0.49\textwidth]{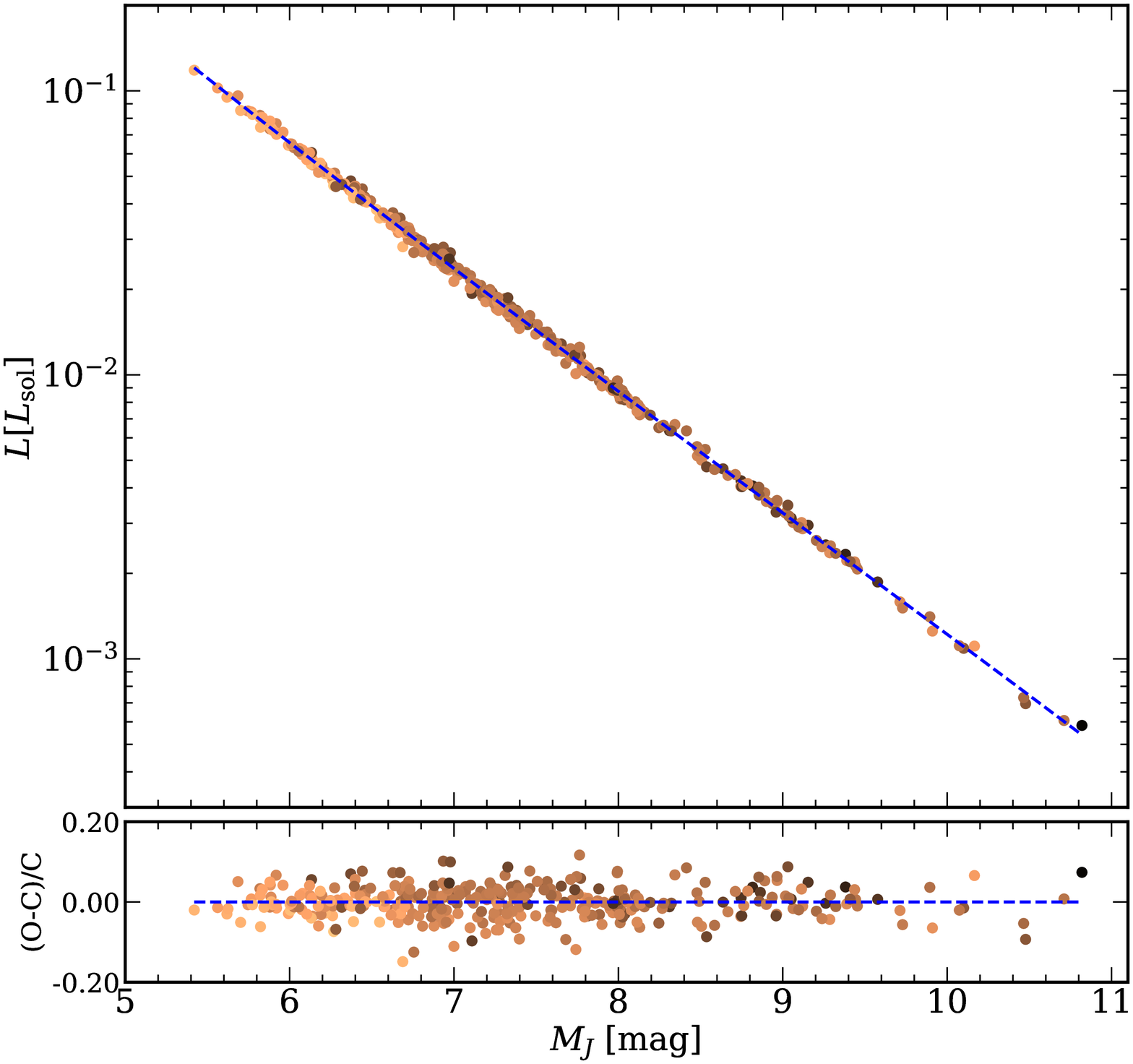}
    \includegraphics[width=0.49\textwidth]{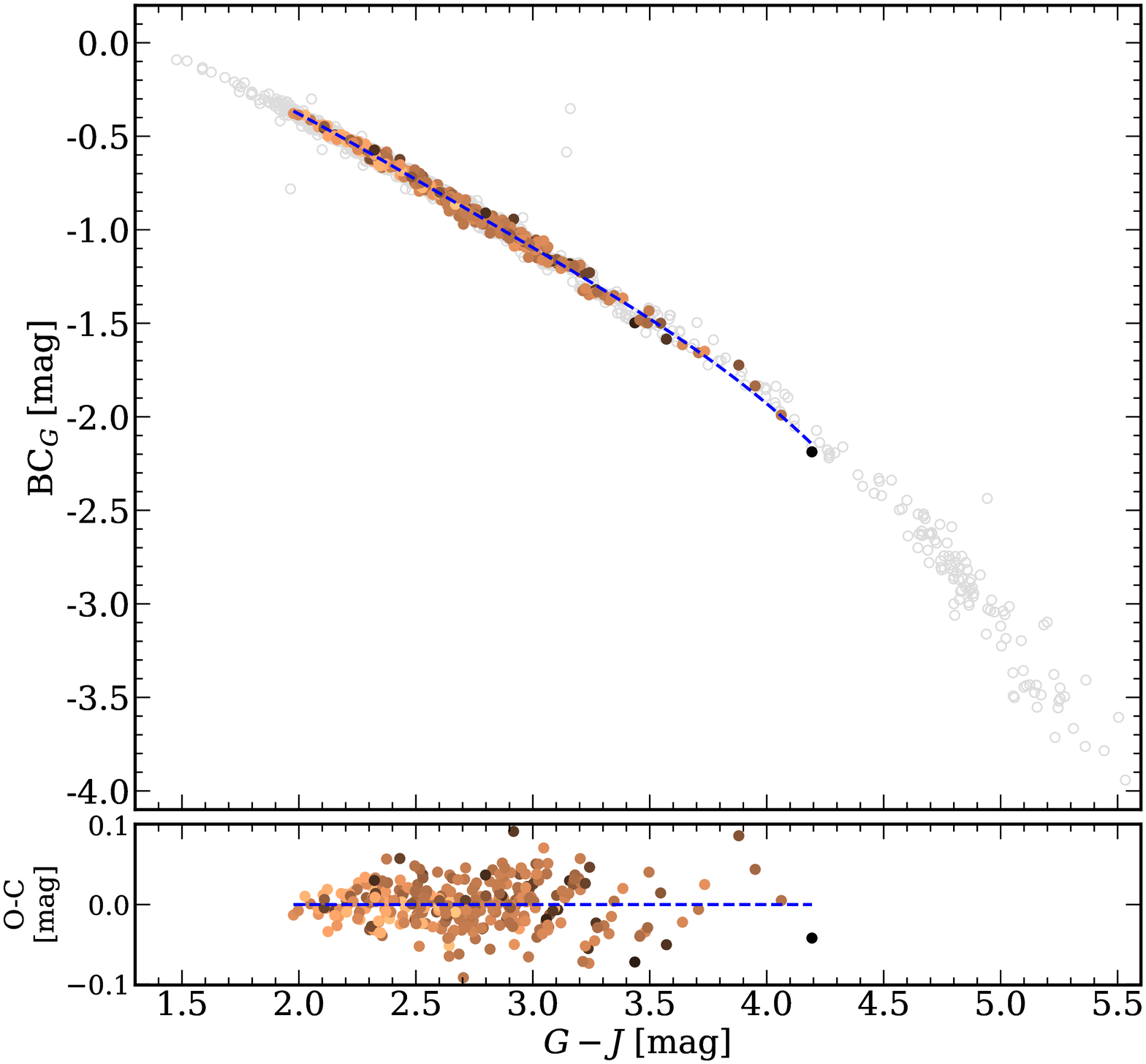}
    \includegraphics[width=0.49\textwidth]{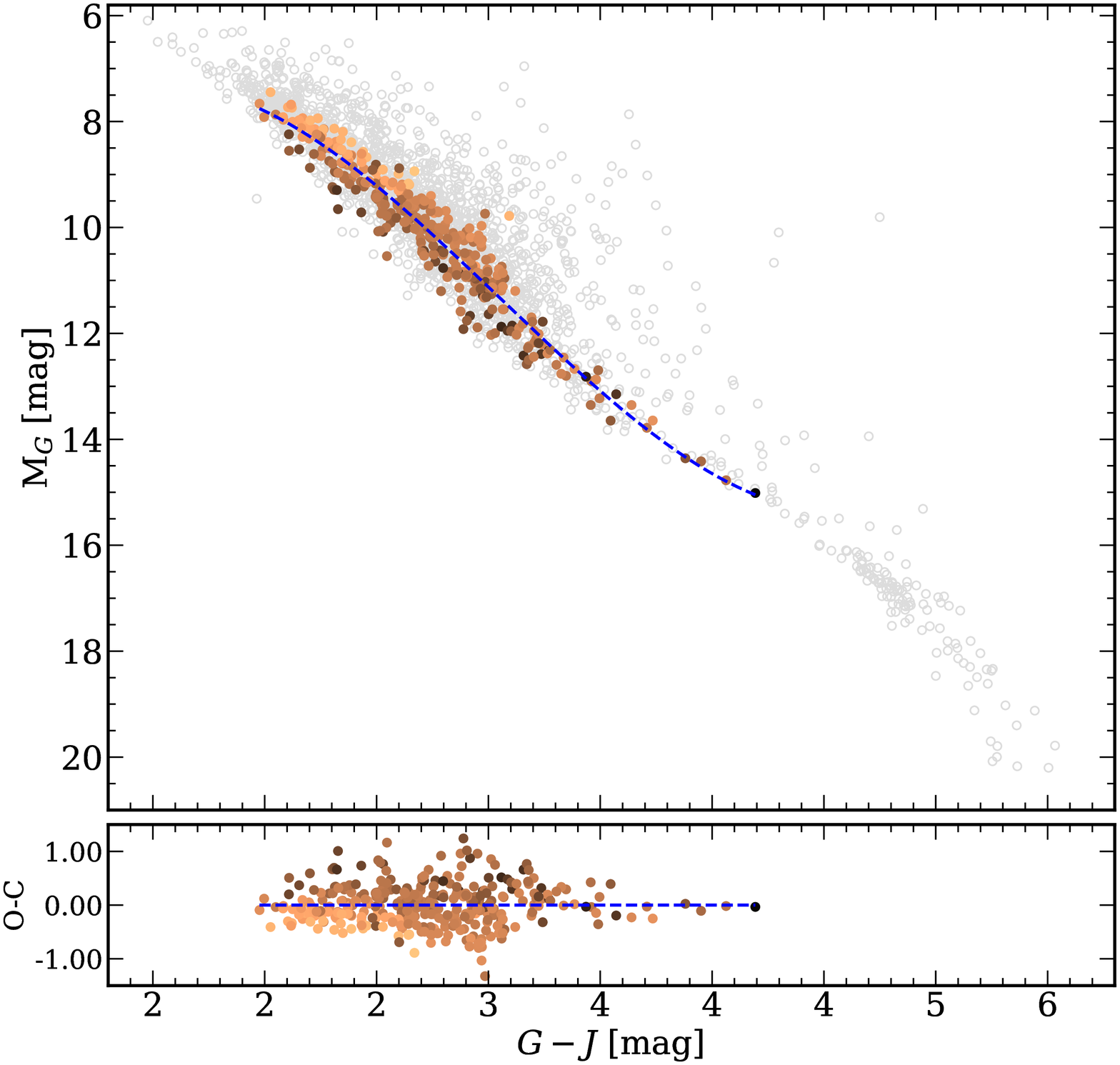}
    \includegraphics[width=0.49\linewidth]{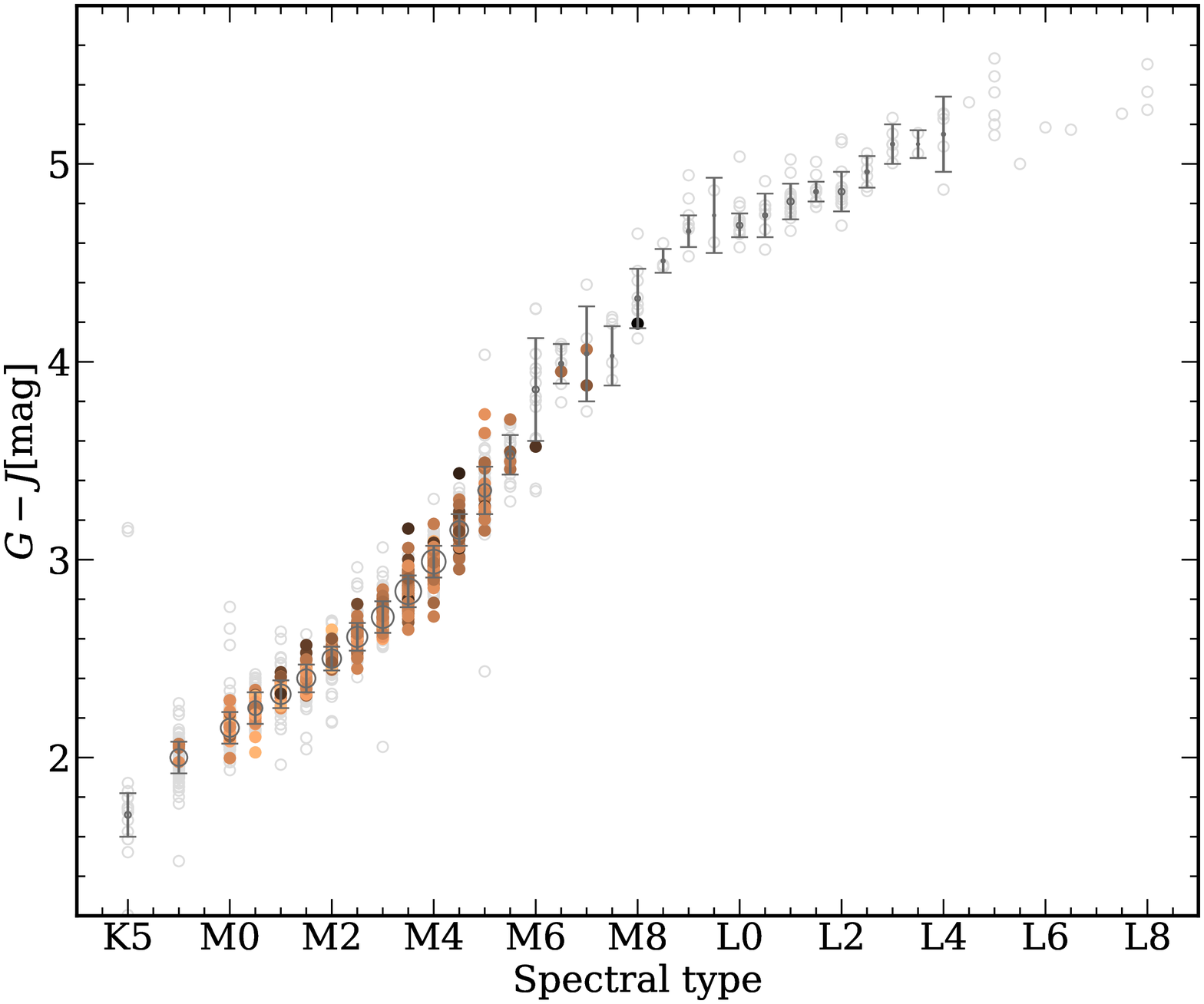}
    \includegraphics[width=0.85\linewidth]{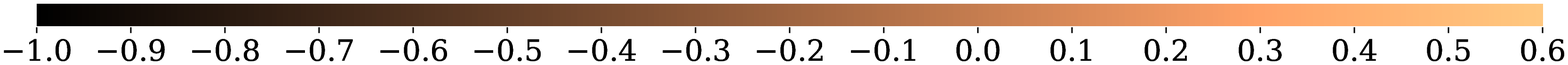}
    \caption{
    {
    Revisiting empirical relations in Table~\ref{tab:eqn_parameters} for stars with [Fe/H] values published in the literature.
    {Top left:} $L_{\rm VOSA,BT-Settl~CIFIST}$ vs. $M_J$. {Top right:} $BC_G$ vs. $G-J$. 
    {Bottom left:} $M_G$ vs. $G-J$.
    The blue dashed lines in the three panels represent new 3-, 4-, and 3-degree polynomial fits, respectively.
    {Bottom right:} $G-J$ vs. spectral type. The grey circles represent the mean values with a symbol size proportional to the sample size in each type. 
    }
        }
    \label{fig:models_meta}
\end{figure*}

Here we compare our $L$, $T_{\rm eff}$, $\mathcal{R}$, $\mathcal{M}$, and photometric data with those in the literature.
Tables~\ref{tab:parameters-1} and~\ref{tab:parameters-2} and Figs.~\ref{fig:Discussion.L} to~\ref{fig:Discussion.colour} illustrate the discussion.
In particular, in Table~\ref{tab:parameters-1} we show average values of $BC_G$, $BC_J$, $L$, $T_{\rm eff}$, $\mathcal{M}$, and $\mathcal{R}$ for single, main-sequence stars with spectral types from K5\,V to L2.0.
The last column, $N$, indicates the number per spectral type bin of well-behaved stars (i.e. with no companions at $\rho <$ 5\,arcsec, no overluminousity due to extreme youth, and of good {\em Gaia} DR2 astrometric and photometric quality).
After applying a $2.5 \sigma$ {clipping}, we calculated three-point rolling medians and standard deviations between M0.0\,V and L2.0 (e.g. tabulated values for M4.0\,V stars are the median and standard deviation of all individual $BC_G$ values of stars with spectral types M3.5, M4.0, and M4.5\,V), and simple medians and standard deviations for K5\,V and K7\,V stars.
With these rolling medians, we conservatively smoothed potential inter-type variability due to the small number of stars per bin at the latest spectral types and the typical uncertainty in M-dwarf spectral type determination, of 0.5\,dex \citep{Haw02,Lep13,Alo15a}.
The correspondingly large standard deviations denote the large natural scatter of the main sequence at the earliest spectral types and the difficulty in determining precise parameters at the latest ones.
The boundary values for K5\,V type were not smoothed and, therefore, must be handled with care.
On the other hand, Table~\ref{tab:parameters-2} complements Table~\ref{table.colours} and lists the average absolute magnitudes of K5\,V to L2.0 objects in the 14 most representative bands (i.e. all except for {\em GALEX} $FUV$ and $NUV$, SDSS9 $u'$, Tycho-2 $B_T$ and $V_T$, and {\em WISE} $W4$).
We applied the same rolling medians and $2.5 \sigma$ clipping as in Table~\ref{tab:parameters-1}.
For each spectral type K5--M7.0\,V, a total of 6.227\,10$^9$ different colours can be determined from the tabulated absolute magnitudes (e.g. $G-J = M_G - M_J$).
For spectral types L0.0--2.0, the number of possible colours is 3\,628\,800.

\paragraph{Luminosities} (Fig.~\ref{fig:Discussion.L}).
First, we compared our $L$ computed with VOSA with those from a number of works in the literature \citep[left panel --][]{Gol04,Vrb04,How10,Kun11,Bon12,Man13b,Gai14a,Aff16,Aff19,Tuo14,New15,Ang16,Ast17a,Dit17,Gil17,Mal17,Sua17a,Sua17b,Gaia18bro,Hir18,Hob18}.
In spite of (or due to) the relatively large published $L$ uncertainties of a few ultracool dwarfs, the agreement is in general excellent, especially in the case of \citet{Gaia18bro}.
Our median $L$ values per spectral type also match those of \citet[][right panel]{Pec13}.
When integrated from a well-calibrated, multi-band spectral energy distribution in a wide wavelength coverage and calculated with the latest {\em Gaia} parallaxes
as in this work,
$L$  can become the most reliable ``observable'' of low-mass stars, instead of the widely used temperature, which is inferred through colours, spectral classification, or expensive, model-dependent, spectral synthesis.

\paragraph{Effective temperatures} (Fig.~\ref{fig:Discussion.Teff}).
Next, we compared our $T_{\rm eff}$ from VOSA with the values from the works referred to in the previous paragraph, except from \citet{Gaia18bro}, plus from \citet{Pas19}, who in turn compared their $T_{\rm eff}$ with those from \cite{Roj12}, \cite{Gai14a}, \cite{Mal15}, \cite{Man15}, \cite{Raj18b}, and \cite{Sch19}.
From the top left panel in Fig.~\ref{fig:Discussion.Teff}, our $T_{\rm eff}$ are cooler than those of the literature by --86\,$\pm$\,82\,K.
This systematic difference is within the grid step size of the theoretical models used by VOSA, of 100\,K {or 50\,K}, but appreciable in the whole $T_{\rm eff}$ = 3000--4000\,K range.
{That VOSA does not interpolate between grid points may partly explain this systematic difference.}
In the empirical $T_{\rm eff}$-spectral type relation shown in the top right panel, $T_{\rm eff}$ from \cite{Raj18b} and \citet{Pas19} are, again, slightly warmer than ours in the late- and early-M domains, respectively.
However, the agreement with the relation of \citet{Pec13} is exquisite.
The K/M and M/L boundaries occur at about 3900\,K and 2300\,K, respectively, in line with the standard values \citep[e.g.][see also Table~\ref{tab:parameters-1}]{Hab81,Kir05}.
In the Hertzsprung-Russell diagram in the bottom left panel, as expected, our targets are significantly less luminous than the very young stars and brown dwarfs of the same $T_{\rm eff}$ tabulated by \citet{Pec13} and \citet{Fah16}, but our main sequence (excluding young targets) matches that of \citet{New15}. 
The most convincing plot is perhaps the $M_J$ versus $T_{\rm eff}$ diagram in the bottom right panel, where our M-dwarf main sequence perfectly overlaps with those defined by \citet{Lep13} and \citet{Gai14a} and extrapolates reasonably well into the ultracool dwarf sequence of \citet{Dah02}.
The absolute magnitude in the vertical axis does not depend on models, Virtual Observatory tools, spectral synthesis, or multi-band photometry, but only on reliable 2MASS $J$-band magnitude and {\em Gaia} parallaxes.

{\paragraph{Metallicity} (Figs.~\ref{fig:models_meta} to~\ref{fig:LT_meta}).
The role of metallicity in the empirical relations between physical parameters of M dwarfs has been the subject of investigation by many teams \citep[e.g.][see Sect.~4.3 in \citealt{Alo15b} for a short review]{Bon05,Woo05,Cas08,Roj12,Boy12,Boe19}.
Of them, \citet{Man15} showed that empirical relations such as absolute magnitude-radius, radius-temperature, or colour-temperature could benefit from incorporating an additional term that accounts for metallicity.
However, mainly because of the limitations of the BT-Settl CIFIST grid of theoretical models stored in the VOSA database, in our work we computed $L$ and $T_{\rm eff}$ assuming a solar metallicity ([Fe/H] = 0)\footnote{Actually, BT-Settl CIFIST models are defined for solar metal abundance, [M/H], but here we used solar iron abundance, [Fe/H], for simplicity.}.

In order to quantify the impact of metallicity within our empirical relations in Table~\ref{tab:eqn_parameters}, first we compiled values of spectroscopically derived iron abundances of 510 single stars in our sample from \citet{Man15,Man19}, \citet{Maj17}, and \citet{Pas19}.
The compiled [Fe/H] values ranged from --1.63\,dex for the mid-M dwarf \object{HD~285190} to +0.59\,dex for the early-M dwarf \object{LP~397--041}, with a mean and dispersion of 
$-0.04\,\pm\,0.26$\,dex.

Figure~\ref{fig:models_meta} displays the relations parametrised in Table~\ref{tab:eqn_parameters}, as well as the colour-spectral type diagram discussed in Sect.~\ref{sssection:diagrams_colour_SpT}, colour-coded by the metallicity values from the literature. 
In either of the top plots ($L$ vs. $M_J$ and $BC_G$ vs. $G-J$), the distribution of residuals did not show any correlation with the metal content of the stars. 
Both representations benefit from the fact that deriving $L$ does not rely on precise [Fe/H] measurements. 
In the bottom panels, the distribution of metallicity values in the $G-J$ versus spectral type diagram shows no significant dependence on metallicity.
This lack of correlation is also apparent in the additional colour diagrams displayed in Appendix~\ref{appsection:tables}.
However, the $M_G$ versus $G-J$ relation exhibits a notable correlation between metallicity and the residuals of the fit: more metallic stars appear brighter than less metallic stars of the same $G-J$ colour or, alternatively or simultaneously, more metallic stars appear redder than less metallic stars of the same $M_G$ absolute magnitude. 
This dependence is most likely the main source of uncertainty for photometric distances, as we noted in Sect.~\ref{sssection:diagrams_absmag_colour}. 
By using standard broad passbands in the red optical or the near infrared, such as $r'$ or $J$, the effect of metallicity can be reduced compared to using wider, bluer passbands, such as $G$, which are more affected by the features that metallicity imprints on the spectra.

In the diagrams involving $T_{\rm eff}$, \cite{Man15} pointed out that the effect of metallicity can be severely masked due to the steeper dependence on the temperature.
This is an important point to underline because the uncertainties in $T_{\rm eff}$ of models are a major source of uncertainties in the final products of the SED fitting.  
In other words, the approximation of near-solar metallicity implies an error that is always within the errors due to temperature uncertainties. 
We argue that, with the exception of absolute magnitude against colours and extreme cases (i.e. very metal-poor stars), the models described in this work can be treated as independent of the metal content of the star.

\begin{figure}[]
    \centering
    \includegraphics[width=.99\linewidth]{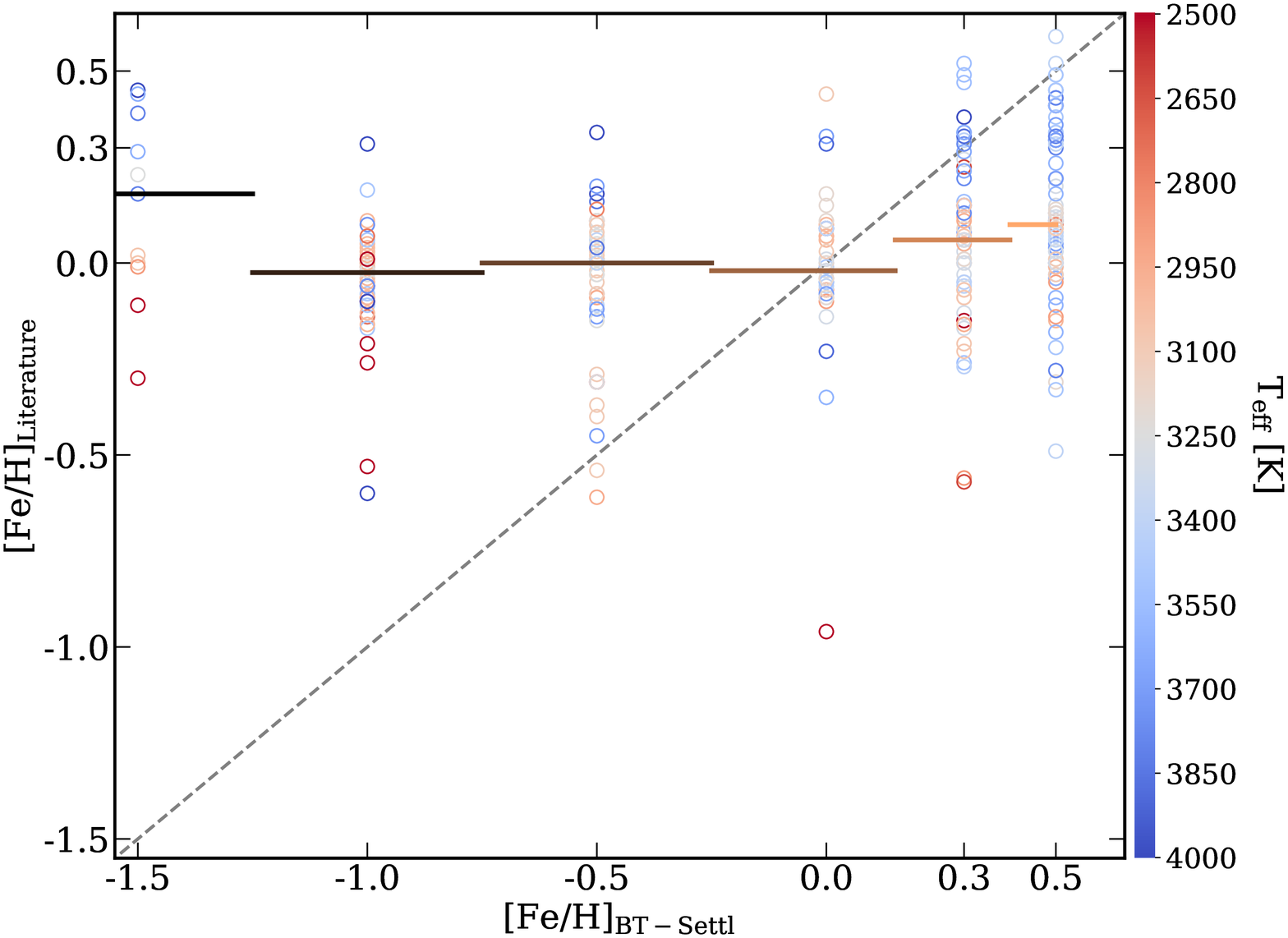}
    \caption{
    {Comparison of metallicites from VOSA BT-Settl fit and from literature for CARMENES GTO M dwarfs, colour-coded by BT-Settl CIFIST $T_{\rm eff}$. 
    Horizontal lines represent the median values in each BT-Settl [Fe/H] bin.
    }
    }
    \label{fig:box_meta}
\end{figure}

\begin{figure*}[]
    \centering
    \includegraphics[width=.49\textwidth]{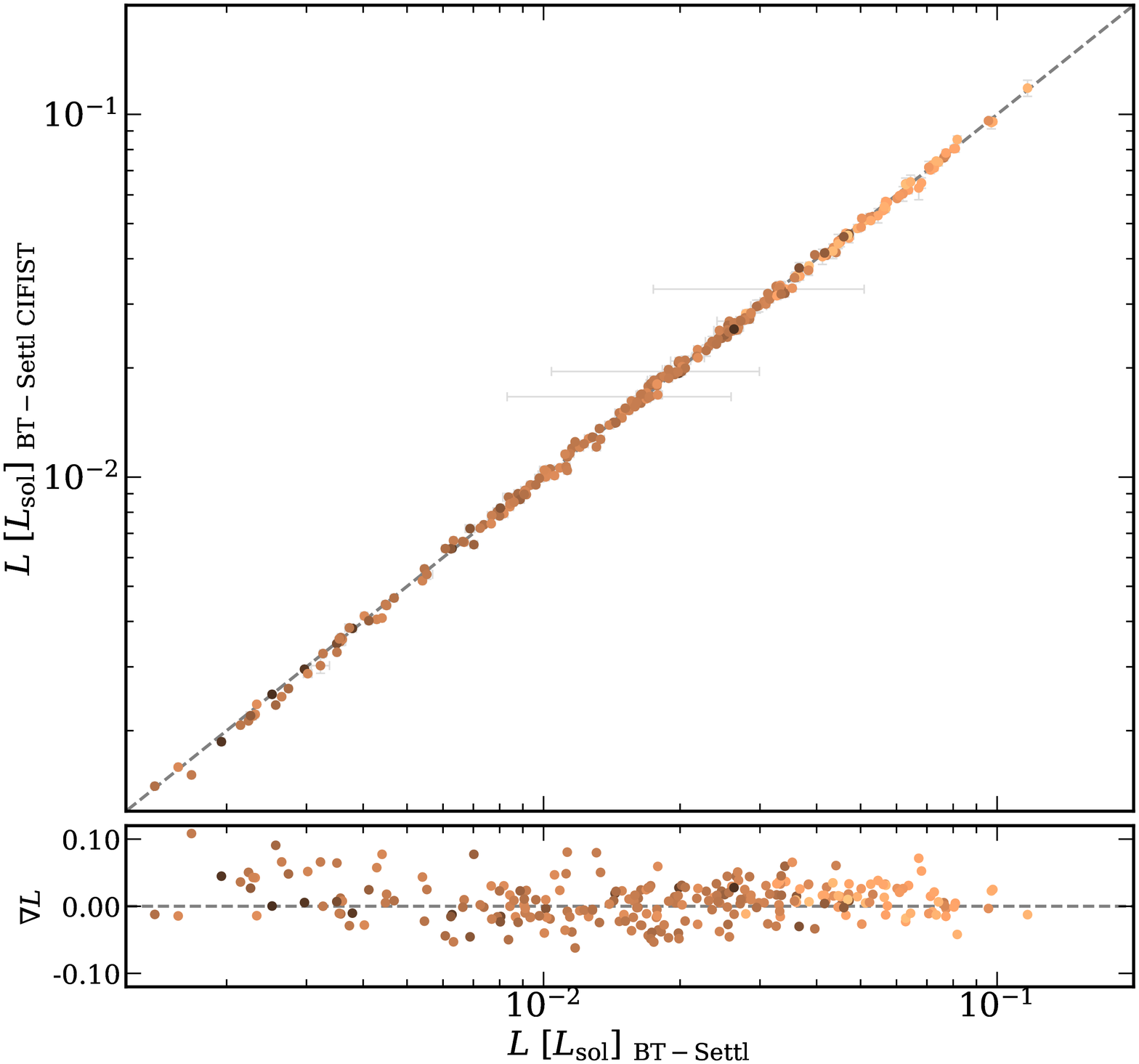}
    \includegraphics[width=.49\textwidth]{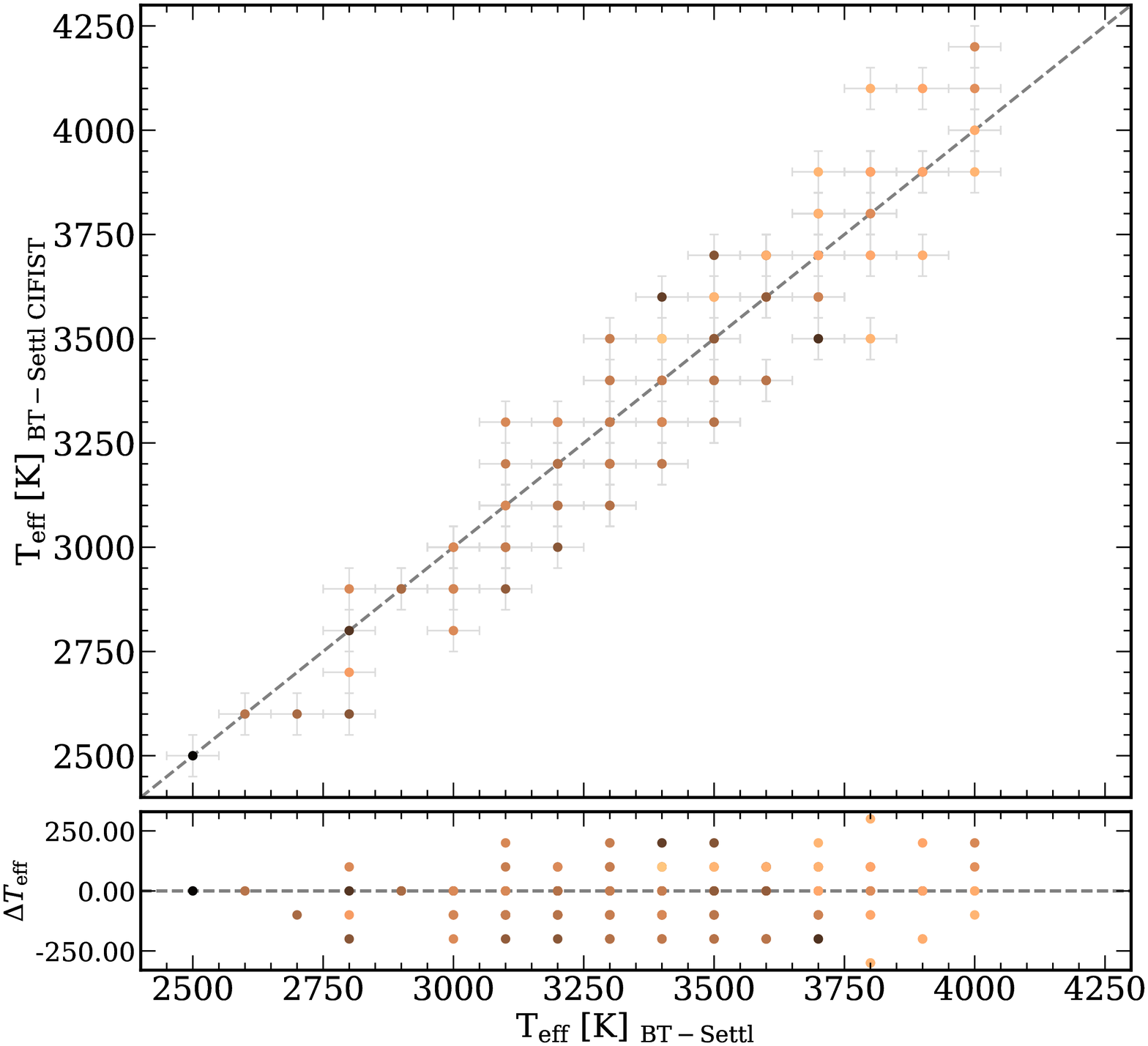}
    \includegraphics[width=.85\linewidth]{cif20_colorbar_FeH.eps}
    \caption{
    {Comparison of previous and recomputed $L$ ({left}) and $T_{\rm eff}$ ({right}) using BT-Settl with [Fe/H] = --1.5 to +0.5 for CARMENES GTO M dwarfs, colour-coded by metallicities published in the literature. 
    The small bottom panels depict $\nabla L = \log{L_{\rm BT-Settl\,CIFIST}} - \log{L_{\rm BT-Settl}}$ ({left}) and $\Delta T_{\rm eff} = T_{\rm eff, BT-Settl\,CIFIST} - T_{\rm eff, BT-Settl}$ ({right}).
    }
    }
    \label{fig:LT_meta}
\end{figure*}

As an additional test, we used VOSA to perform a new SED fit of the CARMENES GTO stars in the sample using the BT-Settl grid of spectra \citep[``no CIFIST'';][]{All12}, which allowed us to explore iron abundances different from [Fe/H] = 0. 
In particular, we let [Fe/H]  vary between --1.5\,dex and +0.5\,dex, with a step size of 0.5\,dex, and constrained $T_{\rm eff}$ and $\log{g}$ as in Table~\ref{tab:VOSA}. 
The [Fe/H] values derived from this new fit are compared to the spectroscopic values from the literature in Fig.~\ref{fig:box_meta}.
While the median of VOSA BT-Settl and published values are in fair agreement 
(--0.097\,dex and +0.033\,dex, respectively),
the scatter of the VOSA [Fe/H] values is much greater than that of the literature ($\sigma_{\rm [Fe/H],VOSA}$ = 0.596\,dex and $\sigma_{\rm [Fe/H],literature}$ = 0.216\,dex). 
From the diagram, VOSA assigned artificially low [Fe/H] to stars with spectroscopically derived solar values, which reinforced our initial approach of setting [Fe/H] = 0.
This is in line with the quality tests carried out by the VOSA team in 2017, in which they compared VOSA metallicities with those derived by \citet{Yee17}, \citet{LH17}, and \citet{Raj18a}.
In particular, they concluded that ``metallicities [...] provided by VOSA are not reliable due to the minor contribution of [this parameter] to the SED shape''\footnote{\url{http://svo2.cab.inta-csic.es/theory/vosa/helpw4.php?otype=star\&action=help\&what=qua\_libraries}}. 

In Fig.~\ref{fig:LT_meta} we compare the $L$ and $T_{\rm eff}$ obtained for the GTO stars using BT-Settl~CIFIST with [Fe/H] = 0 (used throughout this work) and BT-Settl with a free range in metallicity.
While the derivation of bolometric luminosities in K dwarfs, with a normalised difference of $\Delta L /L = -0.0065 \pm 0.0046$, is marginally dependent on metallicity, the derivation in M dwarfs is independent: the normalised differences between $L$ computed with the two methods is 
$\Delta L / L = 0.012 \pm 0.035$, 
consistent with zero.
The $T_{\rm eff}$ difference is also consistent with zero, and its standard deviation is 101\,K, identical to the $T_{\rm eff}$ step size in the M-dwarf domain.}

\begin{figure*}[]
    \centering
    \includegraphics[width=.49\textwidth]{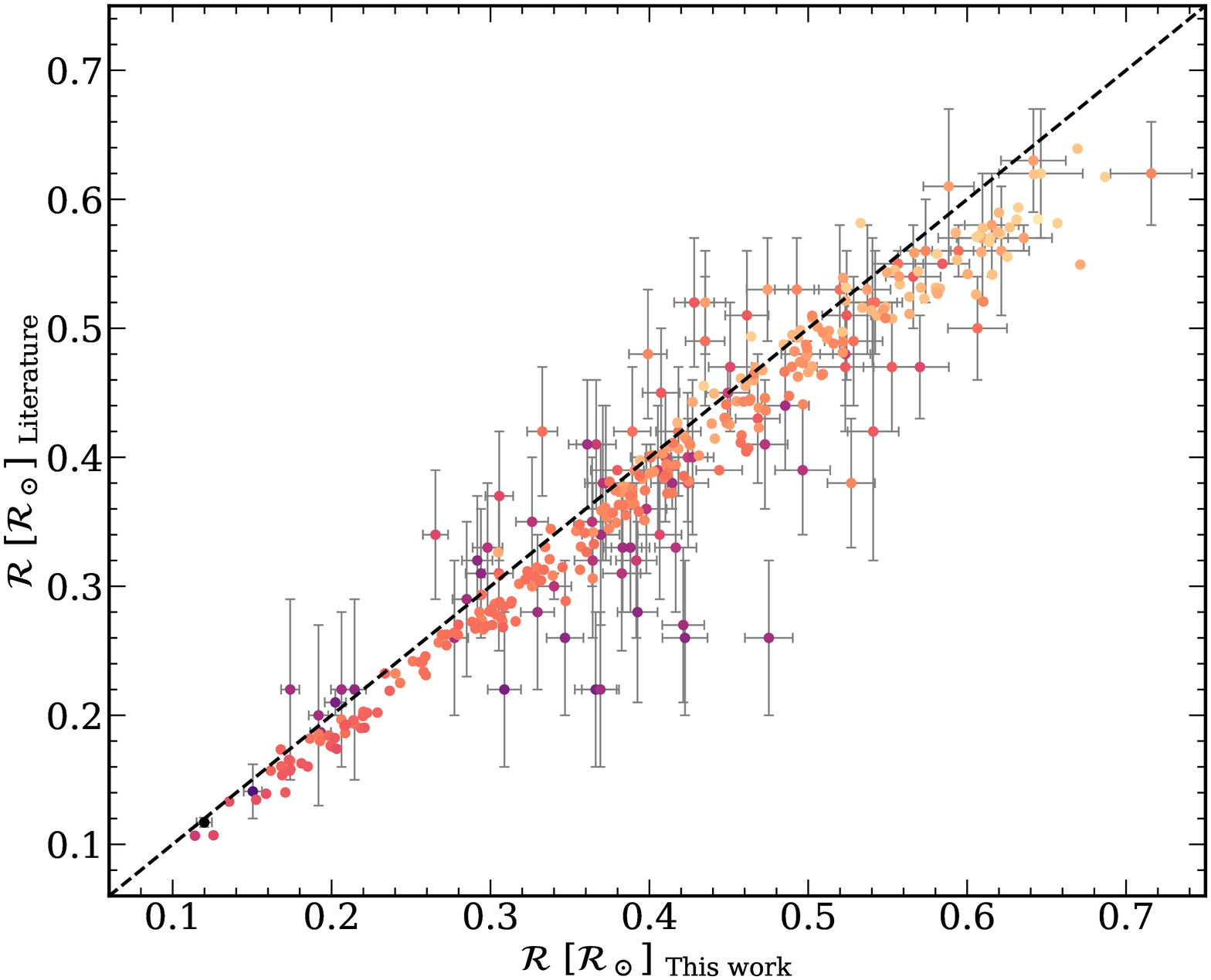}
    \includegraphics[width=.49\textwidth]{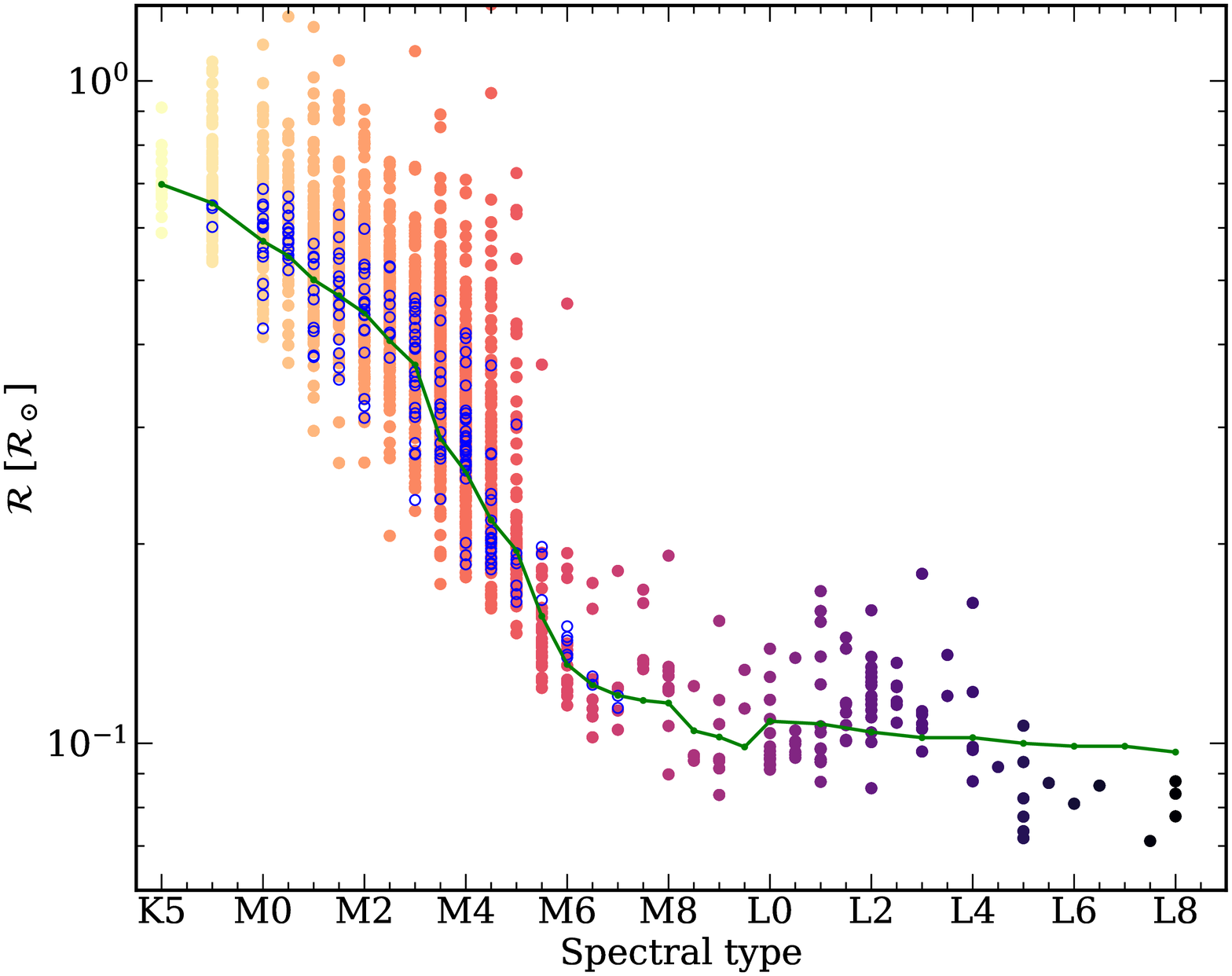}
    \includegraphics[width=.49\textwidth]{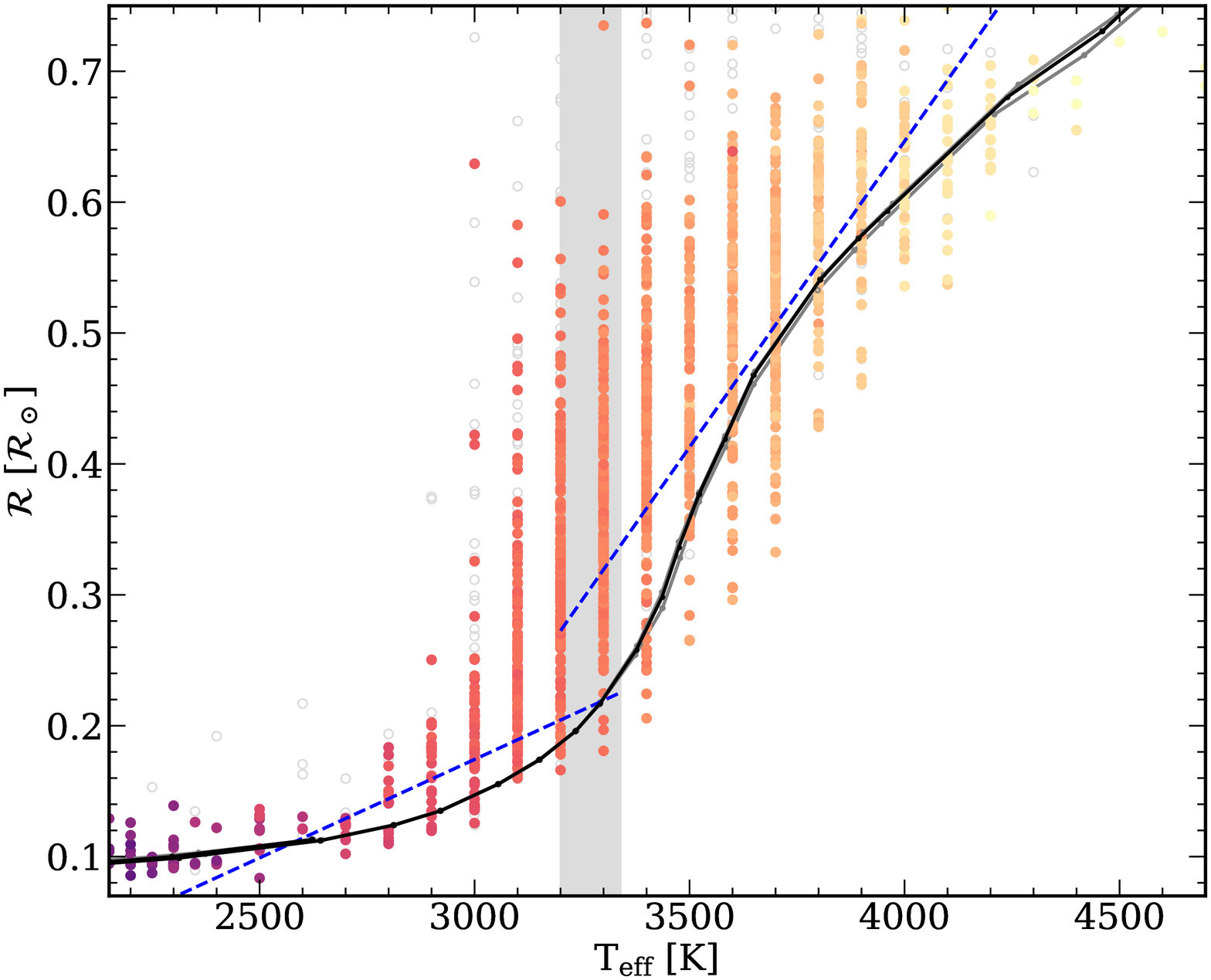}
    \includegraphics[width=.49\textwidth]{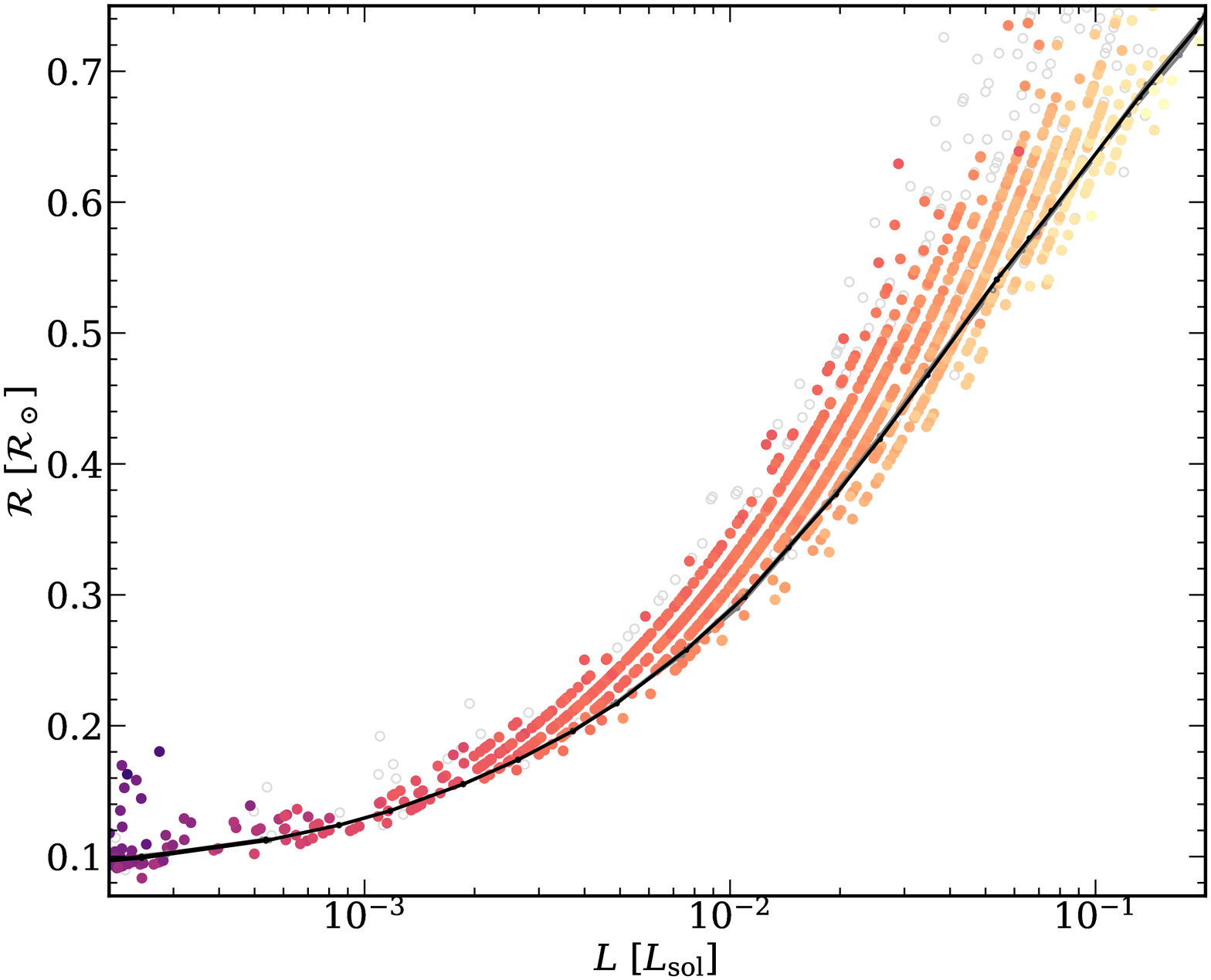}
    \includegraphics[width=.90\linewidth]{cif20_colorbar_SpT_bis.eps}
    \caption{Four representative diagrams involving $\mathcal{R}$.
    In the four panels our investigated stars are represented with filled circles colour-coded by spectral type.
    {Top left:} 
    Comparison of $\mathcal{R}$ from this work and from the literature, including \citealt{Sch19} (with symbol-size error bars).
    {Top right:} 
    Individual (coloured points) and median (black circles) values of $\mathcal{R}$ as a function of the spectral sequence shown in Table~\ref{tab:parameters-1}.
    The green line marks the median values from \cite{Pec13} and the blue circles are stars from \cite{Man15}.
    {Bottom left:} 
    $\mathcal{R}$ vs. $T_{\rm eff}$.
    The black and grey solid lines are the NextGen isochrones of the Lyon group for 1.0, 4.0, and 8.0\,Ga (overlapping), the blue dashed lines are the linear fittings from \cite{Rab19}, and the grey shaded area is the region where they reported a possible discontinuity.
    {Bottom right:} 
    $\mathcal{R}$ vs. $L$.
    The black solid lines are the same isochrones as in the bottom left panel.
    }
    \label{fig:Discussion.R}
\end{figure*}

\paragraph{Radii} (Fig.~\ref{fig:Discussion.R}).
We compared our $\mathcal{R}$, derived from VOSA's $L$ {(BT-Settl CIFIST, [Fe/H] = 0)} and $T_{\rm eff}$ using the Stefan-Bolztmann law, with the same works as in Fig.~\ref{fig:Discussion.L} (top left panel). 
Some of these works in turn compared their results with independent direct radius determinations \citep[e.g. near-infrared interferometry --][]{Boy12,vonB14}.
On average, our $\mathcal{R}$ are larger by 0.022\,$\pm$\,0.037\,$\mathcal{R}_\odot$, meaning they are identical within the dispersion of the data.
However, {the standard deviation includes both random errors (in magnitudes, parallax, SED integration) and systematic errors (in passband $\lambda_{\rm eff}$ and $W_{\rm eff}$, VOSA minimisation procedure, CIFIST models), and the $\mathcal{R}$ difference appears systematically across the whole sample, so it is likely to be significant. 
Furthermore,} because of the $T_{\rm eff}$ shift with respect to \citet{Pas18} and other spectral synthesis works, our $\mathcal{R}$ are also larger by about 5\,\% than those of \cite{Sch19}, who used almost identical $L$ to ours. 
{For that reason, when $T_{\rm eff}$ from spectral synthesis on high-resolution spectra is available, we recommend using it together with our $L$ for determining $\mathcal{R}$ (and $\mathcal{M}$), and use $T_{\rm eff}$ from VOSA when there is no spectral synthesis.}

In spite of the large spread at spectral types earlier than M4.5\,V and some poorly sampled SEDs later than M7.0\,V, the matches with the $\mathcal{R}$-spectral type relation of \cite{Pec13} and the values reported by \cite{Man15} are also good (top right panel).
Our $\mathcal{R}$-$T_{\rm eff}$ diagram (bottom left panel) naturally reproduces the sigmoid shape predicted by the widely used theoretical models of \cite{Bar98}, but shifted by $\sim$100\,K towards cooler $T_{\rm eff}$ (see previous paragraph).
More than two decades after that cornerstone work by the Lyon group, \cite{Rab19} fitted an $\mathcal{R}$-$T_{\rm eff}$ relation using two linear polynomials and identified a discontinuous behaviour that the authors attributed to the transition between partially and fully convective stars at 3200--3340\,K or $\sim$0.23\,$M_\odot$.
Soon after, \cite{Cas19} confirmed this discontinuity, but considered instead the contribution of the electron degeneracy to the gas equation of state as the physical phenomenon behind this behaviour \citep[see also][]{Cha97}.
While the boundary between partially and fully convective stars is better exposed in for example the $NUV-G_{RP}$ versus spectral type diagram (see Fig.~\ref{fig:colourcolour}), in our data we did not find evidence for any discontinuity in the vicinity of 3250\,K in the $\mathcal{R}$-$T_{\rm eff}$ diagram, but just a change of slope, as proven by \citet[][ (see their Fig.~11)]{Sch19}.
The statistics in \cite{Rab19} were poorer than ours: they added around one hundred objects from \citet{Man15} to their sample of 22 low-mass dwarfs, while we have 1031 homogeneously investigated stars with $T_{\rm eff}$ in the 3000--3500\,K interval.
Furthermore, the continuity of $\mathcal{R}$ as a function of $L$ is obvious in the bottom right panel of Fig.~\ref{fig:Discussion.R}. 

\begin{figure*}[]
    \centering
    \includegraphics[width=.49\linewidth]{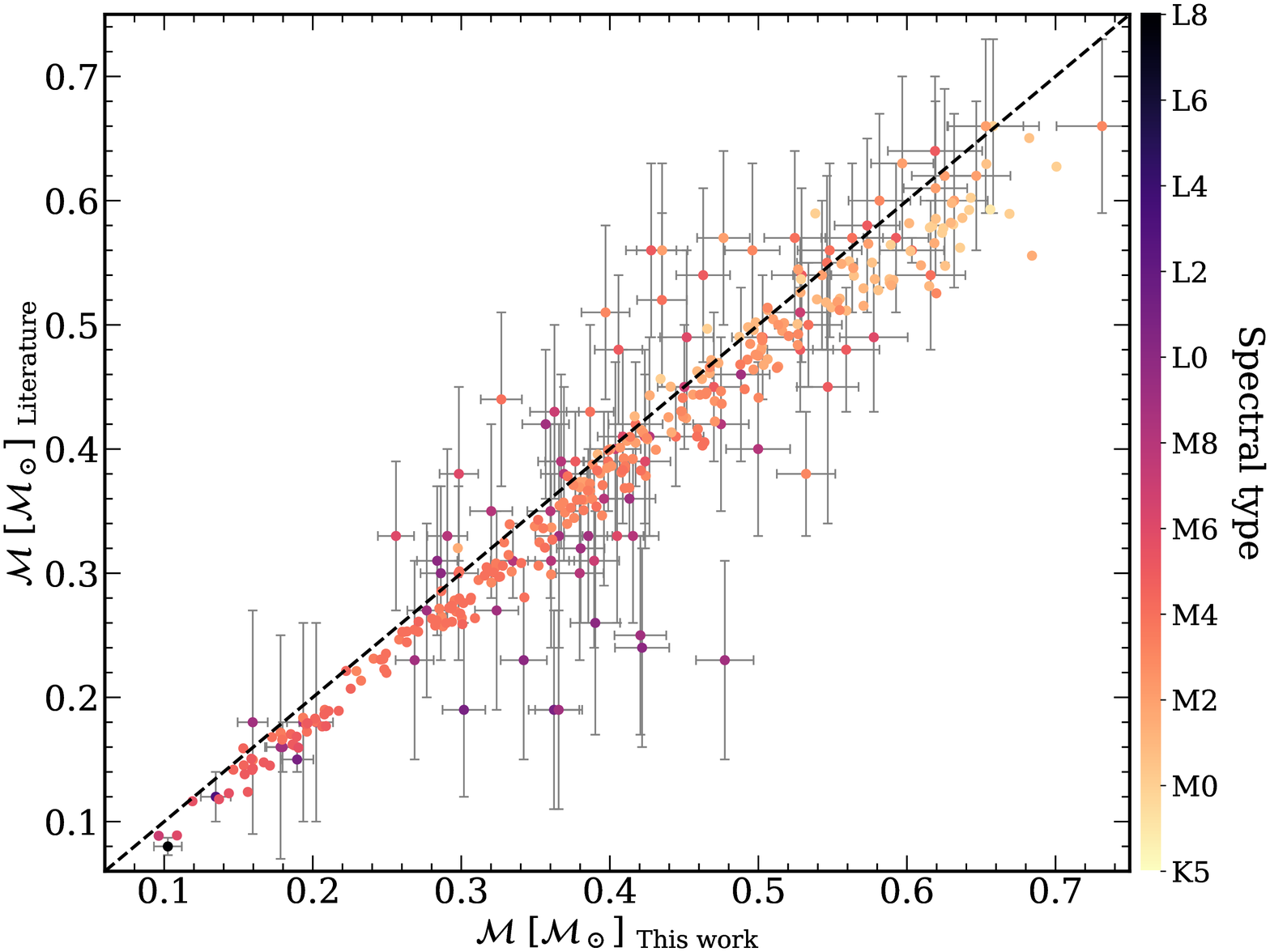}
    \includegraphics[width=.49\linewidth]{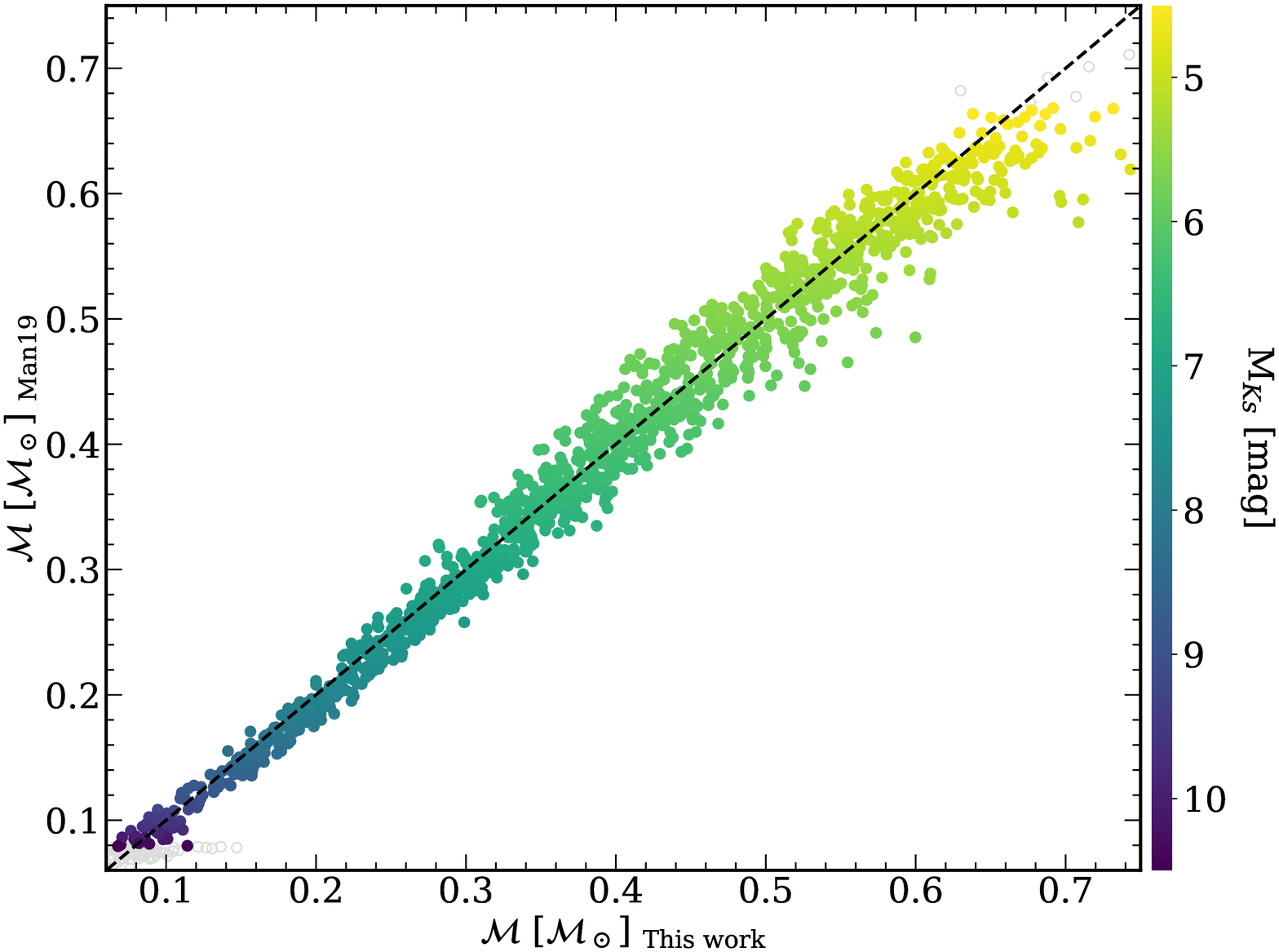}
    \caption{{Left:} Comparison of our masses with those from the literature. 
    {Right:} Comparison with those derived from absolute magnitude $M_{K_s}$ using the metallicity-independent relation from \cite{Man19} only in its validity range (4.5\,mag $< M_{K_s} <$ 10.5\,mag).
    }
    \label{fig:Discussion.M}
\end{figure*}

\paragraph{Masses} (Fig.~\ref{fig:Discussion.M}).
We compared our $\mathcal{M}$, derived from our $\mathcal{R}$ and the $\mathcal{M}$-$\mathcal{R}$ relation of \cite{Sch19}, with those from the literature (same works as in Fig.~\ref{fig:Discussion.L}, left panel).
This comparison is shown in the left panel of Fig.~\ref{fig:Discussion.M}. 
Among our parameters, $\mathcal{M}$ is the one that shows more dissimilarities with respect to published values, although $\mathcal{M_{\rm This~work}}-\mathcal{M}_{\rm lit}$ = 0.025\,$\pm$\,0.081\,$\mathcal{M}_{\odot}$, consistent with a null difference {(but probably significant as in $\mathcal{M}$ when random and systematic errors are taken into account)}.
For example, the two stars for which our $\mathcal{M}$ deviate more than 80\,\% from published values are
\object{LP~229--17} (M3.5\,V, $\mathcal{M}_{\rm This~work}$ = 0.476\,$\pm$\,0.017\,$\mathcal{M}_{\odot}$, $\mathcal{M}_{\rm lit}$ = 0.23\,$\pm$\,0.08\,$\mathcal{M}_{\odot}$) 
and \object{YZ~CMi} (M4.5\,V, $\mathcal{M}_{\rm This~work}$ = 0.368\,$\pm$\,0.008\,$\mathcal{M}_{\odot}$, $\mathcal{M}_{\rm lit}$ = 0.19\,$\pm$\,0.08\,$\mathcal{M}_{\odot}$), both from \citet{Gai14a}.
The former star was tabulated as a spectroscopic binary by \citet{Hou19}, although we do not see any CARMENES radial-velocity variation attributable to binarity \citep[][see also \citealt{Cor17} for a lucky imaging analysis]{Rei18b}, 
while the latter star is a candidate member of the young $\beta$~Pictoris moving group \citep[not in Table~\ref{tab:young} --][]{Mon01,Alo15b} with strong chromospheric activity \citep{Kah82,Kow10,Tal18}, which may partly explain the differences.
In planet-host stars, such changes can translate into significant differences in the published (minimum) masses of M-dwarf planets.

We also compared our values of $\mathcal{M}$ with those calculated from the $\mathcal{M}$-$M_{K}$ relations of 
\cite{Del00}, valid for 4.5\,mag $\le M_{K} \le$ 9.5\,mag, and 
\cite{Ben16}, valid for $M_K \le 10$\,mag,
and the $\mathcal{M}$-$M_{K_s}$ relation of
\cite{Man19}, valid for 4\,mag $\le M_{K_s} \le$ 11\,mag, and ``safe'' for 4.5\,mag $\le M_{K_s} \le$ 10.5\,mag.
For the relations of \cite{Del00}, we converted our 2MASS $K_s$ magnitudes to CIT $K$ values \citep{Eli82} using the colour transformation provided by \cite{Car01}.
The means of the mass differences were:
$\mathcal{M_{\rm This~work}}-\mathcal{M}_{\rm Del00}$ = --0.0080\,$\pm$\,0.0320\,$\mathcal{M}_\odot$,
$\mathcal{M_{\rm This~work}}-\mathcal{M}_{\rm Ben06}$ =
0.0242\,$\pm$\,0.0474\,$\mathcal{M}_\odot$, 
and $\mathcal{M_{\rm This~work}}-\mathcal{M}_{\rm Man19}$ = 0.0042\,$\pm$\,0.0223\,$\mathcal{M}_\odot$ 
Taking into account the standard deviations, \cite{Man19} provided the relation that best matched our $\mathcal{M}$.
In the right panel of Fig.~\ref{fig:Discussion.M} we show this relation, 
valid in a wide mass range from 0.075\,$\mathcal{M}_\odot$ to 0.70\,$\mathcal{M}_\odot$.
Since we fixed [Fe/H] = 0, we used the \cite{Man19} relation independent of metallicity ($f = 0$).
Besides, the authors stated that the impact of [Fe/H] is sufficiently weak for the $f = 0$ relation to be safely used for most stars in the solar neighbourhood.

\begin{figure*}[]
    \centering
    \includegraphics[width=.49\textwidth]{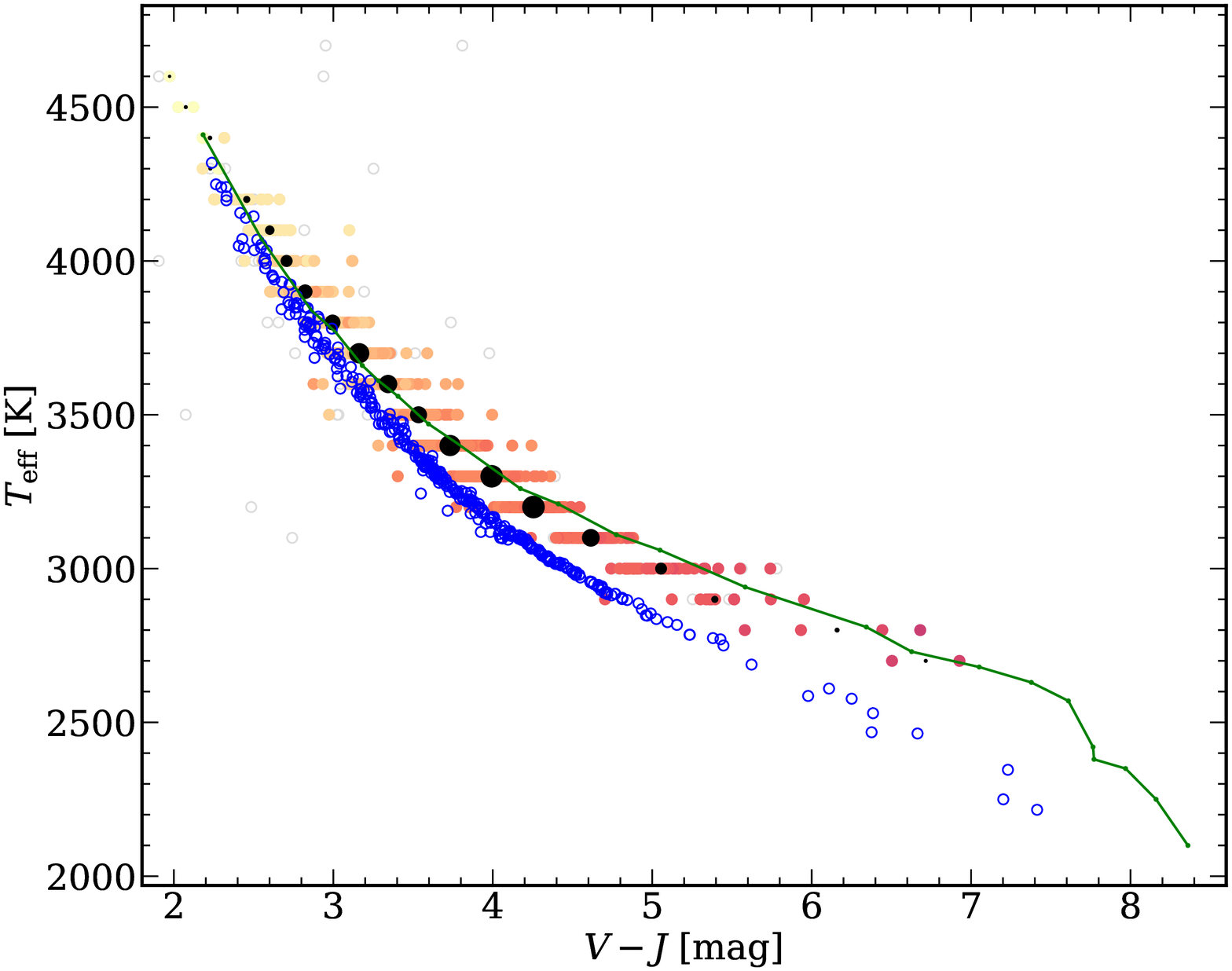}    
    \includegraphics[width=.49\textwidth]{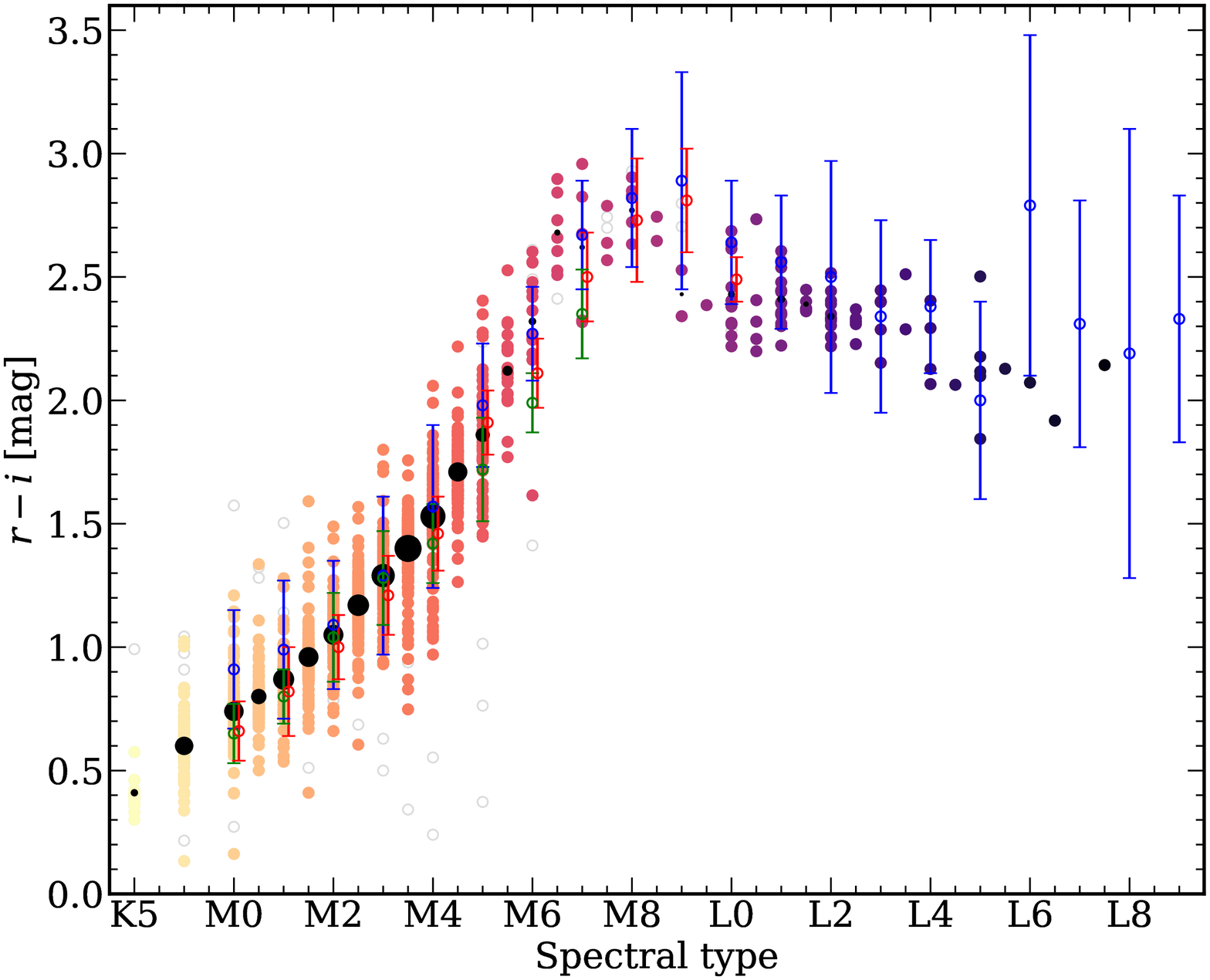}
    \includegraphics[width=.49\textwidth]{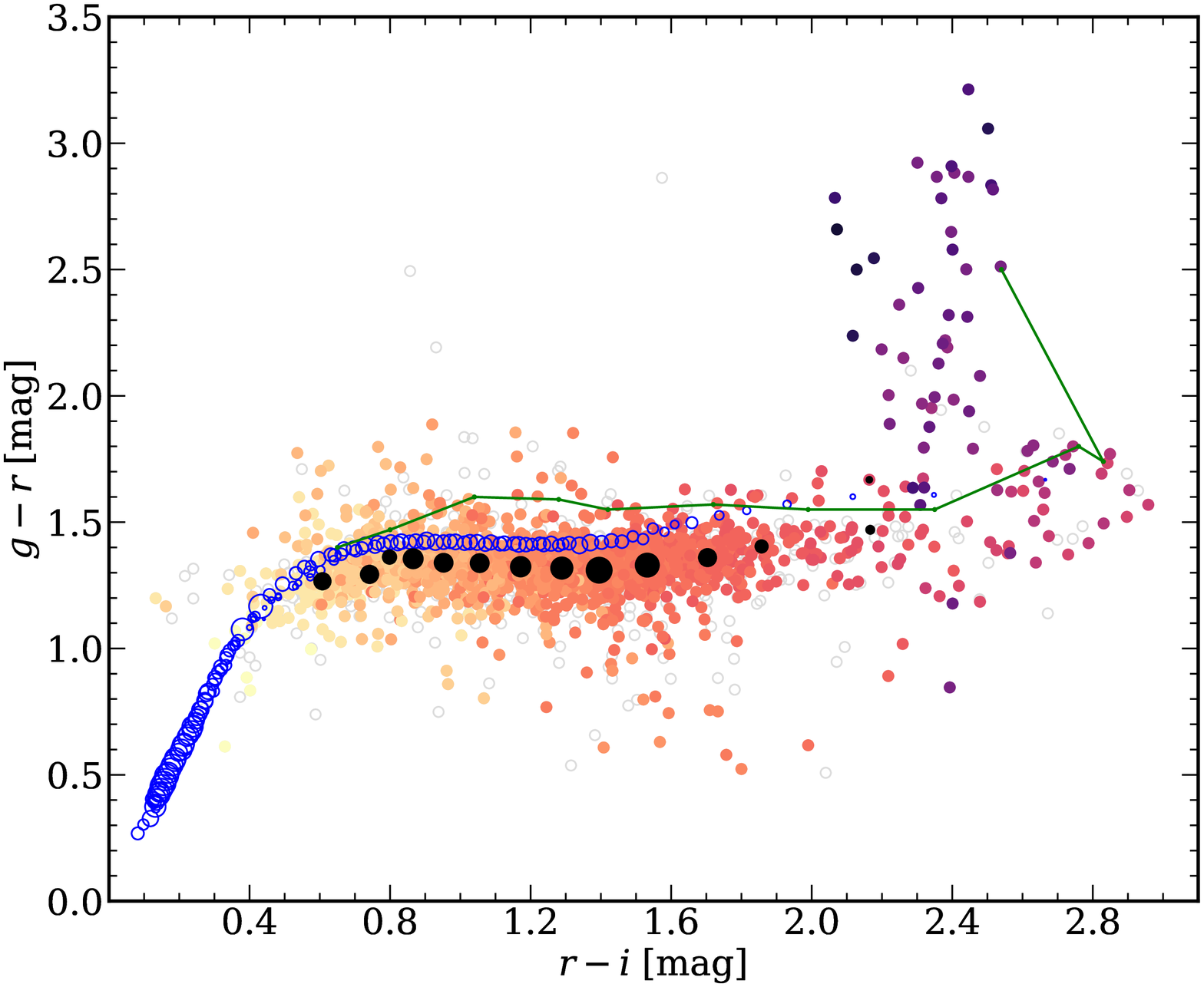}
    \includegraphics[width=.49\textwidth]{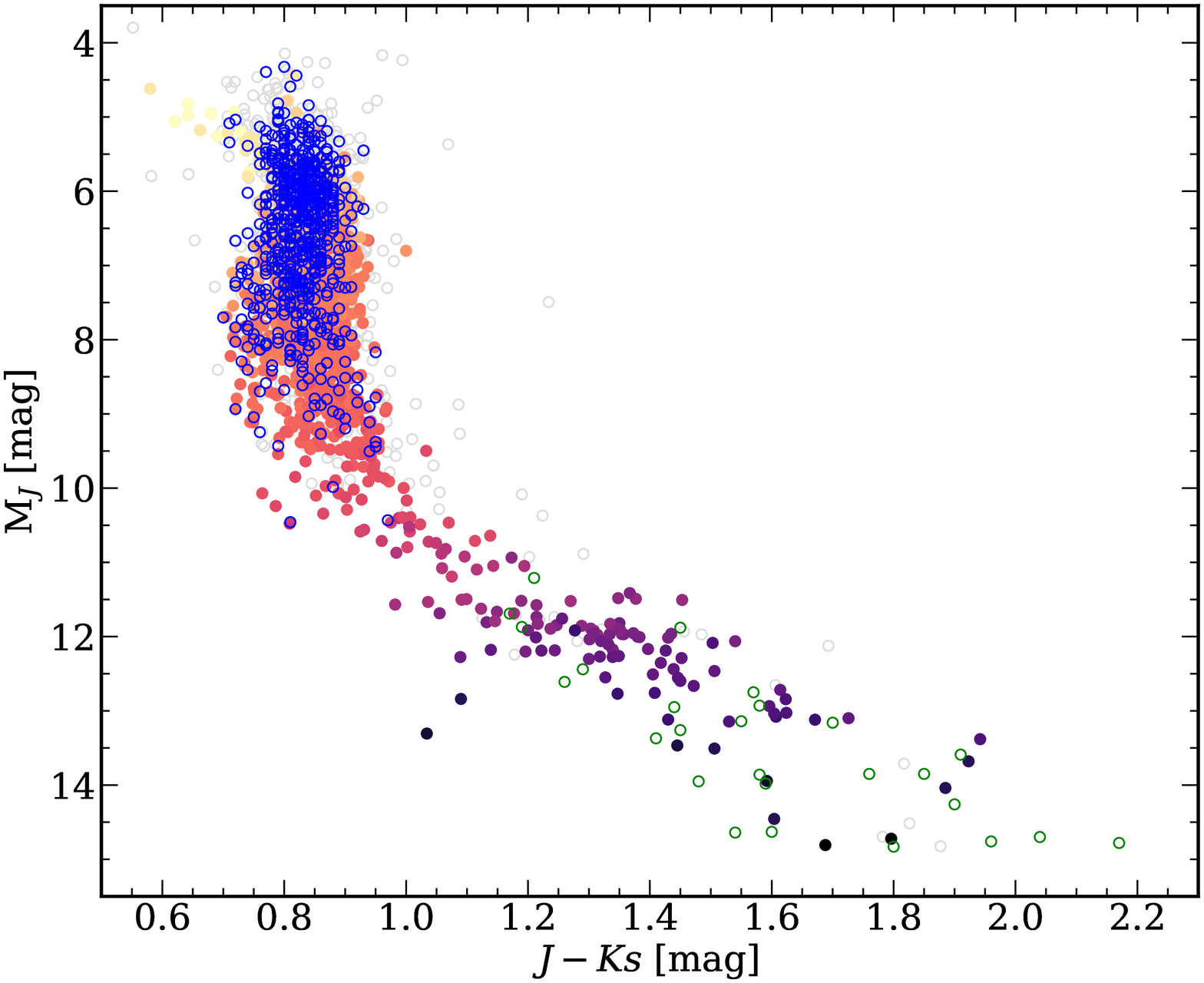}
    \includegraphics[width=.90\linewidth]{cif20_colorbar_SpT_bis.eps}
    \caption{{Top left:} 
    $T_{\rm eff}$ vs. $V-J$. 
    The blue empty circles and green line are data from \cite{Cas08} and \cite{Pec13}, respectively.
    Black filled circles are the mean colours from 2700\,K to 4600\,K in our sample, with a size proportional to the number of stars.
    {Top right:} 
    $r'-i'$ vs. spectral type.
    Open circles are the mean colours of \citet[][blue]{Haw02},  \citet[][``inactive'' colours, green]{Boc07}, and \citet[][red]{Wes08}.
    Black filled circles are the mean colours from K5\,V to L2 in our sample, taken from Table~\ref{table.colours}, with a size proportional to the number of stars.
    The error bars are the standard deviation \citep{Haw02,Boc07} and the intrinsic scatter of the stellar locus \citep{Wes08}.
    {Bottom left:} 
    $g'-r'$ vs. $r'-i'$.
    Black filled circles are the mean colours from K5\,V to L2 in our sample, taken from Table~\ref{table.colours}, with a size proportional to the number of stars. 
    Blue empty circles are the mean colours by \cite{Dav14}, with a size proportional to the numbers of stars, and the green line links the mean ``inactive'' colours by \cite{Boc07} for spectral types M0\,V to L0.
    {Bottom right:} 
    $M_J$ vs. $J-K_s$.
    Green and blue empty circles are data from \cite{Kna04} and \cite{Lep13}, respectively.
    In the four panels, the stars in our sample are colour-coded by spectral type, and the discarded stars are plotted with grey empty circles.
    }
    \label{fig:Discussion.colour}
\end{figure*}

\paragraph{Colours} (Fig.~\ref{fig:Discussion.colour}).
Although with the advent of {\em Gaia} the $V-J$ colour should be replaced by $G-J$, the former had been used extensively in the past.
The match of our mean $V-J$ colours as a function of $T_{\rm eff}$ with those of \cite{Pec13} is once again excellent, but the relation significantly deviates from the values tabulated by \citet{Cas08}.
However, as noted by them, the range of applicability of their  colour-temperature-metallicity relations involving $V-J$ is narrow, between 0.61\,mag and 2.44\,mag.
As a result, from the top left panel, extrapolating the $T_{\rm eff}$ versus $V-J$ relation of \cite{Cas08} beyond 2.44\,mag may result in $T_{\rm eff}$ systematically cooler by more than 300\,K.  
In the top right panel, we revisit the $r'-i'$-spectral type diagram, which is an evolution of that with $R-I$ colour in the Johnson-Cousins passbands \citep{Vee74,Bes79,Leg92,Boy12,Man15,Hou19}.
We reproduce the reversal at M7.0--8.0\,V ($r'-i' \sim 2.8$\,mag) observed by \cite{Haw02}, \cite{Boc07}, and \cite{Wes08}, among many others.
Therefore, we confirm that the $r'-i'$ colour alone cannot be used for spectral {classification} beyond M5.0\,V.
In the optical colour-colour diagram of the bottom left panel, our $g'-r'$ colours  are slightly bluer than those of \citet{Dav14} for a fixed $r'-i'$, and significantly bluer, by about 0.5\,mag, than those of \citet{Boc07}.
Finally, in the bottom right panel, there is a good agreement with the location of the M-dwarf main sequence of \cite{Kna04} in the near-infrared $M_J$ versus $J-K_s$ diagram, but our data show instead the {turnovers} towards bluer and redder $J-K_s$ colours of late-K dwarfs and early-L dwarfs, respectively.


\section{Summary}
\label{section:summary}

\begin{table*}
\caption{Description of the online table.}
\label{tab:description}
\centering
\begin{tabular}{lccl}
    \hline
    \hline 
    \noalign{\smallskip}
    Parameter & Units & Column(s) & Description \\
    \noalign{\smallskip}
    \hline
    \noalign{\smallskip}
{\tt Karmn}             & \ldots & 1 & Carmencita star identifier (JHHMMm+DDd)$^a$ \\
{\tt Name}              & \ldots & 2 & Discovery name or most common name$^b$ \\ 
{\tt RA, DE}            & hms & 3--4 & Right ascension and declination (equinox J2000, epoch J2015.5) \\
{\tt SpType, SpTnum}    & \ldots & 5--6 & Spectral type and its numerical format$^c$ \\
{\tt Ref{\_}SpT}          & \ldots & 7 & Reference for the spectral type \\
{\tt Plx, ePlx}         & mas & 8--9 & Parallax and its uncertainty \\
{\tt Ref{\_}Plx}          & \ldots  & 10 & Reference for the parallax \\
{\tt d{\_}pc, ed{\_}pc} & pc & 11--12 & Distance and its uncertainty \\
{\tt Ref{\_}d}          & \ldots  & 13 & Reference for the distance \\
{\tt Lbol, eLbol}               & L$_\odot$ & 14--15 & Luminosity and its uncertainty from VOSA \\
{\tt Teff}                  & K & 16 & Effective temperature from VOSA$^d$ \\
{\tt logg}                          & dex & 17 & Surface gravity from VOSA$^d$ \\
{\tt Radius, eRadius}   & $\mathcal{R}_\odot$ & 18--19 &  Radius and its uncertainty \\
{\tt Mass, eMass}               & $\mathcal{M}_\odot$ & 20--21 &  Mass and its uncertainty \\
{\tt NN{\_}mag, eNN{\_}mag }& mag & 22--97 & Magnitude and its uncertainty for the {\tt NN} passband$^e$ \\
{\tt Qf{\_}NN, Ref{\_}NN} & mag & 22--97 & Quality flag (if available) and reference for the {\tt NN} passband$^e$ \\
{\tt Gaia{\_}id{\_}1}   & \ldots & 98 & {\em Gaia} DR2 identifier of single or primary star \\      
{\tt Gaia{\_}id{\_}2}   & \ldots & 99 & {\em Gaia} DR2 identifier of secondary star in close binary system \\  
{\tt Multiple}          & \ldots & 100 &  Boolean index for close multiple stars \\
{\tt Young}                     & \ldots & 101 &  Boolean index for overluminous young stars \\
{\tt RUWE}                      & \ldots & 102 &  Boolean index for stars with {\em Gaia} RUWE $>$ 1.41 \\
{\tt Excess}                    & \ldots & 103 & Boolean index for stars with photometric flux excess in {\em Gaia}  $G_{B_P}$ and $G_{R_P}$ passbands \\
\noalign{\smallskip}
\hline
\end{tabular}
\tablefoot{
\tablefoottext{a}{For the K dwarfs, we tabulate the SUPERBLINK catalogue identifier \citep{Lep05,Lep13}.}
\tablefoottext{b}{For the ultracool dwarfs, we tabulate the {\em Gaia} UltraCool Dwarf Catalogue identifier \citep{Sma17,Sma19}.}
\tablefoottext{c}{{\tt SpTnum} = --2 for K5\,V, --1 for K7\,V, 0.0 for M0.0\,V, 0.5 for M0.5\,V... 10.0 for L0.0, etc.}
\tablefoottext{d}{VOSA uncertainties are 50\,K for $T_{\rm eff}$ (25\,K for $T_{\rm eff} \lesssim$ 2400\,K) and 0.5\,dex for $\log{g}$.}
\tablefoottext{e}{{\tt FUV}, {\tt NUV}: GALEX DR5 $FUV$ and $NUV$; 
{\tt BP}, {\tt GG}, {\tt RP}: $G_{B_P}$, $G$, and $G_{R_P}$ from {\em Gaia} DR2; 
{\tt BT}, {\tt VT}: $B_T$ and $V_T$ from Tycho-2;
{\tt B}, {\tt V}: $B$ and $V$ from UCAC4 or APASS9;
{\tt u}, {\tt g}, {\tt r}, {\tt i}: $u'$, $g'$, $r'$, and $i'$ from SDSS9, UCAC4, APASS9, PanSTARRS-1 and/or CMC15; 
{\tt J}, {\tt H}, {\tt Ks}: $J$, $H$, and $Ks$ from 2MASS;
{\tt W1}, {\tt W2}, {\tt W3}, {\tt W4}: $W1$, $W2$, $W3$, and $W4$ from AllWISE or WISE.}
}
\end{table*}

Here we present the most comprehensive photometric analysis to date of M dwarfs in the close solar neighbourhood, many of which are being followed-up by current radial-velocity and transit exoplanet surveys.
We started with the latest version of the CARMENES input catalogue of 2194 M dwarfs, dubbed Carmencita, to which we added 168 and 117 single, nearby, bright K and ultracool dwarfs, respectively. 
Although our main objective was investigating luminosities, colours, and spectral energy distributions of M dwarfs, our sample contains stars and ultracool dwarfs as early as K5\,V and as late as L8, which may avoid any boundary value problem. 
From public all-sky surveys, we collected 40\,094 photometric magnitudes for the 2479 stars and ultracool dwarfs in 20 different passbands from the far ultraviolet, through the blue and red optical and near infrared, to the mid infrared.
Except for the bluest passbands, the completeness of high-quality data is of the order of 97\,\%.
Thanks especially to {\em Gaia}, we could collect parallactic distances for 97.8\,\% of the sample and identified close multiple systems unresolved by ground all-sky surveys and {\em WISE}. 
From the new data, we estimated spectro-photometric distances for 31 single stars without parallactic distance, and tabulated angular separations and position angles for 40 new close multiple systems and candidates. 

Next, we computed bolometric luminosities, effective temperatures, and surface gravities for 1843 stars and ultracool dwarfs with parallactic distance and no physical companions at less than 5\,arcsec or less than 5\,mag fainter in {\em Gaia} $G$ than our target.
For that, we used VOSA and all high-quality photometric data redder than SDSS $u'$.
Because of the limitations of the BT-Settl CIFIST models implemented in VOSA, we set the metallicity to solar.
However, except for a few stars with poorly sampled spectral energy distributions, the  luminosities are independent of models at least at the 99.5\,\% level. 
They supersede any pre-{\em Gaia} determination.
From their loci in the Hertzsprung-Russell diagram, we identified 36 overluminous stars that had been previously assigned to young stellar kinematic groups and associations. 

We examined colour-spectral type, colour-colour, colour-magnitude, luminosity-magnitude, and bolometric correction-colour diagrams.
After discarding stars with young ages, close companions, and bad photometric or astrometric quality flags (i.e. {\em Gaia} {\tt phot\_bp\_rp\_excess\_factor} and {\tt RUWE}), we fitted empirical relations of absolute magnitude-colour, bolometric correction-colour, and luminosity-absolute magnitudes including widely available $G$, $r'$, and $J$ magnitudes and {\em Gaia} DR2 parallaxes.
In addition, we also used the Stefan-Boltzman law and the $\mathcal{M}$-$\mathcal{R}$ relation of \cite{Sch19} to derive radii and masses of all well-behaved stars in our sample. 
Finally, we tabulated median $G$- and $J$-band bolometric corrections, $L$, $T_{\rm eff}$, $\mathcal{R}$, and $\mathcal{M}$, as well as absolute magnitudes in 14 passbands, for stars and ultracool dwarfs with spectral types from K5\,V to L2.0.

We provide a summary table with the compiled astro-photometric data and derived stellar parameters of all our targets.  
The assembled catalogue in comma separated value ({\tt csv}) format is available in its entirety in the electronic edition.
As described in Table~\ref{tab:description}, for each star or ultracool dwarf we tabulate its identifiers, equatorial coordinates, spectral type (and reference), parallax and distance (and reference), all magnitudes and their uncertainties, origin, quality flags (when available), $L$, $T_{\rm eff}$, and $\log{g}$ from VOSA, $\mathcal{R}$ and $\mathcal{M}$ from the Stefan-Boltzmann law and the $\mathcal{M}$-$\mathcal{R}$ relation, {\em Gaia} DR2 identifier for primary and secondary sources (in the case of binary sources), four Boolean indices for close multiplicity ($\rho <$ 5\,arcsec), astrometric and photometric quality of the {\em Gaia} solution, and youth.
Finally, most of the {\tt Python} code developed by us for determining the parameters or preparing the plots shown in this work is available at GitHub\footnote{\url{https://github.com/ccifuentesr/CARMENES-V}}.

There are many ways of improving our $L$, $\mathcal{R}$, and $\mathcal{M}$ determinations.
CatWISE \citep{Eis20}, a recent NeoWISE enhanced and contributed product \citep{Mai11}, represents a step forward with respect to the AllWISE mid-infrared photometry used here.
The {\em Gaia} DR3, previously scheduled for the second half of 2021, will improve $G$, $G_{BP}$, and $G_{RP}$ photometry and, especially, astrometry, with which we will have more accurate parallax and close multiplicity determinations.
Soon after, in early 2022, 
the LSST, with its spectacularly large etendue and multi-band photometry in $u'g'r'i'z'y'$ passbands, will start operations.
The first LSST data release will supersede all previous UCAC, SDSS, and Pan-STARRS optical datasets \citep[but see also J-PAS,][]{Ben14}.
The ESA {\em Euclid} mission will complement LSST in the near infrared at Galactic latitudes far from the ecliptic, especially for the latest M dwarfs.
By that time, new grids of theoretical atmospheric models, with a much wider range of metallicities, will be available for VOSA, which will also be upgraded.
Thanks to the Transiting Exoplanet Survey Satellite ({\em TESS}) and the discovery of new detached M-dwarf eclipsing binaries, the $\mathcal{M}$-$\mathcal{R}$ relation will be refined and probably determined for different intervals of age.

There are even more ways to improve our results. 
To name a few:
new volume-limited samples including all M dwarfs known in the solar vicinity, not only limited to the Calar Alto sky, plus more intermediate- and late-K dwarfs and ultracool dwarfs, especially in the M6.0--9.5\,V range;
new spectral-synthesis determinations of $T_{\rm eff}$, $\log{g}$, and [Fe/H] in late-type stars for calibration (e.g. with the equivalent width method or with deep learning -- \citealt{Mar18}; V.\,M.~Passegger et~al., in~prep.); 
discovery of low-mass spectroscopic binaries;
or new studies that link kinematics, activity, and youth (and, therefore, radius and surface gravity).
All these will be taken into account by the CARMENES consortium to improve our knowledge of M dwarfs and their planets.


\begin{acknowledgements}

We thank the anonymous referee for the comments that helped to improve the quality of this paper, and also E.~E.~Mamajek and E.~Gaidos for their valuable insights and suggestions.

CARMENES is an instrument for the Centro Astron\'omico Hispano-Alem\'an de Calar Alto (CAHA, Almer\'{\i}a, Spain).

CARMENES is funded by the German Max-Planck-Gesellschaft (MPG), the Spanish Consejo Superior de Investigaciones Cient\'{\i}ficas (CSIC), the European Union through FEDER/ERF FICTS-2011-02 funds, and the members of the CARMENES Consortium (Max-Planck-Institut f\"ur Astronomie, Instituto de Astrof\'{\i}sica de Andaluc\'{\i}a, Landessternwarte K\"onigstuhl, Institut de Ci\`encies de l'Espai, Insitut f\"ur Astrophysik G\"ottingen,Universidad Complutense de Madrid, Th\"uringer Landessternwarte Tautenburg, Instituto de Astrof\'{\i}sica de Canarias, Hamburger Sternwarte, Centro de Astrobiolog\'{\i}a and Centro Astron\'omico Hispano-Alem\'an), with additional contributions by the Spanish Ministry of Economy, the German Science Foundation through the Major Research Instrumentation Programme and DFG Research Unit FOR2544 ``Blue Planets around Red Stars'', the Klaus Tschira Stiftung, the states of Baden-W\"urttemberg and Niedersachsen, and by the Junta de Andaluc\'{\i}a.

This work was partly financed by the Spanish Ministry of Science and Innovation through grants AYA2016-79425-C3-1/2/3-P 
and BES-2017-080769, 
and by NASA through grant NNX17AG24G. 
C.C., F.J.A.P., R.D., and G.H. compiled data for this work during their MSc theses at the Universidad Complutense de Madrid.
This publication made use of VOSA and the Filter Profile Service, developed and maintained by the Spanish Virtual Observatory {through grant AYA2017-84089},
the SIMBAD database, the Aladin sky atlas, and the VizieR catalogue access tool developed at CDS, Strasbourg Observatory, France, 
and the Python libraries {\tt Matplotlib, NumPy, SciPy} and collection of software packages {\tt AstroPy}.

\end{acknowledgements}

\bibliographystyle{aa} 
\bibliography{bibliography} 


\appendix
\section{Long tables and additional diagrams}
\label{appsection:tables}

\begin{landscape} 
{\fontsize{7}{10}\selectfont 
\begin{longtable}{lllccccccccc}
\caption{\label{tab:multiplicity} Star candidates belonging to multiple systems not tabulated by the Washington Double Star Catalog (WDS).}
\\
        \noalign{\smallskip}
        \noalign{\smallskip}
        \noalign{\smallskip}
\hline\hline
        \noalign{\smallskip}
Identifier       & Name\tablefootmark{a}      & Spectral             & $\alpha$  & $\delta$  & $\pi$        & $\mu_{\alpha}\cos{\delta}$        & $\mu_{\delta}$         & $\mu_{\rm total}$    & $G$            & $\theta$        & $\rho$       \\
        \noalign{\smallskip}
            &             & type                & (J2015.5) & (J2015.5)  & {[}mas{]}   & {[}mas a$^{-1}${]}       & {[}mas a$^{-1}${]}       & {[}mas a$^{-1}${]}           & {[}mag{]}        & {[}deg{]}     & {[}arcsec{]} \\ 
        \noalign{\smallskip}
\hline
        \noalign{\smallskip}
\endfirsthead
\caption{continued.}\\
\hline\hline
        \noalign{\smallskip}
Identifier       & Name\tablefootmark{a}         & Spectral          & $\alpha$  & $\delta$ & $\pi$     & $\mu_{\alpha}\cos{\delta}$        & $\mu_{\delta}$         & $\mu_{\rm total}$                   & $G$          & $\theta$          & $\rho$     \\
        \noalign{\smallskip}
            &            & type            & (J2015.5) & (J2015.5) & {[}mas{]}    & {[}mas a$^{-1}${]}       & {[}mas a$^{-1}${]}       & {[}mas a$^{-1}${]}          & {[}mag{]}       & {[}deg{]}      & {[}arcsec{]} \\ 
        \noalign{\smallskip}
\hline
        \noalign{\smallskip}
\endhead
\hline
\endfoot
        \noalign{\smallskip}
J00026+383  & 2M J00024011+3821453  & M4.0\,V  & 00:02:40.00  & +38:21:44.1  & 24.54 $ \pm $ 0.24 & -70.31 $ \pm $ 0.27  & -22.34 $ \pm $ 0.19  & 73.77 $ \pm $ 0.27  & 13.1900 $ \pm $ 0.0012 & 34.0  & 1.419 \\
            &                       &                         & 00:02:40.06  & +38:21:45.4  & 24.16 $ \pm $ 0.38 & -57.16 $ \pm $ 0.54  & -35.59 $ \pm $ 0.20  & 67.33 $ \pm $ 0.47  & 13.3648 $ \pm $ 0.0014 &       &       \\
I01007+2356 & PM J01007+2356        & K7\,V    & 01:00:46.85  & +23:56:54.4  & 24.75 $ \pm $ 0.04 & 129.88 $ \pm $ 0.06  & 8.41 $ \pm $ 0.06    & 130.15 $ \pm $ 0.06 & 10.7186 $ \pm $ 0.0007 & 4.7   & 1.480 \\
            &                       &                         & 01:00:46.85  & +23:56:55.9  &           \ldots         &           \ldots           &           \ldots           &           \ldots          & 14.6491 $ \pm $ 0.0089 &       &       \\
J01074-025  & RAVE J010727.5-023326 & K5\,V    & 01:07:27.46  & --02:33:27.4  & 6.77 $ \pm $ 0.05  & -54.38 $ \pm $ 0.09  & -62.33 $ \pm $ 0.05  & 82.72 $ \pm $ 0.07  & 12.1930 $ \pm $ 0.0003 & 165.2 & 1.363 \\
            &                       &                         & 01:07:27.48  & --02:33:28.8  & 6.42 $ \pm $ 0.08  & -54.13 $ \pm $ 0.17  & -61.36 $ \pm $ 0.07  & 81.82 $ \pm $ 0.12  & 14.5322 $ \pm $ 0.0020 &       &       \\
J02026+105  & RX J0202.4+1034       & M4.5\,V  & 02:02:28.15  & +10:34:51.9  & 70.43 $ \pm $ 0.53 & -54.60 $ \pm $ 1.07  & -96.95 $ \pm $ 0.77  & 111.27 $ \pm $ 0.85 & 11.8652 $ \pm $ 0.0012 &    25.3   & 0.904 \\
            &                       &                         & 02:02:28.18  & +10:34:52.7  & 68.79 $ \pm $ 1.20 & -101.45 $ \pm $ 2.02 & -58.95 $ \pm $ 1.43  & 117.33 $ \pm $ 1.89 & 12.3296 $ \pm $ 0.0064 &       &       \\
J02287+156  & BPM 85139             & M2.0\,V  & 02:28:47.14  & +15:38:53.6  & 28.53 $ \pm $ 0.11 & 170.91 $ \pm $ 0.18  & -9.17 $ \pm $ 0.17   & 171.15 $ \pm $ 0.18 & 11.5139 $ \pm $ 0.0037 & 147.7 & 0.814 \\
            &                       &                         & 02:28:47.17  & +15:38:52.9  &           \ldots         &           \ldots           &           \ldots           &           \ldots           & 13.0258 $ \pm $ 0.0055 &       &       \\
J02289+226  & BPM 85140             & M2.0\,V  & 02:28:58.41  & +22:36:24.5  & 17.16 $ \pm $ 0.04 & 148.78 $ \pm $ 0.08  & -48.74 $ \pm $ 0.07  & 156.56 $ \pm $ 0.08 & 11.3170 $ \pm $ 0.0006 & 147.4 & 3.022 \\
            &                       &                         & 02:28:58.52  & +22:36:21.9  & 17.19 $ \pm $ 0.53 & 134.47 $ \pm $ 1.13  & -36.20 $ \pm $ 0.99  & 139.26 $ \pm $ 1.12 & 12.0611 $ \pm $ 0.0003 &       &       \\
J03207+397  & LP 198-637            & M1.5\,V  & 03:20:45.41  & +39:42:59.4  & 31.60 $ \pm $ 0.51 & 126.71 $ \pm $ 1.33  & -129.25 $ \pm $ 0.86 & 181.00 $ \pm $ 1.12 & 10.9868 $ \pm $ 0.0020 & 278.1 & 0.783 \\
            &                       &                         & 03:20:45.35  & +39:42:59.7  &           \ldots         &           \ldots           &           \ldots           &           \ldots          & 11.2553 $ \pm $ 0.0065 &       &       \\
I03276+0956 & GJ 3226               & K7\,V    & 03:27:38.21  & +09:56:05.3  & 22.80 $ \pm $ 0.09 & 77.33 $ \pm $ 0.15   & -24.81 $ \pm $ 0.14  & 81.22 $ \pm $ 0.15  & 10.5483 $ \pm $ 0.0014 & 296.8 & 1.541 \\
            &                       &                         & 03:27:38.12  & +09:56:06.0  & 23.82 $ \pm $ 0.12 & 57.31 $ \pm $ 0.19   & -13.65 $ \pm $ 0.21  & 58.91 $ \pm $ 0.19  & 10.5825 $ \pm $ 0.0008 &       &       \\
J03284+352  & LSPM J0328+3515       & M2.0\,V  & 03:28:29.35  & +35:15:18.7  & 20.90 $ \pm $ 0.09 & 99.21 $ \pm $ 0.15   & -121.12 $ \pm $ 0.08 & 156.56 $ \pm $ 0.11 & 12.1366 $ \pm $ 0.0007 &   204.2    & 1.230 \\
            &                       &                         & 03:28:29.31  & +35:15:17.5  & 21.14 $ \pm $ 0.09 & 95.48 $ \pm $ 0.15   & -108.47 $ \pm $ 0.08 & 144.51 $ \pm $ 0.12 & 12.1711 $ \pm $ 0.0010 &       &       \\
J03544-091  & StKM 1-430            & M1.0\,V  & 03:54:25.52  & --09:09:29.2  & 47.39 $ \pm $ 0.04 & -95.44 $ \pm $ 0.06  & 110.84 $ \pm $ 0.05  & 146.27 $ \pm $ 0.06 & 10.5351 $ \pm $ 0.0006 & 153.2 & 3.177 \\
            &                       &                         & 03:54:25.62  & --09:09:32.1  & 47.40 $ \pm $ 0.06 & -96.46 $ \pm $ 0.09  & 98.93 $ \pm $ 0.08   & 138.17 $ \pm $ 0.09 & 11.8800 $ \pm $ 0.0012 &       &       \\
J05530+047  & G 106-007             & M1.5\,V  & 05:53:04.74  & +04:43:02.7  & 24.70 $ \pm $ 0.10 & 258.57 $ \pm $ 0.28  & -295.56 $ \pm $ 0.20 & 392.71 $ \pm $ 0.24 & 11.3303 $ \pm $ 0.0011 & 278.1 & 1.517 \\
            &                       &                         & 05:53:04.64  & +04:43:02.9  &           \ldots         &           \ldots           &           \ldots           &           \ldots           & 15.8285 $ \pm $ 0.0108 &       &       \\
I07245+1836 & PM J07245+1836        & K7\,V    & 07:24:32.30  & +18:36:31.3  & 19.99 $ \pm $ 0.05 & 53.55 $ \pm $ 0.09   & -36.35 $ \pm $ 0.08  & 64.72 $ \pm $ 0.09  & 10.8267 $ \pm $ 0.0005 & 326.9 & 1.836 \\
            &                       &                         & 07:24:32.23  & +18:36:32.9  &           \ldots         &           \ldots           &           \ldots           &           \ldots          & 12.6023 $ \pm $ 0.0024 &       &       \\
J07418+050$^b$  & G 050-001             & M2.5\,V+ & 07:41:52.56  & +05:02:23.1  & 36.09 $ \pm $ 0.07 & -248.35 $ \pm $ 0.13 & -87.34 $ \pm $ 0.10  & 263.26 $ \pm $ 0.12 & 11.6216 $ \pm $ 0.0009 & 136.1 & 1.006 \\
            &                       &                         & 07:41:52.61  & +05:02:22.4  &           \ldots         &           \ldots           &           \ldots           &           \ldots          & 16.1257 $ \pm $ 0.0263 &       &       \\
J07545-096  & 2M J07543272-0941478  & M3.5\,V  & 07:54:32.61  & --09:41:47.9  & 27.81 $ \pm $ 0.10 & -91.49 $ \pm $ 0.16  & -13.16 $ \pm $ 0.11  & 92.43 $ \pm $ 0.16  & 12.6737 $ \pm $ 0.0014 & 129.9 & 1.233 \\
            &                       &                         & 07:54:32.67  & --09:41:48.7  &           \ldots         &           \ldots           &           \ldots           &           \ldots          & 13.9321 $ \pm $ 0.0029 &       &       \\
I08192+5752 & PM J08192+5752        & K7\,V    & 08:19:14.01  & +57:52:26.8  & 19.67 $ \pm $ 0.04 & 39.66 $ \pm $ 0.06   & -79.76 $ \pm $ 0.06  & 89.08 $ \pm $ 0.06  & 10.7381 $ \pm $ 0.0005 & 93.3  & 1.475 \\
            &                       &                         & 08:19:14.19  & +57:52:26.6  & 19.71 $ \pm $ 0.14 & 40.66 $ \pm $ 0.23   & -72.23 $ \pm $ 0.46  & 82.89 $ \pm $ 0.42  & 14.1599 $ \pm $ 0.0067 &       &       \\
J09050+028  & LP 546-48             & M1.5\,V  & 09:05:04.12  & +02:50:03.8  & 42.58 $ \pm $ 0.25 & -312.21 $ \pm $ 0.39 & 29.17 $ \pm $ 0.42   & 313.57 $ \pm $ 0.39 & 10.9288 $ \pm $ 0.0021 & 253.0 & 1.214 \\
            &                       &                         & 09:05:04.04  & +02:50:03.5  &           \ldots         &           \ldots           &           \ldots           &           \ldots           & 12.1972 $ \pm $ 0.0021 &       &       \\
J09527+554  & G 195-043             & M1.5\,V  & 09:52:45.24  & +55:28:16.3  & 28.51 $ \pm $ 0.36 & 298.92 $ \pm $ 0.59  & -201.23 $ \pm $ 0.64 & 360.34 $ \pm $ 0.61 & 11.3624 $ \pm $ 0.0008 & 331.1 & 2.716 \\
            &                       &                         & 09:52:45.14  & +55:28:18.9  & 27.15 $ \pm $ 0.12 & 285.07 $ \pm $ 0.23  & -190.75 $ \pm $ 0.15 & 343.00 $ \pm $ 0.21 & 16.4864 $ \pm $ 0.0034 &       &       \\
I10526+0029 & PM J10526+0029        & K7\,V    & 10:52:39.52  & +00:29:01.5  & 25.18 $ \pm $ 0.08 & -91.32 $ \pm $ 0.10  & -31.13 $ \pm $ 0.08  & 96.48 $ \pm $ 0.09  & 10.1687 $ \pm $ 0.0016 & 251.0 & 1.594 \\
            &                       &                         & 10:52:39.42  & +00:29:01.0  &           \ldots         &           \ldots           &           \ldots           &           \ldots          & 12.7067 $ \pm $ 0.0042 &       &       \\
I11585+4626$^c$ & PM J11585+4626        & K7\,V    & 11:58:33.82  & +46:26:28.9  & 16.20 $ \pm $ 0.06 & -129.71 $ \pm $ 0.07 & 1.77 $ \pm $ 0.06    & 129.72 $ \pm $ 0.07 & 10.9651 $ \pm $ 0.0011 & 333.9 & 1.540 \\
            &                       &                         & 11:58:33.77  & +46:26:30.4  & 14.50 $ \pm $ 0.14 & -141.05 $ \pm $ 0.31 & 1.26 $ \pm $ 0.17    & 141.05 $ \pm $ 0.31 & 11.5845 $ \pm $ 0.0015 &       &       \\
J12191+318$^b$  & LP 320-626            & M4.0\,V+ & 12:19:05.57  & +31:50:43.6  &           \ldots         &           \ldots           &           \ldots           &           \ldots          & 11.1940 $ \pm $ 0.0006 & 225.2 & 1.764 \\
            &                       &                         & 12:19:05.48  & +31:50:42.2  & 35.13 $ \pm $ 0.09 & -295.73 $ \pm $ 0.10 & 5.02 $ \pm $ 0.11    & 295.77 $ \pm $ 0.10 & 13.9526 $ \pm $ 0.0023 &       &       \\
J12390+470  & G 123-049             & M2.0\,V  & 12:39:05.24  & +47:02:21.4  &           \ldots         &           \ldots           &           \ldots           &           \ldots          & 11.1336 $ \pm $ 0.0044 & 110.4 & 0.463 \\
            &                       &                         & 12:39:05.28  & +47:02:21.2  & 43.50 $ \pm $ 0.05 & 384.45 $ \pm $ 0.07  & -118.41 $ \pm $ 0.08 & 402.27 $ \pm $ 0.07 & 11.2091 $ \pm $ 0.0009 &       &       \\
J12513+221  & GJ 1166A              & M3.0\,V  & 12:51:23.72  & +22:06:15.7  & 30.32 $ \pm $ 0.51 & -177.34 $ \pm $ 0.98 & 50.54 $ \pm $ 0.79   & 184.40 $ \pm $ 0.96 & 12.1313 $ \pm $ 0.0019 &   91.8    & 1.263 \\
            &                       &                         & 12:51:23.81  & +22:06:15.6  &           \ldots         &           \ldots           &           \ldots           &           \ldots          & 13.3117 $ \pm $ 0.0038 &       &       \\
J13282+300  & BD+30 2400            & M0.0\,V  & 13:28:17.54  & +30:02:43.1  & 25.33 $ \pm $ 0.08 & -186.41 $ \pm $ 0.25 & -183.87 $ \pm $ 0.13 & 261.84 $ \pm $ 0.20 & 10.5043 $ \pm $ 0.0006 & 320.1 & 1.243 \\
            &                       &                         & 13:28:17.48  & +30:02:44.1  &           \ldots         &           \ldots           &           \ldots           &           \ldots          & 14.1569 $ \pm $ 0.0103 &       &       \\
J13445+249  & LP 379-098            & M1.0\,V  & 13:44:33.39  & +24:57:03.7  & 22.31 $ \pm $ 0.33 & -245.17 $ \pm $ 0.54 & -96.38 $ \pm $ 0.39  & 263.43 $ \pm $ 0.52 & 11.5057 $ \pm $ 0.0021 & 355.8 & 0.879 \\
            &                       &                         & 13:44:33.39  & +24:57:04.6  &           \ldots         &           \ldots           &           \ldots           &           \ldots          & 11.8891 $ \pm $ 0.0098 &       &       \\
J13490+026  & Wolf 1495             & M1.5\,V  & 13:49:01.18  & +02:47:23.3  &           \ldots         &           \ldots           &           \ldots           &           \ldots          & 10.7706 $ \pm $ 0.0210 & 315.7 & 0.680 \\
            &                       &                         & 13:49:01.15  & +02:47:23.8  & 55.78 $ \pm $ 0.75 & 149.68 $ \pm $ 1.64  & -333.14 $ \pm $ 1.56 & 365.22 $ \pm $ 1.57 & 10.8252 $ \pm $ 0.0075 &       &       \\
I15380+3224$^c$ & PM J15380+3224        & K7\,V    & 15:38:04.49  & +32:24:31.9  & 16.48 $ \pm $ 0.23 & -63.14 $ \pm $ 0.31  & -78.46 $ \pm $ 0.38  & 100.71 $ \pm $ 0.36 & 11.2639 $ \pm $ 0.0021 & 143.8 & 1.061 \\
            &                       &                         & 15:38:04.53  & +32:24:31.0  & 15.10 $ \pm $ 0.25 & -74.75 $ \pm $ 0.39  & -83.43 $ \pm $ 0.48  & 112.02 $ \pm $ 0.44 & 11.3307 $ \pm $ 0.0019 &       &       \\
J16573+271  & 2M J16572235+2708304  & M2.0\,V  & 16:57:22.27  & +27:08:31.1  & 27.12 $ \pm $ 0.12 & -34.09 $ \pm $ 0.21  & 44.34 $ \pm $ 0.26   & 55.93 $ \pm $ 0.24  & 12.3752 $ \pm $ 0.0011 & 117.6 & 1.048 \\
            &                       &                         & 16:57:22.34  & +27:08:30.6  &           \ldots         &           \ldots           &           \ldots           &           \ldots          & 13.2959 $ \pm $ 0.0025 &       &       \\
I17068+3212 & PM J17068+3212        & K7\,V    & 17:06:48.88  & +32:11:59.3  & 31.93 $ \pm $ 0.02 & 53.18 $ \pm $ 0.03   & -74.71 $ \pm $ 0.04  & 91.70 $ \pm $ 0.04  & 10.7788 $ \pm $ 0.0003 & 31.4  & 3.279 \\
            &                       &                         & 17:06:49.00  & +32:12:02.2  & 31.93 $ \pm $ 0.03 & 46.05 $ \pm $ 0.09   & -82.76 $ \pm $ 0.06  & 94.71 $ \pm $ 0.07  & 12.6244 $ \pm $ 0.0004 &       &       \\
J18116+061  & NLTT 46076            & M3.0\,V  & 18:11:36.49  & +06:06:27.8  &           \ldots         &           \ldots           &           \ldots           &           \ldots          & 11.9662 $ \pm $ 0.0074 & 139.6 & 0.625 \\
            &                       &                         & 18:11:36.51  & +06:06:27.3 &           \ldots         &           \ldots           &           \ldots           &           \ldots          & 13.5146 $ \pm $ 0.0077 &       &       \\
J18400+726  & LP 044-334            & M6.5\,V  & 18:40:02.20  & +72:40:57.1  & 51.04 $ \pm $ 0.52 & -43.74 $ \pm $ 0.83  & 184.49 $ \pm $ 1.09  & 189.60 $ \pm $ 1.08 & 15.3854 $ \pm $ 0.0114 & 110.3 & 0.821 \\
            &                       &                         & 18:40:02.32  & +72:40:56.5  &           \ldots         &           \ldots           &           \ldots           &           \ldots          & 15.7040 $ \pm $ 0.0035 &       &       \\
I18447+6241 & PM J18447+6241        & K7\,V    & 18:44:47.49  & +62:41:08.3  & 22.52 $ \pm $ 0.03 & -33.82 $ \pm $ 0.06  & 56.65 $ \pm $ 0.05   & 65.98 $ \pm $ 0.05  & 10.7927 $ \pm $ 0.0006 & 277.5 & 1.342 \\
            &                       &                         & 18:44:47.30  & +62:41:08.7  &           \ldots         &           \ldots           &           \ldots           &           \ldots          & 14.7260 $ \pm $ 0.0167 &       &       \\
I21088+1247 & BD+12 4554            & K7\,V    & 21:08:51.84  & +12:47:36.9  & 23.81 $ \pm $ 0.04 & 85.71 $ \pm $ 0.06   & -67.99 $ \pm $ 0.05  & 109.40 $ \pm $ 0.06 & 10.4390 $ \pm $ 0.0005 & 2.5   & 1.833 \\
            &                       &                         & 21:08:51.85  & +12:47:38.7  &           \ldots         &           \ldots           &           \ldots           &           \ldots          & 14.4780 $ \pm $ 0.0077 &       &       \\
I21415+4925 & PM J21415+4925        & K7\,V    & 21:41:31.36  & +49:25:38.1  & 29.93 $ \pm $ 0.02 & 33.99 $ \pm $ 0.04   & -85.44 $ \pm $ 0.04  & 91.95 $ \pm $ 0.04  & 9.9125 $ \pm $ 0.0003  & 308.8 & 1.609 \\
            &                       &                         & 21:41:31.26  & +49:25:39.3  & 30.10 $ \pm $ 0.10 & 50.32 $ \pm $ 0.60   & -88.08 $ \pm $ 0.83  & 101.44 $ \pm $ 0.78 & 13.2790 $ \pm $ 0.0044 &       &       \\
J22012+323  & TYC 2723-908-1        & M1.5\,V  & 22:01:14.12  & +32:23:13.9  & 32.55 $ \pm $ 0.08 & 118.50 $ \pm $ 0.09  & 62.63 $ \pm $ 0.15   & 134.03 $ \pm $ 0.11 & 11.4391 $ \pm $ 0.0012 & 239.8 & 1.295 \\
            &                       &                         & 22:01:14.04  & +32:23:13.1  & 32.10 $ \pm $ 0.20 & 107.88 $ \pm $ 0.27  & 53.40 $ \pm $ 0.31   & 120.37 $ \pm $ 0.28 & 12.8902 $ \pm $ 0.0022 &       &       \\
I22142+1712 & PM J22142+1712        & K7\,V    & 22:14:12.84  & +17:12:24.4  & 17.18 $ \pm $ 0.05 & -10.52 $ \pm $ 0.08  & -84.77 $ \pm $ 0.07  & 85.42 $ \pm $ 0.07  & 10.7551 $ \pm $ 0.0006 & 268.2 & 1.510 \\
            &                       &                         & 22:14:12.73  & +17:12:24.4  &           \ldots         &           \ldots           &           \ldots           &           \ldots          & 15.4819 $ \pm $ 0.0194 &       &       \\
I22569+0031 & PM J22569+0031        & K7\,V    & 22:56:54.65  & +00:31:23.6  &           \ldots         &           \ldots           &           \ldots           &           \ldots          & 10.9107 $ \pm $ 0.0145 & 22.4  & 0.907 \\
            &                       &                         & 22:56:54.67  & +00:31:24.4  & 17.57 $ \pm $ 0.68 & 13.04 $ \pm $ 1.16   & -82.42 $ \pm $ 0.87  & 83.44 $ \pm $ 0.88  & 11.0021 $ \pm $ 0.0058 &       &       \\
I22596+2154 & PM J22596+2154        & K7\,V    & 22:59:41.42  & +21:54:05.8  & 26.30 $ \pm $ 0.05 & 127.97 $ \pm $ 0.09  & -59.09 $ \pm $ 0.06  & 140.96 $ \pm $ 0.09 & 10.1878 $ \pm $ 0.0006 & 37.3  & 2.093 \\
            &                       &                         & 22:59:41.51  & +21:54:07.6  &           \ldots         &           \ldots           &           \ldots           &           \ldots          & 13.3417 $ \pm $ 0.0037 &       &       \\
J23051+452  & LSPM J2305+4517       & M3.5\,V  & 23:05:08.99  & +45:17:32.9  & 21.94 $ \pm $ 0.17 & 184.58 $ \pm $ 0.26  & 67.37 $ \pm $ 0.27   & 196.49 $ \pm $ 0.26 & 12.3370 $ \pm $ 0.0021 & 80.7  & 0.785 \\
            &                       &                         & 23:05:09.06  & +45:17:33.1  &           \ldots         &           \ldots           &           \ldots           &           \ldots          & 14.3079 $ \pm $ 0.0029 &       &       \\
J23489+098  & {[}R78b{]} 377        & M1.0\,V  & 23:48:58.97  & +09:51:53.4  & 21.08 $ \pm $ 0.05 & 147.68 $ \pm $ 0.09  & -52.74 $ \pm $ 0.05  & 156.82 $ \pm $ 0.08 & 11.3848 $ \pm $ 0.0009 & 20.1  & 1.945 \\
            &                       &                         & 23:48:59.02  & +09:51:55.2  & 20.76 $ \pm $ 0.08 & 141.19 $ \pm $ 0.18  & -43.69 $ \pm $ 0.06  & 147.80 $ \pm $ 0.17 & 13.9181 $ \pm $ 0.0013 &       &       \\
J23590+208  & G 129-051             & M2.5\,V  & 23:59:00.73  & +20:51:37.3  & 14.96 $ \pm $ 0.79 & 228.92 $ \pm $ 1.25  & -104.85 $ \pm $ 0.57 & 251.79 $ \pm $ 1.16 & 12.0177 $ \pm $ 0.0072 & 170.7 & 0.521 \\
            &                       &                         & 23:59:00.73  & +20:51:36.7  &           \ldots         &           \ldots           &           \ldots           &           \ldots          & 12.3273 $ \pm $ 0.0186 &       &       \\
\end{longtable}
\tablefoot{
\tablefoottext{a}{Primaries ``A'' are always brighter than secondaries ``B'' in the $G$ band.}
\tablefoottext{b}{Previously identified as spectroscopic binaries in \cite{Rei12} and \cite{Jef18}.} 
\tablefoottext{c}{Common proper motion pairs with $\Delta \pi >$ 5\,\% labelled in Fig.~\ref{fig:binaries_distance}.}
}
}
\end{landscape}

\begin{landscape} 
{\fontsize{7}{10}\selectfont 
\begin{longtable}{lccccccccccc}
\caption{\label{table.colours} Average colours for K5\,V to L8 sources. The number in parentheses indicates the number of useful data points.}\\

\hline\hline
	\noalign{\smallskip}
Spectral       &            $FUV-NUV$        &  $NUV-u$   &  $u-B_T$      & $B_T-B$         & $B-g$       & $g-G_{BP}$     & $G_{BP}-V_T$   &  $V_T-V$  &  $V-G$  &  $G-r$  \\
        \noalign{\smallskip}
type            &           [mag]        & [mag] & [mag]   & [mag]  & [mag]   & [mag]   & [mag]    & [mag] & [mag] & [mag] \\ 
        \noalign{\smallskip}
\hline
        \noalign{\smallskip}
\endfirsthead

\caption{continued.}\\
\hline\hline
        \noalign{\smallskip}
Spectral       &            $r-i$        &  $i-G_{RP}$   &  $G_{RP}-J$      & $J-H$         & $H-Ks$       & $Ks-W1$     & $W1-W2$   &  $W2-W3$  &  $W3-W4$   \\
        \noalign{\smallskip}
type            &           [mag]        & [mag] & [mag]   & [mag]  & [mag]   & [mag]   & [mag]    & [mag] & [mag] \\ 
        \noalign{\smallskip}
\hline
        \noalign{\smallskip}
\endhead

\hline
\endfoot
        \noalign{\smallskip}
K5\,V  &  3.59 $\pm$ 0.95 (2)  &  5.37   (1)  &  \ldots  &  0.24 $\pm$ 0.40 (6)  &  0.63 $\pm$ 0.23 (11)  &  0.32 $\pm$ 0.11 (11)  &  0.04 $\pm$ 0.08 (8)  &  0.26 $\pm$ 0.07 (4)  &  0.45 $\pm$ 0.04 (9)  &  --0.03 $\pm$ 0.21 (13) \\ K7\,V  &  3.26 $\pm$ 0.65 (7)  &  4.23 $\pm$ 0.75 (36)  &  2.66 $\pm$ 0.83 (59)  &  0.49 $\pm$ 0.43 (98)  &  0.62 $\pm$ 0.16 (107)  &  0.48 $\pm$ 0.13 (101)  &  0.02 $\pm$ 0.12 (94)  &  0.21 $\pm$ 0.16 (96)  &  0.61 $\pm$ 0.10 (100)  &  --0.06 $\pm$ 0.12 (103) \\ 
M0.0\,V  &  2.83 $\pm$ 0.61 (22)  &  4.05 $\pm$ 0.83 (37)  &  2.87 $\pm$ 0.73 (54)  &  0.46 $\pm$ 0.36 (108)  &  0.65 $\pm$ 0.14 (113)  &  0.51 $\pm$ 0.12 (117)  &  0.08 $\pm$ 0.09 (113)  &  0.15 $\pm$ 0.11 (109)  &  0.71 $\pm$ 0.09 (115)  &  --0.17 $\pm$ 0.14 (120) \\ 
M0.5\,V  &  2.40 $\pm$ 0.42 (14)  &  4.01 $\pm$ 0.96 (22)  &  2.85 $\pm$ 0.82 (36)  &  0.43 $\pm$ 0.36 (70)  &  0.66 $\pm$ 0.13 (72)  &  0.53 $\pm$ 0.11 (72)  &  0.07 $\pm$ 0.14 (69)  &  0.17 $\pm$ 0.16 (68)  &  0.77 $\pm$ 0.10 (72)  &  --0.18 $\pm$ 0.11 (71) \\ 
M1.0\,V  &  2.29 $\pm$ 0.44 (11)  &  4.72 $\pm$ 0.82 (43)  &  2.48 $\pm$ 0.70 (42)  &  0.35 $\pm$ 0.40 (86)  &  0.72 $\pm$ 0.14 (135)  &  0.51 $\pm$ 0.10 (140)  &  0.09 $\pm$ 0.16 (87)  &  0.15 $\pm$ 0.16 (87)  &  0.82 $\pm$ 0.09 (135)  &  --0.23 $\pm$ 0.10 (140) \\ 
M1.5\,V  &  2.24 $\pm$ 0.49 (13)  &  4.95 $\pm$ 0.98 (43)  &  2.53 $\pm$ 0.66 (32)  &  0.28 $\pm$ 0.47 (63)  &  0.72 $\pm$ 0.13 (115)  &  0.51 $\pm$ 0.15 (121)  &  0.12 $\pm$ 0.17 (65)  &  0.10 $\pm$ 0.20 (65)  &  0.89 $\pm$ 0.11 (115)  &  --0.30 $\pm$ 0.13 (121) \\ 
M2.0\,V  &  1.90 $\pm$ 0.66 (12)  &  5.30 $\pm$ 0.78 (36)  &  2.08 $\pm$ 0.69 (29)  &  0.36 $\pm$ 0.59 (54)  &  0.76 $\pm$ 0.05 (118)  &  0.49 $\pm$ 0.11 (124)  &  0.15 $\pm$ 0.24 (55)  &  0.11 $\pm$ 0.26 (53)  &  0.95 $\pm$ 0.10 (117)  &  --0.37 $\pm$ 0.10 (124) \\ 
M2.5\,V  &  1.66 $\pm$ 0.65 (9)  &  5.37 $\pm$ 1.07 (37)  &  2.35 $\pm$ 0.85 (21)  &  0.35 $\pm$ 0.55 (30)  &  0.79 $\pm$ 0.07 (138)  &  0.48 $\pm$ 0.09 (145)  &  0.12 $\pm$ 0.18 (31)  &  0.16 $\pm$ 0.18 (29)  &  1.04 $\pm$ 0.09 (137)  &  --0.47 $\pm$ 0.11 (146) \\ 
M3.0\,V  &  1.33 $\pm$ 0.45 (12)  &  5.17 $\pm$ 0.93 (38)  &  1.82 $\pm$ 0.37 (8)  &  0.42 $\pm$ 0.37 (20)  &  0.80 $\pm$ 0.11 (157)  &  0.47 $\pm$ 0.09 (173)  &  0.09 $\pm$ 0.12 (20)  &  0.17 $\pm$ 0.11 (20)  &  1.13 $\pm$ 0.09 (156)  &  --0.55 $\pm$ 0.11 (173) \\ 
M3.5\,V  &  1.31 $\pm$ 0.36 (23)  &  4.72 $\pm$ 1.13 (46)  &  2.18 $\pm$ 0.80 (7)  &  0.20 $\pm$ 0.59 (14)  &  0.85 $\pm$ 0.07 (217)  &  0.46 $\pm$ 0.09 (237)  &  0.13 $\pm$ 0.19 (12)  &  0.14 $\pm$ 0.19 (13)  &  1.24 $\pm$ 0.09 (215)  &  --0.66 $\pm$ 0.10 (237) \\ 
M4.0\,V  &  1.49 $\pm$ 0.56 (29)  &  3.91 $\pm$ 1.36 (36)  &  2.43 $\pm$ 0.63 (2)  &  0.05 $\pm$ 0.50 (5)  &  0.90 $\pm$ 0.07 (167)  &  0.46 $\pm$ 0.08 (203)  &  0.23 $\pm$ 0.24 (6)  &  0.04 $\pm$ 0.24 (6)  &  1.38 $\pm$ 0.09 (165)  &  --0.77 $\pm$ 0.09 (203) \\ 
M4.5\,V  &  1.04 $\pm$ 0.65 (15)  &  2.96 $\pm$ 0.90 (18)  &  \ldots  &  \ldots  &  0.97 $\pm$ 0.08 (95)  &  0.47 $\pm$ 0.10 (116)  &  \ldots  &  \ldots  &  1.54 $\pm$ 0.10 (96)  &  --0.92 $\pm$ 0.10 (116) \\ 
M5.0\,V  &  1.12 $\pm$ 0.42 (8)  &  2.75 $\pm$ 0.77 (12)  &  \ldots  &  \ldots  &  0.98 $\pm$ 0.19 (35)  &  0.46 $\pm$ 0.16 (60)  &  \ldots  &  \ldots  &  1.71 $\pm$ 0.16 (38)  &  --1.10 $\pm$ 0.16 (60) \\ 
M5.5\,V  &  1.52 $\pm$ 0.34 (2)  &  2.80 $\pm$ 0.58 (7)  &  \ldots  &  \ldots  &  1.13 $\pm$ 0.21 (14)  &  0.49 $\pm$ 0.13 (26)  &  \ldots  &  \ldots  &  2.04 $\pm$ 0.12 (14)  &  --1.37 $\pm$ 0.10 (26) \\
M6.0\,V  &  0.84 $\pm$ 0.15 (2)  &  0.89 $\pm$ 1.78 (5)  &  \ldots  &  \ldots  &  1.24 $\pm$ 0.04 (2)  &  0.48 $\pm$ 0.20 (14)  &  \ldots  &  \ldots  &  2.01   (1)  &  --1.73 $\pm$ 0.24 (14) \\ 
M6.5\,V  &  \ldots  &  \ldots  &  \ldots  &  \ldots  &  1.07 $\pm$ 0.20 (2)  &  0.49 $\pm$ 0.15 (7)  &  \ldots  &  \ldots  &  2.72 $\pm$ 0.21 (2)  &  --2.00 $\pm$ 0.17 (7) \\ 
M7.0\,V  &  \ldots  &  2.90   (1)  &  \ldots  &  \ldots  &  1.30   (1)  &  0.46 $\pm$ 0.19 (5)  &  \ldots  &  \ldots  &  2.80   (1)  &  --2.08 $\pm$ 0.28 (5) \\
M7.5\,V  &  \ldots  &  \ldots  &  \ldots  &  \ldots  &  \ldots  &  0.42 $\pm$ 0.16 (3)  &  \ldots  &  \ldots  &  \ldots  &  --2.06 $\pm$ 0.12 (3) \\ 
M8.0\,V  &  \ldots  &  \ldots  &  \ldots  &  \ldots  &  \ldots  &  0.56 $\pm$ 0.15 (4)  &  \ldots  &  \ldots  &  \ldots  &  --2.23 $\pm$ 0.08 (6) \\ 
M8.5\,V  &  \ldots  &  \ldots  &  \ldots  &  \ldots  &  \ldots  &  0.52 $\pm$ 0.01 (2)  &  \ldots  &  \ldots  &  \ldots  &  --2.16 $\pm$ 0.13 (3) \\ 
M9.0\,V  &  \ldots  &  \ldots  &  \ldots  &  \ldots  &  \ldots  &  0.72   (1)  &  \ldots  &  \ldots  &  \ldots  &  --2.01 $\pm$ 0.07 (2) \\ 
M9.5\,V  &  \ldots  &  \ldots  &  \ldots  &  \ldots  &  \ldots  &  \ldots  &  \ldots  &  \ldots  &  \ldots  &  --1.96   (1) \\ 
L0.0  &  \ldots  &  \ldots  &  \ldots  &  \ldots  &  \ldots  &  \ldots  &  \ldots  &  \ldots  &  \ldots  &  --1.98 $\pm$ 0.16 (10) \\ 
L0.5  &  \ldots  &  \ldots  &  \ldots  &  \ldots  &  \ldots  &  \ldots  &  \ldots  &  \ldots  &  \ldots  &  --1.92 $\pm$ 0.18 (5) \\ 
L1.0  &  \ldots  &  \ldots  &  \ldots  &  \ldots  &  \ldots  &  \ldots  &  \ldots  &  \ldots  &  \ldots  &  --1.98 $\pm$ 0.12 (14) \\ 
L1.5  &  \ldots  &  \ldots  &  \ldots  &  \ldots  &  \ldots  &  \ldots  &  \ldots  &  \ldots  &  \ldots  &  --1.99 $\pm$ 0.08 (4) \\ 
L2.0  &  \ldots  &  \ldots  &  \ldots  &  \ldots  &  \ldots  &  \ldots  &  \ldots  &  \ldots  &  \ldots  &  --1.88 $\pm$ 0.08 (11) \\ 
L2.5  &  \ldots  &  \ldots  &  \ldots  &  \ldots  &  \ldots  &  \ldots  &  \ldots  &  \ldots  &  \ldots  &  --1.90 $\pm$ 0.04 (4) \\ 
L3.0  &  \ldots  &  \ldots  &  \ldots  &  \ldots  &  \ldots  &  \ldots  &  \ldots  &  \ldots  &  \ldots  &  --1.88 $\pm$ 0.12 (3) \\ 
L3.5  &  \ldots  &  \ldots  &  \ldots  &  \ldots  &  \ldots  &  \ldots  &  \ldots  &  \ldots  &  \ldots  &  --2.02 $\pm$ 0.15 (2) \\ 
L4.0  &  \ldots  &  \ldots  &  \ldots  &  \ldots  &  \ldots  &  \ldots  &  \ldots  &  \ldots  &  \ldots  &  --1.86 $\pm$ 0.10 (3) \\ 
L4.5  &  \ldots  &  \ldots  &  \ldots  &  \ldots  &  \ldots  &  \ldots  &  \ldots  &  \ldots  &  \ldots  &  \ldots \\ 
L5.0  &  \ldots  &  \ldots  &  \ldots  &  \ldots  &  \ldots  &  \ldots  &  \ldots  &  \ldots  &  \ldots  &  --2.01   (1) \\ 
L5.5  &  \ldots  &  \ldots  &  \ldots  &  \ldots  &  \ldots  &  \ldots  &  \ldots  &  \ldots  &  \ldots  &  --2.04   (1) \\ 
L6.0  &  \ldots  &  \ldots  &  \ldots  &  \ldots  &  \ldots  &  \ldots  &  \ldots  &  \ldots  &  \ldots  &  --2.09   (1) \\ 
L6.5  &  \ldots  &  \ldots  &  \ldots  &  \ldots  &  \ldots  &  \ldots  &  \ldots  &  \ldots  &  \ldots  &  \ldots \\ 
L7.0  &  \ldots  &  \ldots  &  \ldots  &  \ldots  &  \ldots  &  \ldots  &  \ldots  &  \ldots  &  \ldots  &  \ldots \\ 
L7.5  &  \ldots  &  \ldots  &  \ldots  &  \ldots  &  \ldots  &  \ldots  &  \ldots  &  \ldots  &  \ldots  &  --2.57   (1) \\ 
L8.0  &  \ldots  &  \ldots  &  \ldots  &  \ldots  &  \ldots  &  \ldots  &  \ldots  &  \ldots  &  \ldots  &  \ldots \\
\newpage
K5\,V  &  0.41 $\pm$ 0.07 (13)  &  0.40 $\pm$ 0.20 (12)  &  0.95 $\pm$ 0.06 (13)  &  0.57 $\pm$ 0.03 (14)  &  0.14 $\pm$ 0.03 (14)  &  0.08 $\pm$ 0.06 (14)  &  0.04 $\pm$ 0.19 (15)  &  --0.02 $\pm$ 0.20 (15)  &  0.01 $\pm$ 0.13 (10) \\ 
K7\,V  &  0.60 $\pm$ 0.12 (106)  &  0.34 $\pm$ 0.12 (99)  &  1.12 $\pm$ 0.05 (104)  &  0.63 $\pm$ 0.03 (112)  &  0.17 $\pm$ 0.03 (112)  &  0.11 $\pm$ 0.04 (109)  &  --0.03 $\pm$ 0.06 (109)  &  0.05 $\pm$ 0.06 (112)  &  0.06 $\pm$ 0.11 (112) \\ 
M0.0\,V  &  0.74 $\pm$ 0.14 (119)  &  0.36 $\pm$ 0.11 (117)  &  1.22 $\pm$ 0.05 (118)  &  0.63 $\pm$ 0.04 (120)  &  0.19 $\pm$ 0.03 (119)  &  0.12 $\pm$ 0.05 (116)  &  0.02 $\pm$ 0.09 (117)  &  0.02 $\pm$ 0.08 (119)  &  0.10 $\pm$ 0.09 (118) \\ 
M0.5\,V  &  0.80 $\pm$ 0.13 (70)  &  0.36 $\pm$ 0.08 (71)  &  1.28 $\pm$ 0.05 (72)  &  0.63 $\pm$ 0.03 (72)  &  0.21 $\pm$ 0.03 (71)  &  0.12 $\pm$ 0.05 (70)  &  0.05 $\pm$ 0.09 (70)  &  0.03 $\pm$ 0.07 (71)  &  0.11 $\pm$ 0.09 (73) \\ 
M1.0\,V  &  0.87 $\pm$ 0.11 (140)  &  0.36 $\pm$ 0.09 (139)  &  1.32 $\pm$ 0.04 (139)  &  0.62 $\pm$ 0.04 (140)  &  0.21 $\pm$ 0.03 (138)  &  0.14 $\pm$ 0.04 (138)  &  0.05 $\pm$ 0.09 (138)  &  0.06 $\pm$ 0.06 (138)  &  0.09 $\pm$ 0.11 (141) \\ 
M1.5\,V  &  0.96 $\pm$ 0.14 (124)  &  0.37 $\pm$ 0.11 (121)  &  1.37 $\pm$ 0.05 (119)  &  0.63 $\pm$ 0.04 (121)  &  0.21 $\pm$ 0.03 (121)  &  0.14 $\pm$ 0.04 (122)  &  0.06 $\pm$ 0.08 (121)  &  0.06 $\pm$ 0.06 (123)  &  0.09 $\pm$ 0.11 (125) \\ 
M2.0\,V  &  1.05 $\pm$ 0.12 (125)  &  0.38 $\pm$ 0.11 (126)  &  1.44 $\pm$ 0.05 (127)  &  0.61 $\pm$ 0.05 (128)  &  0.23 $\pm$ 0.03 (128)  &  0.15 $\pm$ 0.04 (126)  &  0.09 $\pm$ 0.07 (126)  &  0.07 $\pm$ 0.05 (126)  &  0.10 $\pm$ 0.12 (122) \\ 
M2.5\,V  &  1.17 $\pm$ 0.11 (147)  &  0.40 $\pm$ 0.10 (144)  &  1.51 $\pm$ 0.05 (147)  &  0.59 $\pm$ 0.04 (150)  &  0.24 $\pm$ 0.03 (147)  &  0.16 $\pm$ 0.05 (147)  &  0.11 $\pm$ 0.05 (149)  &  0.08 $\pm$ 0.04 (149)  &  0.12 $\pm$ 0.13 (144) \\ 
M3.0\,V  &  1.29 $\pm$ 0.11 (173)  &  0.41 $\pm$ 0.10 (172)  &  1.58 $\pm$ 0.05 (172)  &  0.59 $\pm$ 0.04 (172)  &  0.25 $\pm$ 0.03 (171)  &  0.16 $\pm$ 0.04 (173)  &  0.13 $\pm$ 0.05 (172)  &  0.09 $\pm$ 0.05 (172)  &  0.13 $\pm$ 0.14 (162) \\ 
M3.5\,V  &  1.40 $\pm$ 0.13 (240)  &  0.44 $\pm$ 0.12 (235)  &  1.66 $\pm$ 0.05 (237)  &  0.58 $\pm$ 0.05 (242)  &  0.26 $\pm$ 0.03 (241)  &  0.17 $\pm$ 0.04 (241)  &  0.15 $\pm$ 0.05 (239)  &  0.10 $\pm$ 0.05 (239)  &  0.15 $\pm$ 0.15 (226) \\ 
M4.0\,V  &  1.53 $\pm$ 0.15 (205)  &  0.47 $\pm$ 0.13 (201)  &  1.76 $\pm$ 0.06 (202)  &  0.58 $\pm$ 0.04 (206)  &  0.27 $\pm$ 0.03 (206)  &  0.18 $\pm$ 0.04 (201)  &  0.17 $\pm$ 0.04 (200)  &  0.12 $\pm$ 0.04 (199)  &  0.19 $\pm$ 0.17 (169) \\ 
M4.5\,V  &  1.71 $\pm$ 0.13 (116)  &  0.49 $\pm$ 0.09 (116)  &  1.87 $\pm$ 0.06 (116)  &  0.56 $\pm$ 0.04 (116)  &  0.29 $\pm$ 0.03 (116)  &  0.20 $\pm$ 0.04 (114)  &  0.19 $\pm$ 0.03 (114)  &  0.15 $\pm$ 0.04 (115)  &  0.23 $\pm$ 0.21 (88) \\ 
M5.0\,V  &  1.85 $\pm$ 0.21 (59)  &  0.59 $\pm$ 0.14 (59)  &  2.02 $\pm$ 0.12 (61)  &  0.58 $\pm$ 0.03 (62)  &  0.31 $\pm$ 0.03 (62)  &  0.21 $\pm$ 0.03 (62)  &  0.20 $\pm$ 0.03 (61)  &  0.16 $\pm$ 0.04 (60)  &  0.28 $\pm$ 0.28 (43) \\ 
M5.5\,V  &  2.12 $\pm$ 0.15 (26)  &  0.64 $\pm$ 0.13 (26)  &  2.14 $\pm$ 0.08 (27)  &  0.57 $\pm$ 0.04 (27)  &  0.33 $\pm$ 0.02 (26)  &  0.23 $\pm$ 0.03 (26)  &  0.21 $\pm$ 0.05 (27)  &  0.17 $\pm$ 0.05 (27)  &  0.23 $\pm$ 0.16 (21) \\ 
M6.0\,V  &  2.32 $\pm$ 0.25 (13)  &  0.83 $\pm$ 0.12 (12)  &  2.33 $\pm$ 0.16 (12)  &  0.60 $\pm$ 0.08 (15)  &  0.37 $\pm$ 0.03 (15)  &  0.23 $\pm$ 0.02 (15)  &  0.21 $\pm$ 0.04 (15)  &  0.21 $\pm$ 0.07 (15)  &  0.33 $\pm$ 0.22 (5) \\ 
M6.5\,V  &  2.68 $\pm$ 0.14 (7)  &  0.82 $\pm$ 0.10 (7)  &  2.49 $\pm$ 0.07 (7)  &  0.61 $\pm$ 0.03 (7)  &  0.37 $\pm$ 0.02 (7)  &  0.24 $\pm$ 0.04 (7)  &  0.20 $\pm$ 0.03 (7)  &  0.23 $\pm$ 0.03 (7)  &  0.10 $\pm$ 0.15 (4) \\ 
M7.0\,V  &  2.62 $\pm$ 0.26 (5)  &  0.99 $\pm$ 0.15 (5)  &  2.52 $\pm$ 0.17 (5)  &  0.58 $\pm$ 0.06 (5)  &  0.38 $\pm$ 0.04 (5)  &  0.26 $\pm$ 0.09 (5)  &  0.21 $\pm$ 0.03 (5)  &  0.23 $\pm$ 0.08 (5)  &  0.21 $\pm$ 0.07 (3) \\ 
M7.5\,V  &  2.66 $\pm$ 0.09 (3)  &  0.91 $\pm$ 0.05 (3)  &  2.52 $\pm$ 0.09 (3)  &  0.64 $\pm$ 0.01 (3)  &  0.37 $\pm$ 0.03 (3)  &  0.25 $\pm$ 0.05 (3)  &  0.21 $\pm$ 0.01 (3)  &  0.27 $\pm$ 0.06 (3)  &  \ldots \\ 
M8.0\,V  &  2.77 $\pm$ 0.10 (6)  &  1.03 $\pm$ 0.10 (6)  &  2.75 $\pm$ 0.12 (9)  &  0.66 $\pm$ 0.03 (9)  &  0.41 $\pm$ 0.03 (9)  &  0.25 $\pm$ 0.04 (9)  &  0.24 $\pm$ 0.05 (9)  &  0.28 $\pm$ 0.04 (8)  &  \ldots \\ 
M8.5\,V  &  2.70 $\pm$ 0.05 (2)  &  1.12 $\pm$ 0.02 (2)  &  2.93 $\pm$ 0.05 (4)  &  0.64 $\pm$ 0.05 (4)  &  0.46 $\pm$ 0.01 (4)  &  0.33 $\pm$ 0.04 (4)  &  0.28 $\pm$ 0.05 (4)  &  0.35 $\pm$ 0.08 (4)  &  0.32 $\pm$ 0.05 (3) \\ 
M9.0\,V  &  2.43 $\pm$ 0.09 (2)  &  1.17 $\pm$ 0.02 (3)  &  3.06 $\pm$ 0.08 (5)  &  0.67 $\pm$ 0.06 (5)  &  0.46 $\pm$ 0.04 (5)  &  0.35 $\pm$ 0.03 (5)  &  0.29 $\pm$ 0.03 (5)  &  0.49 $\pm$ 0.08 (5)  &  0.30   (1) \\ 
M9.5\,V  &  2.39   (1)  &  1.18   (1)  &  3.12 $\pm$ 0.11 (2)  &  0.84 $\pm$ 0.07 (2)  &  0.57 $\pm$ 0.03 (2)  &  0.34 $\pm$ 0.02 (2)  &  0.28 $\pm$ 0.02 (2)  &  0.45 $\pm$ 0.08 (2)  &  \ldots \\ 
L0.0  &  2.43 $\pm$ 0.16 (10)  &  1.16 $\pm$ 0.03 (9)  &  3.07 $\pm$ 0.05 (11)  &  0.75 $\pm$ 0.05 (12)  &  0.50 $\pm$ 0.05 (12)  &  0.33 $\pm$ 0.04 (12)  &  0.27 $\pm$ 0.04 (12)  &  0.49 $\pm$ 0.15 (12)  &  \ldots \\ 
L0.5  &  2.38 $\pm$ 0.19 (5)  &  1.17 $\pm$ 0.02 (4)  &  3.15 $\pm$ 0.06 (6)  &  0.73 $\pm$ 0.06 (7)  &  0.53 $\pm$ 0.06 (7)  &  0.33 $\pm$ 0.02 (7)  &  0.26 $\pm$ 0.02 (7)  &  0.45 $\pm$ 0.07 (7)  &  \ldots \\ 
L1.0  &  2.41 $\pm$ 0.10 (14)  &  1.20 $\pm$ 0.04 (14)  &  3.18 $\pm$ 0.10 (15)  &  0.82 $\pm$ 0.05 (15)  &  0.51 $\pm$ 0.06 (15)  &  0.36 $\pm$ 0.05 (15)  &  0.24 $\pm$ 0.04 (15)  &  0.46 $\pm$ 0.10 (14)  &  \ldots \\ 
L1.5  &  2.39 $\pm$ 0.03 (5)  &  1.23 $\pm$ 0.06 (4)  &  3.23 $\pm$ 0.06 (7)  &  0.80 $\pm$ 0.08 (8)  &  0.52 $\pm$ 0.04 (8)  &  0.40 $\pm$ 0.09 (8)  &  0.27 $\pm$ 0.03 (8)  &  0.55 $\pm$ 0.11 (6)  &  \ldots \\ 
L2.0  &  2.34 $\pm$ 0.08 (12)  &  1.16 $\pm$ 0.03 (11)  &  3.24 $\pm$ 0.10 (12)  &  0.85 $\pm$ 0.09 (14)  &  0.53 $\pm$ 0.07 (14)  &  0.40 $\pm$ 0.06 (14)  &  0.27 $\pm$ 0.02 (14)  &  0.54 $\pm$ 0.15 (14)  &  \ldots \\ 
L2.5  &  2.31 $\pm$ 0.05 (5)  &  1.20 $\pm$ 0.04 (4)  &  3.31 $\pm$ 0.07 (5)  &  0.91 $\pm$ 0.04 (6)  &  0.54 $\pm$ 0.08 (6)  &  0.45 $\pm$ 0.06 (6)  &  0.26 $\pm$ 0.03 (6)  &  0.56 $\pm$ 0.32 (6)  &  \ldots \\ 
L3.0  &  2.34 $\pm$ 0.11 (5)  &  1.25 $\pm$ 0.01 (2)  &  3.48 $\pm$ 0.09 (3)  &  1.00 $\pm$ 0.08 (6)  &  0.64 $\pm$ 0.07 (6)  &  0.58 $\pm$ 0.09 (6)  &  0.33 $\pm$ 0.07 (6)  &  0.43 $\pm$ 0.17 (6)  &  \ldots \\ 
L3.5  &  2.40 $\pm$ 0.11 (2)  &  1.35 $\pm$ 0.04 (2)  &  3.37 $\pm$ 0.03 (2)  &  0.93 $\pm$ 0.05 (2)  &  0.58 $\pm$ 0.05 (2)  &  0.58 $\pm$ 0.06 (2)  &  0.30 $\pm$ 0.01 (2)  &  0.53 $\pm$ 0.13 (2)  &  \ldots \\ 
L4.0  &  2.22 $\pm$ 0.13 (4)  &  1.31 $\pm$ 0.07 (3)  &  3.50 $\pm$ 0.15 (4)  &  0.90 $\pm$ 0.09 (5)  &  0.57 $\pm$ 0.07 (5)  &  0.53 $\pm$ 0.09 (5)  &  0.27 $\pm$ 0.02 (5)  &  0.32 $\pm$ 0.14 (5)  &  \ldots \\ 
L4.5  &  2.06   (1)  &  \ldots  &  \ldots  &  1.13   (1)  &  0.69   (1)  &  0.82   (1)  &  0.37   (1)  &  0.66   (1)  &  \ldots \\ 
L5.0  &  2.15 $\pm$ 0.21 (5)  &  1.53   (1)  &  3.75   (1)  &  1.01 $\pm$ 0.16 (5)  &  0.59 $\pm$ 0.14 (5)  &  0.70 $\pm$ 0.09 (5)  &  0.38 $\pm$ 0.11 (6)  &  0.52 $\pm$ 0.33 (6)  &  \ldots \\ 
L5.5  &  2.13   (1)  &  \ldots  &  \ldots  &  0.88   (1)  &  0.56   (1)  &  0.67   (1)  &  0.28   (1)  &  0.62   (1)  &  \ldots \\ 
L6.0  &  2.07   (1)  &  1.71   (1)  &  3.49   (1)  &  0.69   (1)  &  0.34   (1)  &  0.75   (1)  &  0.34   (1)  &  0.76   (1)  &  \ldots \\ 
L6.5  &  1.92   (1)  &  \ldots  &  \ldots  &  1.00   (1)  &  0.59   (1)  &  0.76   (1)  &  0.37   (1)  &  0.61   (1)  &  \ldots \\ 
L7.0  &  \ldots  &  \ldots  &  \ldots  &  \ldots  &  \ldots  &  \ldots  &  \ldots  &  \ldots  &  \ldots \\ 
L7.5  &  2.14   (1)  &  \ldots  &  \ldots  &  1.18   (1)  &  0.70   (1)  &  0.88   (1)  &  0.47   (1)  &  1.10   (1)  &  \ldots \\ 
L8.0  &  \ldots  &  \ldots  &  \ldots  &  \ldots  &  \ldots  &  \ldots  &  \ldots  &  \ldots  &  \ldots \\
\end{longtable}
}
\end{landscape}

\begin{figure*}[]
    \centering
    \includegraphics[width=.49\linewidth]{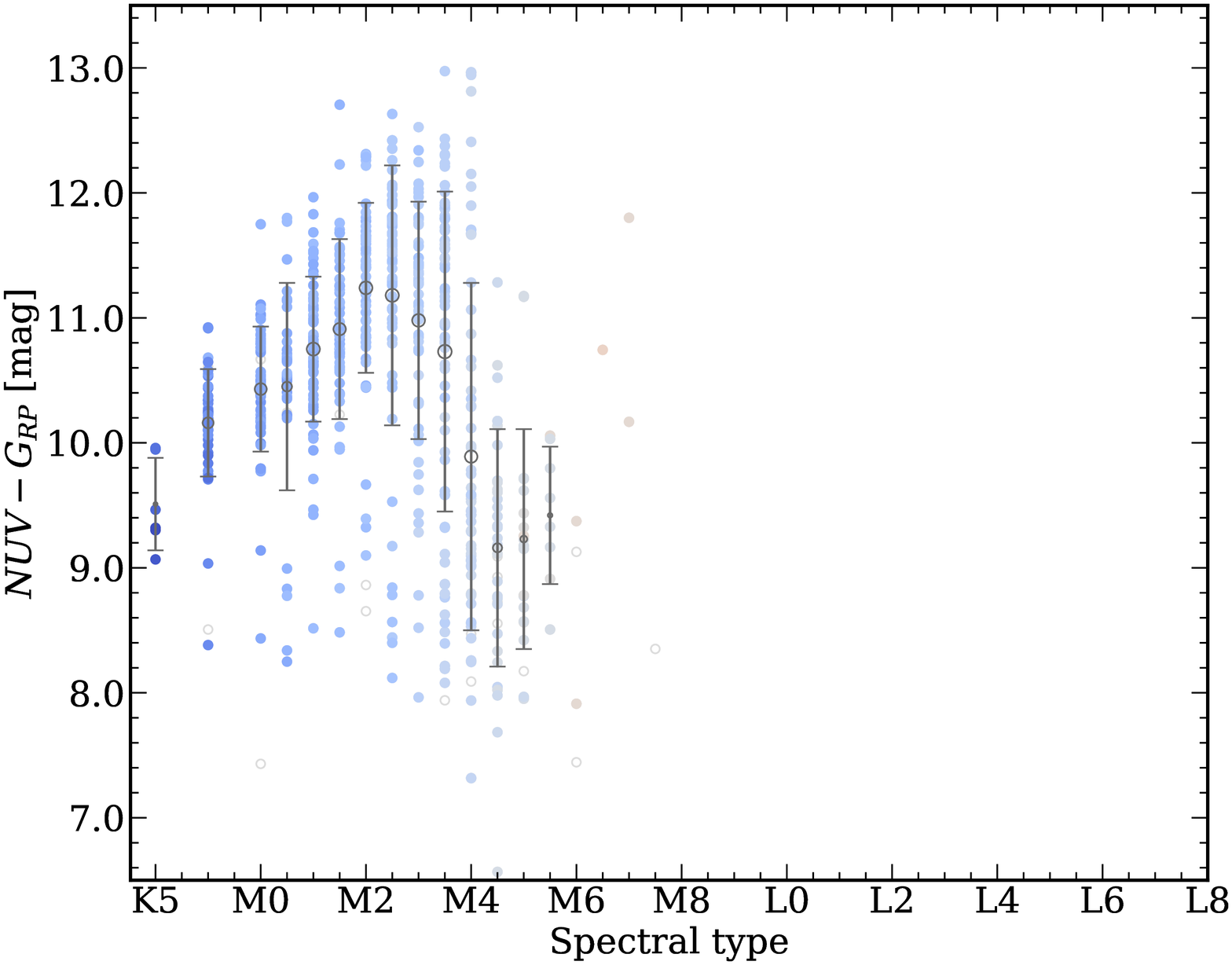}
    \includegraphics[width=.49\linewidth]{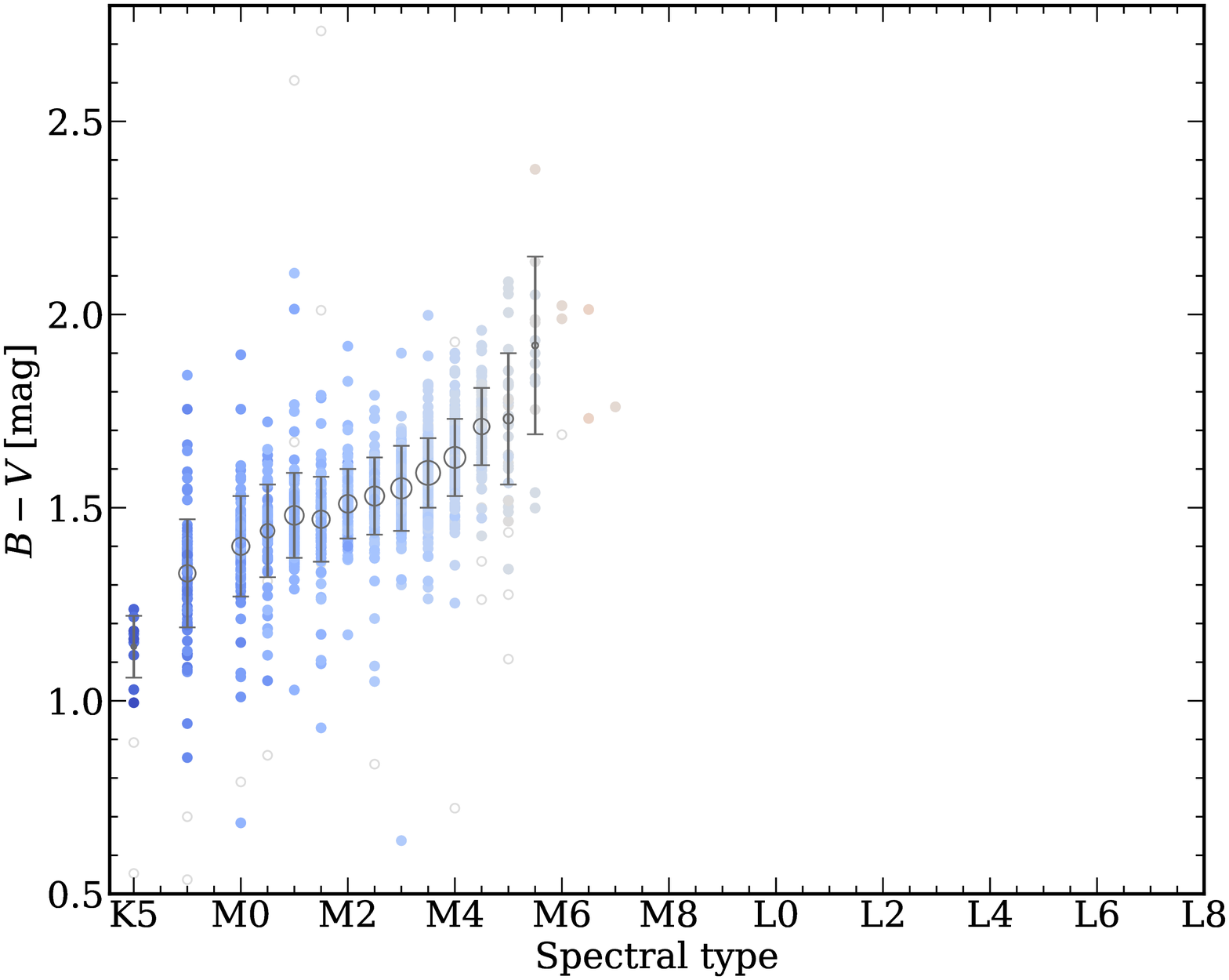}

    \includegraphics[width=.49\linewidth]{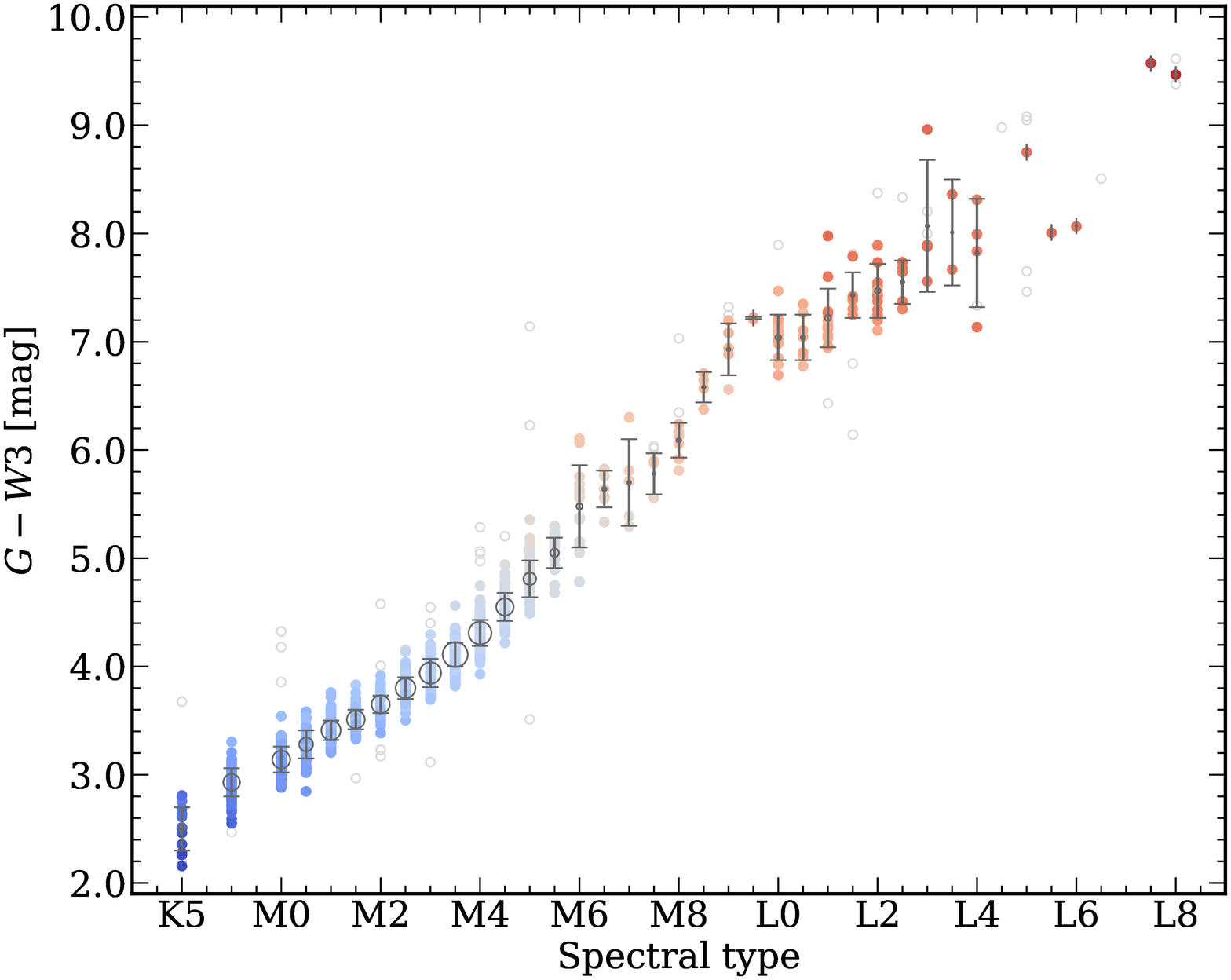}
    \includegraphics[width=.49\linewidth]{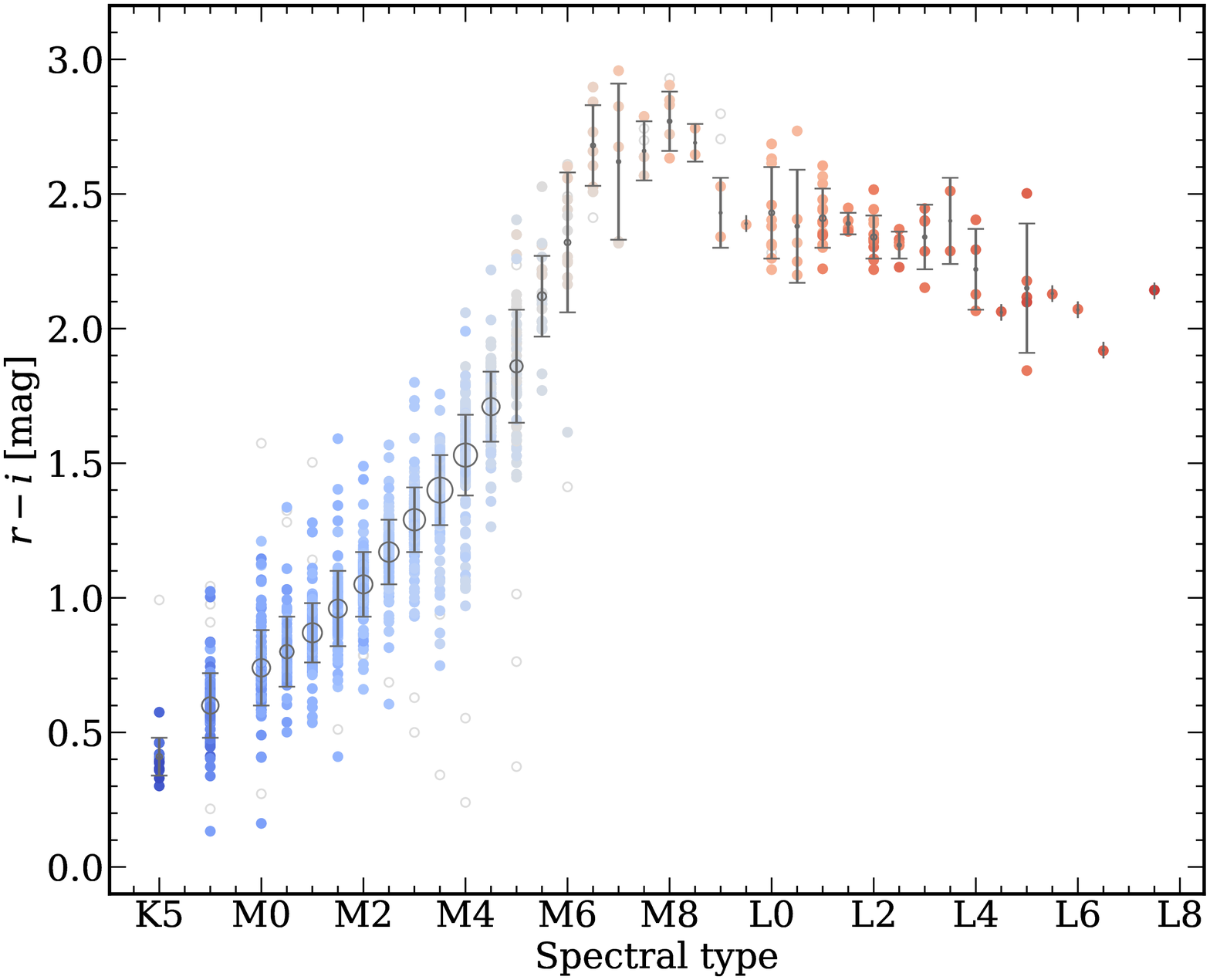}

    \includegraphics[width=.49\linewidth]{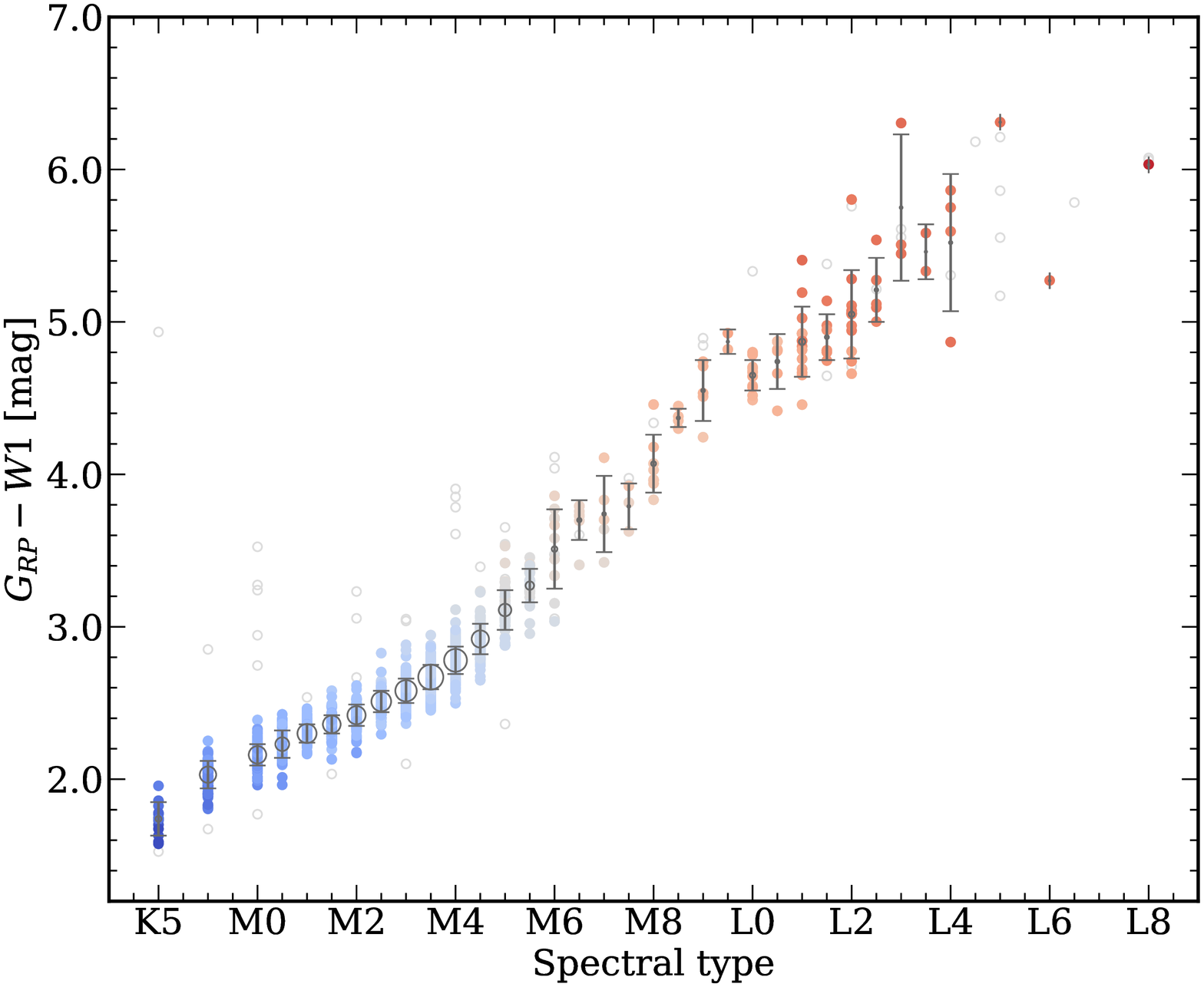}
    \includegraphics[width=.49\linewidth]{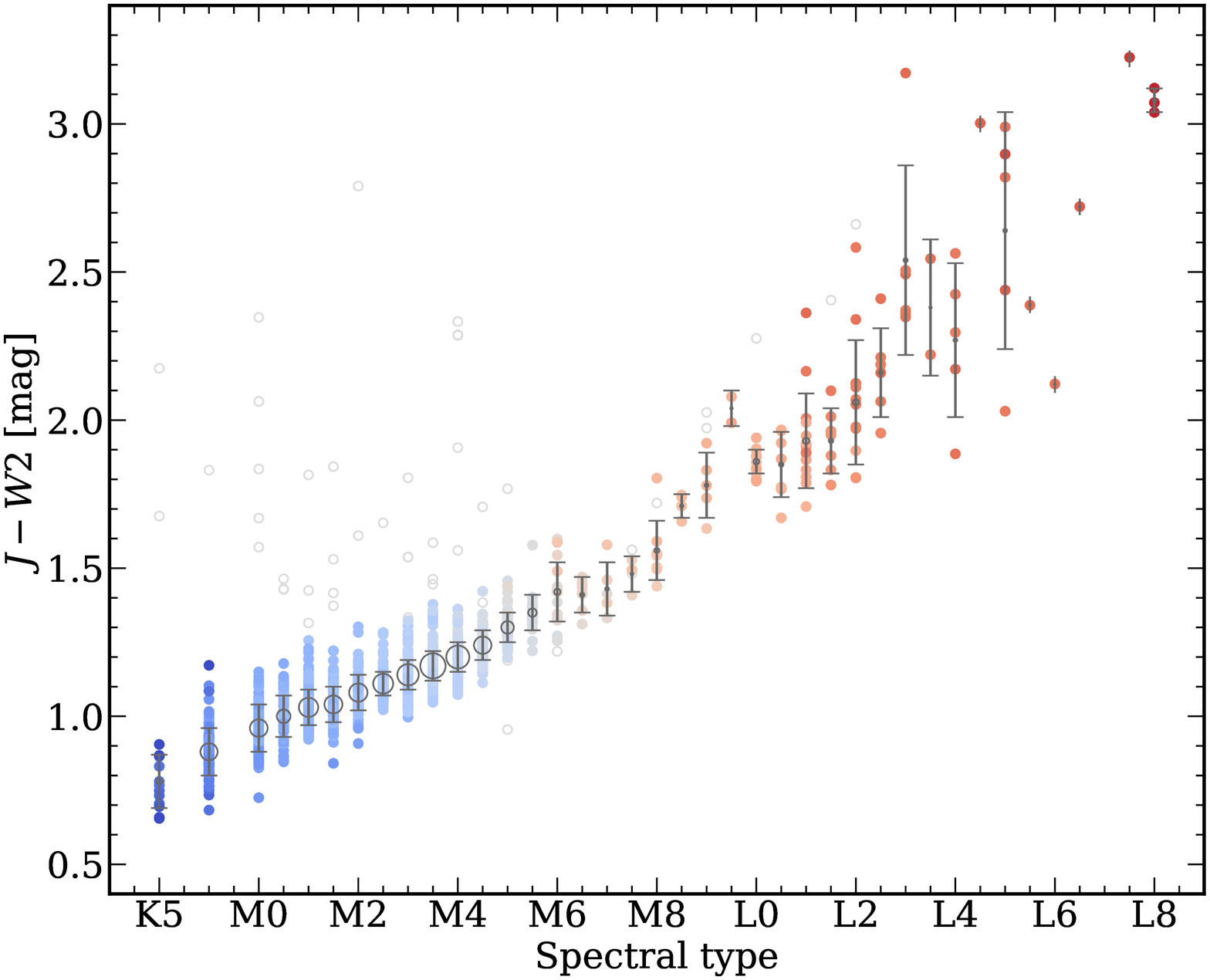}

    \includegraphics[width=.90\linewidth]{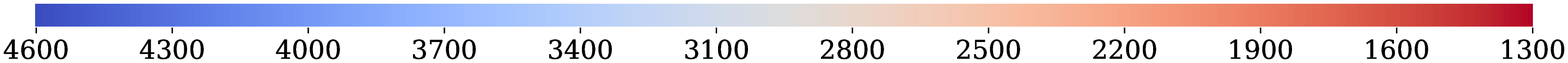}
    \caption{Six representative colour-spectral type diagrams, colour-coded by effective temperature.}
    \label{fig:colourSpT}
\end{figure*}

\begin{figure*}[]
    \centering
    \includegraphics[width=.49\linewidth]{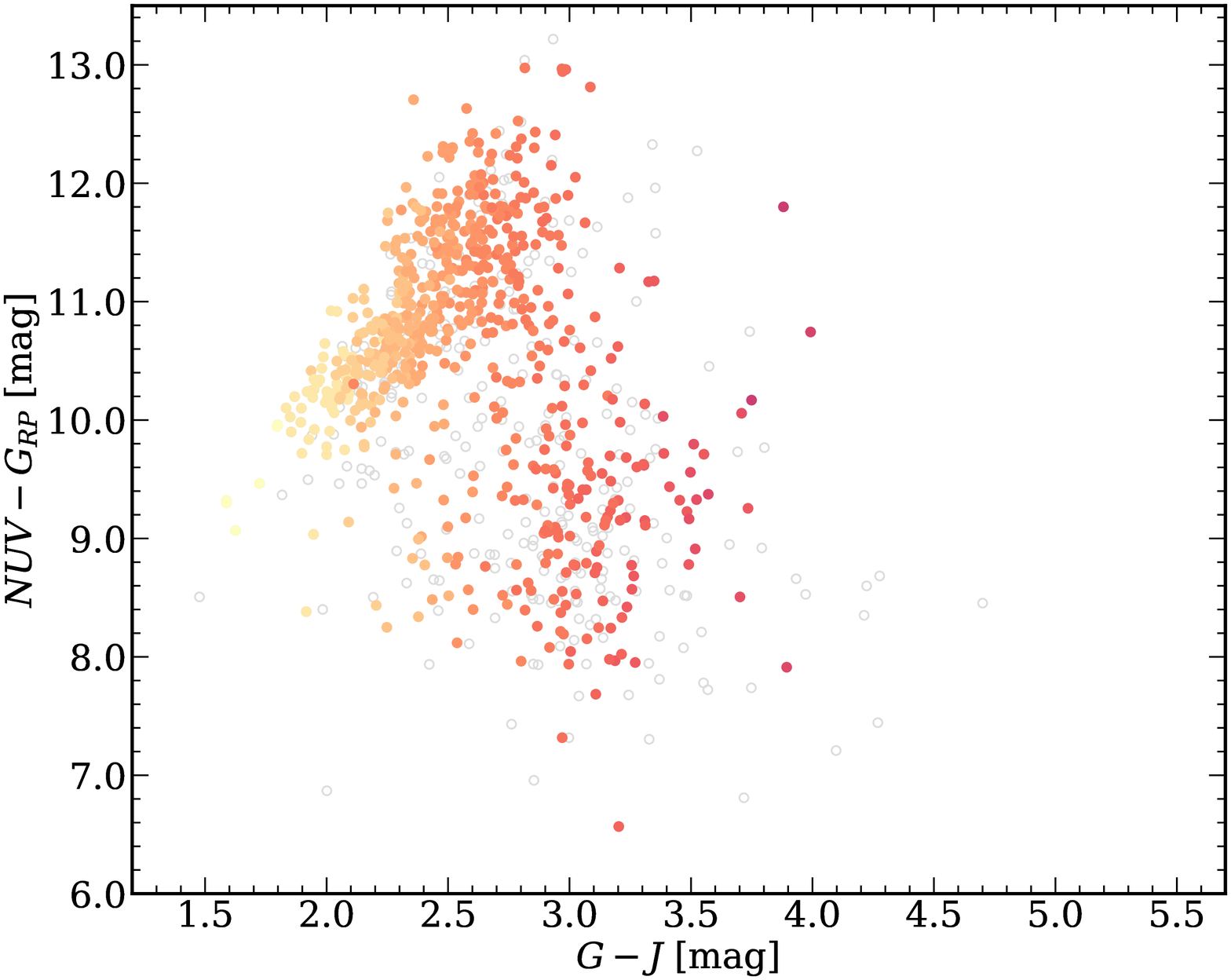}
    \includegraphics[width=.49\linewidth]{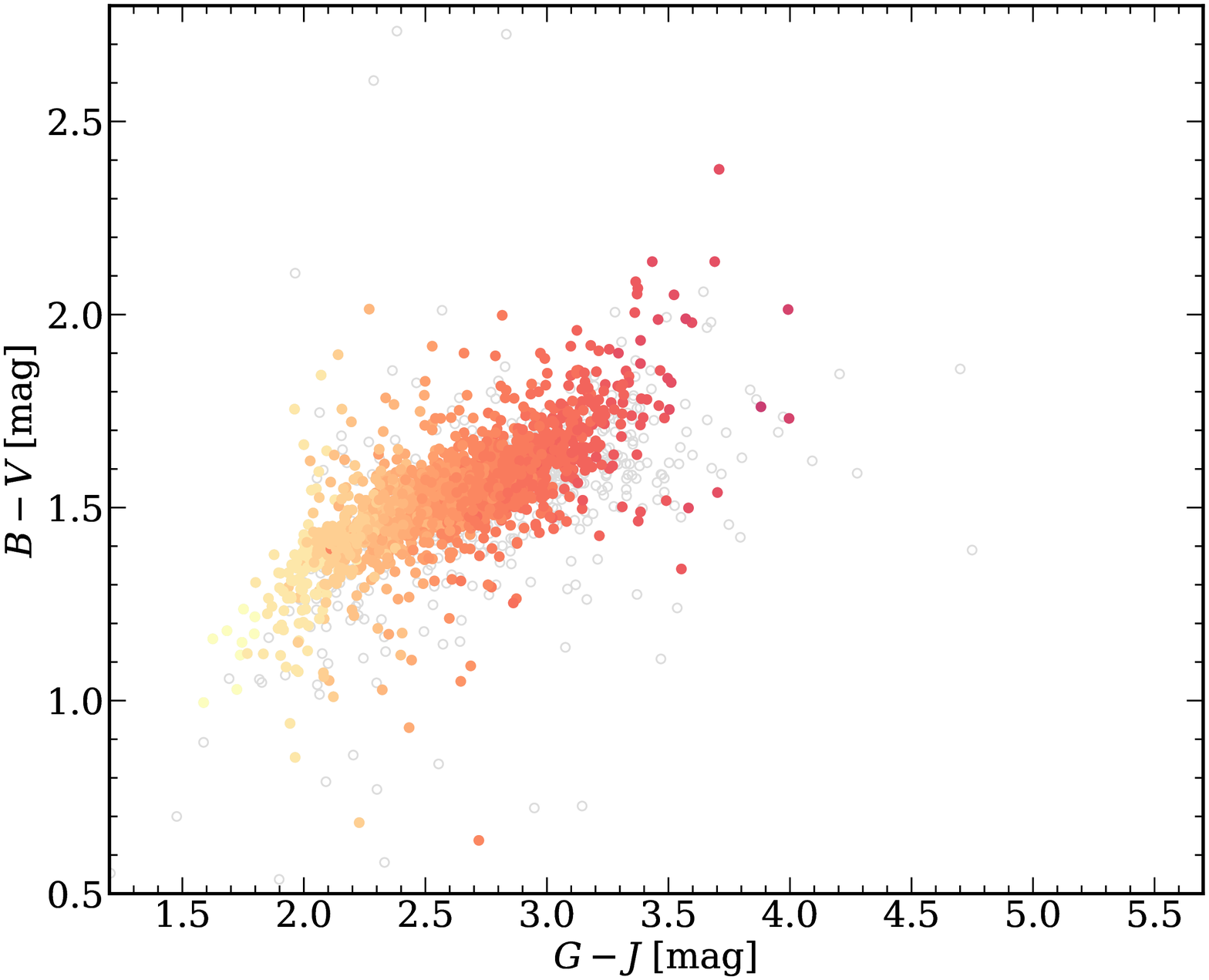}

    \includegraphics[width=.49\linewidth]{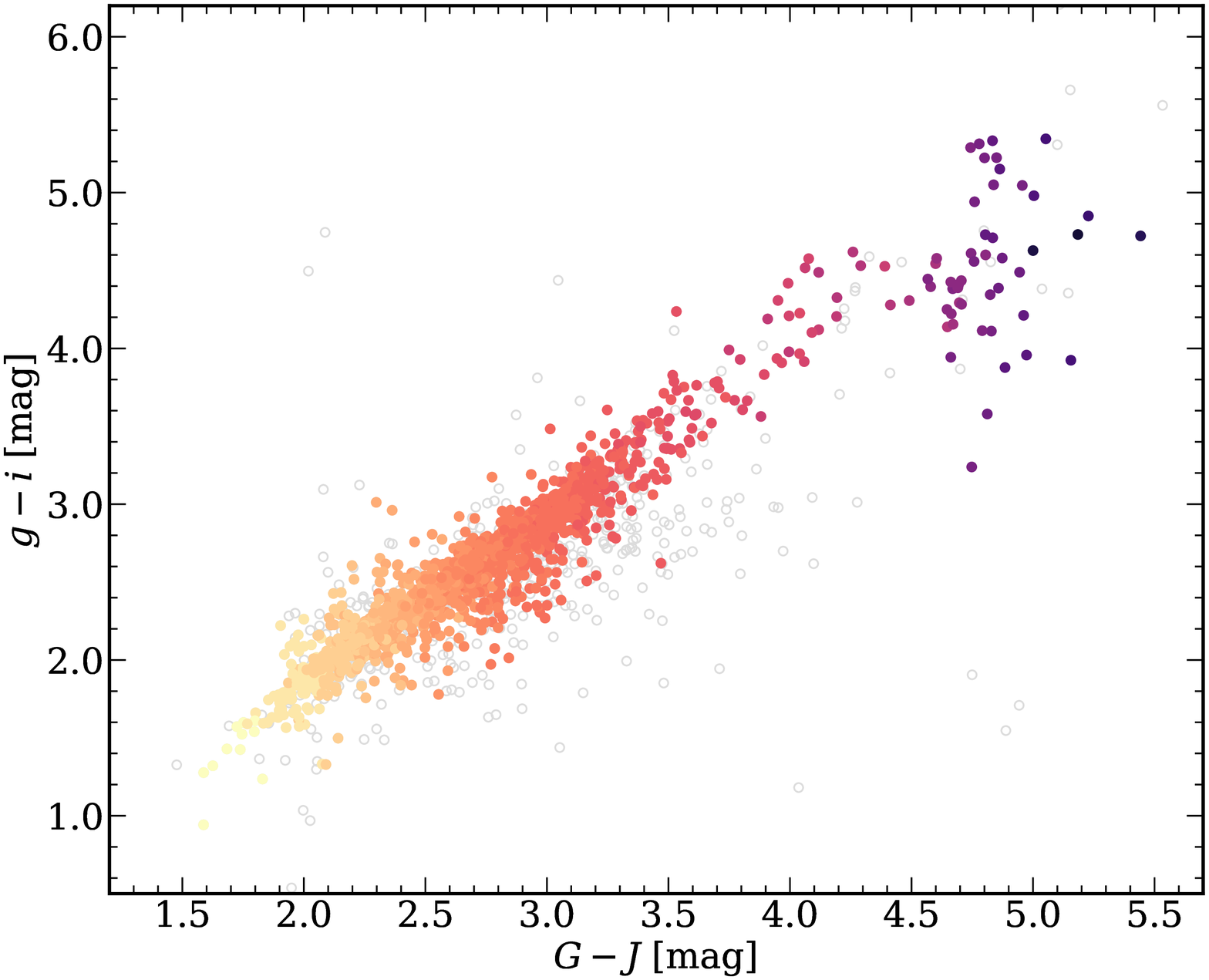}
    \includegraphics[width=.49\linewidth]{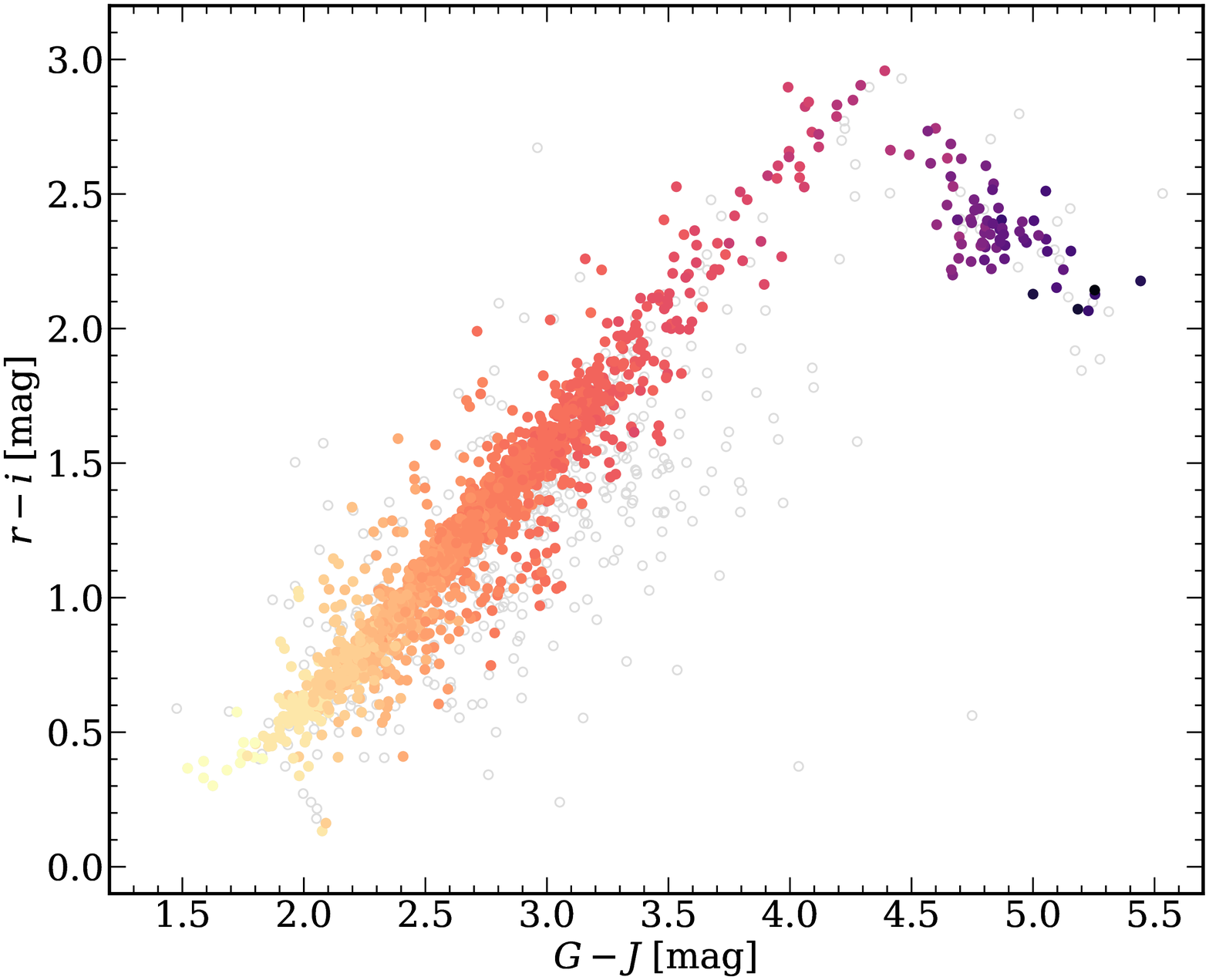}

    \includegraphics[width=.49\linewidth]{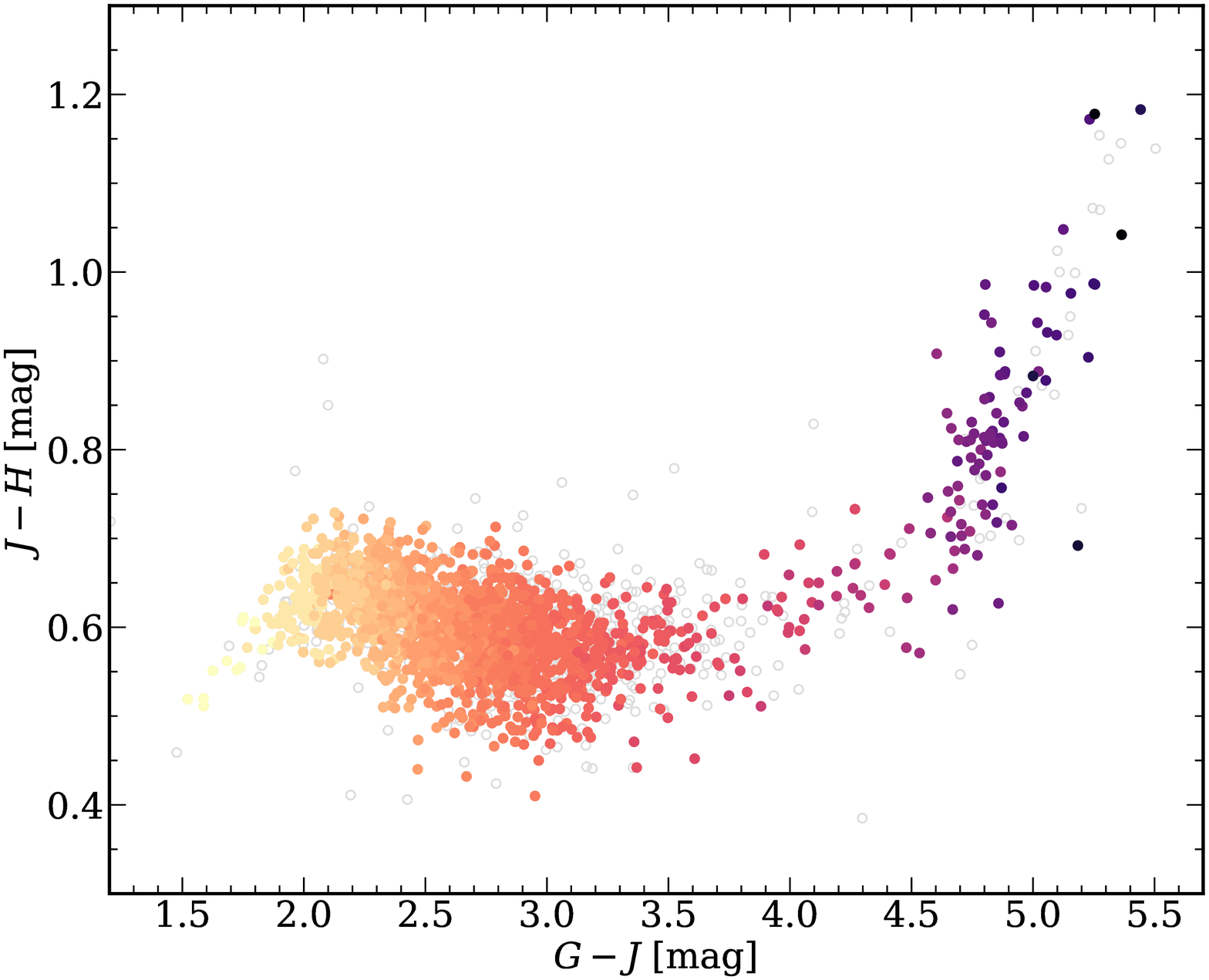}
    \includegraphics[width=.49\linewidth]{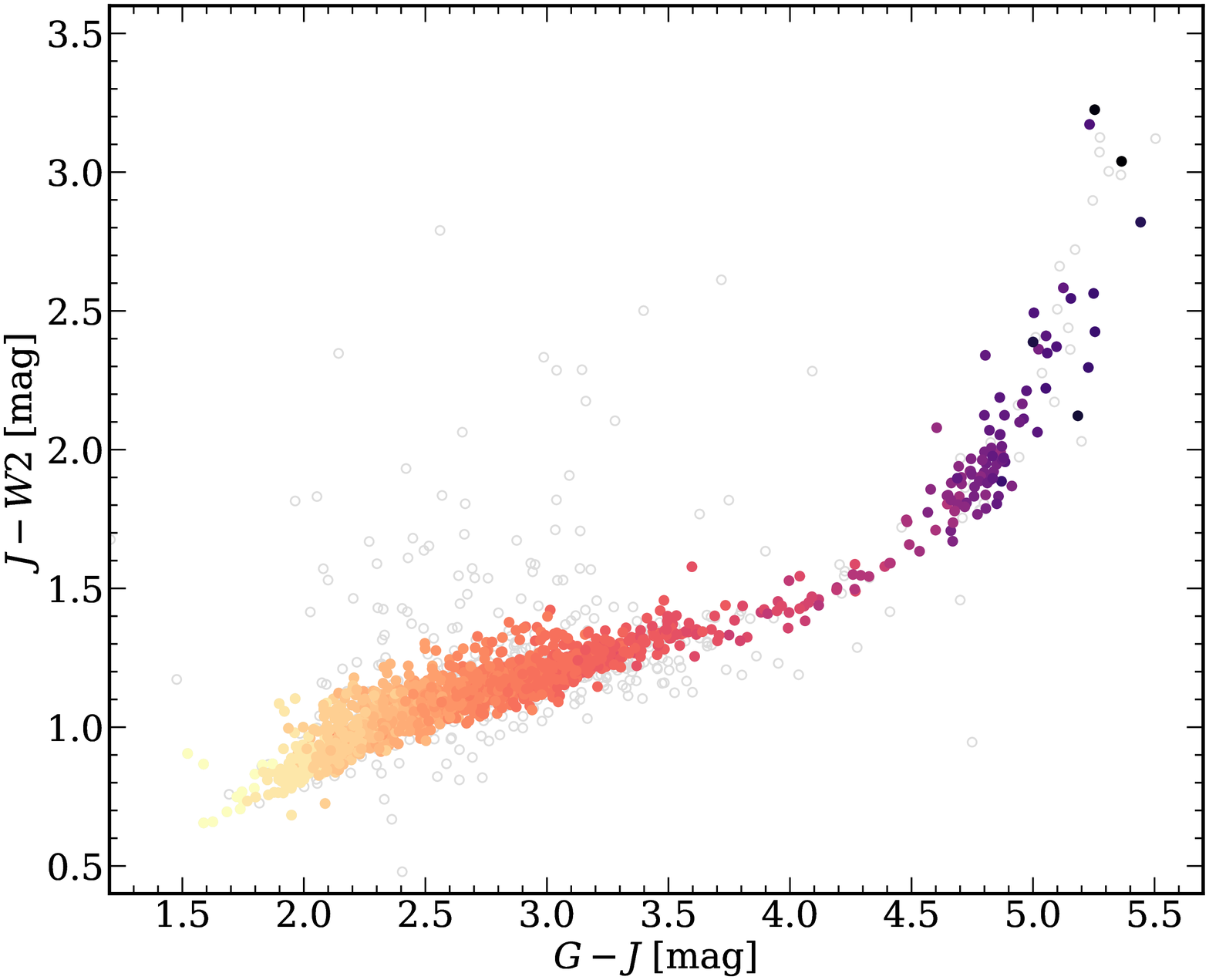}

    \includegraphics[width=.90\linewidth]{cif20_colorbar_SpT_bis.eps}
    \caption{Six representative colour-colour diagrams, colour-coded by spectral type.}
    \label{fig:colourcolour}
\end{figure*}

\end{document}